\documentclass[useAMS,usenatbib]{mn2e}
\usepackage[usenames,dvipsnames]{color}
\usepackage{graphicx}
\usepackage{natbib}
\usepackage{subfigure}

\long\def\symbolfootnote[#1]#2{\begingroup%
\def\thefootnote{\fnsymbol{footnote}}\footnote[#1]{#2}\endgroup}

\def\nABSa{503} 
\def\nABSb{464}
\def\nABSc{402}
\def\nABSd{0.06}
\def\nABSe{1.46} 
\def\nABSfi{0.4} 
\def\nABSf{14.7}
\def\nABSg{255} 
\def\nABSh{356}    
\def\nABSi{261} 
\def\nABSj{10}
\def\nABSk{1.0} 
\def\nABSl{67}
\def\nABSm{5}
\def\nABSn{131} 
\def\nABSo{2} 
\def\nABSp{27}  
\def\nABSq{78}
\def\nABSr{40} 
\def\nABSs{105} 

\def\nINTRa{5,776} 
\def\nINTRb{3,675} 
\def\nINTRc{993} 

\def\nZNEWa{154} 
\def\nZNEWb{415} 
\def\nZNEWc{90} 
\def\nZNEWd{11}
\def\nZNEWe{366}
\def\nZNEWf{30} 
\def\nZNEWg{224} 

\def\nZNEWh{35} 
\def\nZNEWi{265}

\def\nZARCa{42} 

\def\nZARCb{1,721} 
\def\nZARCc{69} 
\def\nZARCd{574}
\def\nZARCe{51} 

\def\nZARCf{5} 
\def\nZARCg{265} 
\def\nZARCh{493} 
\def\nZARCi{142} 
\def\nZARCk{112} 
\def\nZARCm{345}

\def\nZARCn{4} 
\def\nZARCo{11} 
\def\nZARCp{6} 
\def\nZARCq{8} 

\def\nZARCr{229} 
\def\nZARCs{210} 

\def\nQCONTa{484}
\def\nQCONTb{1,151}
\def\nQCONTc{82} 
\def\nQCONTd{310} 
\def\nQCONTe{330} 
\def\nQCONTf{766}
\def\nQCONTg{415}
\def\nQCONTh{19} 
\def\nQCONTi{17} 
\def\nQCONTj{81}
\def\nQCONTk{306}

\def\nQCONTl{43} 
 
\def\nQCONTm{13} 
\def\nQCONTn{103} 

\def\nCCATa{255}
\def\nCCATb{94} 
\def\nCCATc{two}
\def\nCCATd{one} 
\def\nCCATi{127}
\def\nCCATj{111} 
\def\nCCATk{74}
\def\nCCATl{80} 
\def\nCCATm{35} 
\def\nCCATn{19}  
\def\nCCATo{19}
\def\nCCATp{3}
\def\nCCATq{219} 
\def\nCCATr{265}
\def\nCCATs{224}
\def\nCCATt{574}
\def\nCCATu{51} 

\def\nZCOMPa{39}
\def\nZCOMPb{138} 
\def\nZCOMPc{14}

\def\nCCATe{248} 
\def\nCCATf{67}

\def\nCCATg{147} 
\def\nCCATh{346} 

\def\nDISCa{467} 
\def\nDISCb{87} 
\def\nDISCd{34}

\def\nDISCe{387} 
\def\nDISCf{83}	
\def\nDISCg{51}

\def\nDISCw{23}		

\def\nDISCh{16} 
\def\nDISCi{10} 
\def\nDISCj{11} 

\def\nDISCk{0.5} 
\def\nDISCl{31}
\def\nDISCm{25} 
\def\nDISCn{39} 
\def\nDISCo{158} 

\def\nDISCp{27} 

\def\nDISCq{35}

\def\nDISCs{100} 
\def\nDISCt{156} 
 
\def\nDISCu {36}
\def\nDISCv {69} 
 
\title[The \textit{XMM} Cluster Survey: Optical analysis methodology and the first data release]{The \textit{XMM} Cluster Survey: Optical analysis methodology and the first data release}
\author[N. Mehrtens et al. (XCS collaboration)]
{Nicola Mehrtens$^{1}$\thanks{E-mail:n.mehrtens@sussex.ac.uk},
A.~Kathy Romer$^{1,2}$,
E.~J.~Lloyd-Davies$^{1}$,
Matt Hilton$^{3,4}$, 
\newauthor
Christopher~J.~Miller$^{5}$, 
S.~A.~Stanford$^{6,7}$, 
Mark Hosmer$^{1}$, 
Ben Hoyle$^{8,9,10}$,
\newauthor
Chris~A.~Collins$^{11}$,
Andrew~R.~Liddle$^{1,2}$, 
Pedro~T. P.~Viana$^{12,13}$,
\newauthor
Robert C. Nichol$^{2,9}$, 
John~P.~Stott$^{11,14}$,
E.~Naomi Dubois$^{1,2}$,
Scott~T.~Kay$^{15}$, 
\newauthor
Martin~Sahl{\'e}n$^{16}$,
Owain Young$^{1}$,
C.~J.~Short$^{1}$,
L.~Christodoulou$^{1}$,
\newauthor
William A.~Watson$^{1,2}$,
Michael Davidson$^{17}$,
Craig D.~Harrison$^{5}$,
\newauthor
Leon~Baruah$^{1}$,
Mathew~Smith$^{9,18}$,
Claire~Burke$^{11}$,
\newauthor
Paul-James Deadman$^{1}$,
Philip J.~Rooney$^{1}$,
Edward M.~Edmondson$^{9}$,
\newauthor
Michael West$^{19}$,
Heather C.~Campbell$^{1,9}$,
Alastair C.~Edge$^{14}$,
\newauthor
Robert G.~Mann$^{17}$,
David Wake$^{14,20}$,
Christophe Benoist$^{21,22}$,
\newauthor
Luiz da Costa$^{22,23}$,
Marcio A.~G.~Maia$^{22,23}$,
and Ricardo Ogando$^{22,23}$.\\\\
$^{1}$~Astronomy Centre, University of Sussex, Falmer, Brighton, BN1 9QH, UK\\
$^{2}$~SEPnet, South East Physics Network, (www.sepnet.ac.uk)\\
$^{3}$~Astrophysics \& Cosmology Research Unit, School of Mathematical
Sciences, University of KwaZulu-Natal, \\
Private Bag X54001, Durban 4000, South Africa\\
$^{4}$~University of Nottingham, School of Physics \& Astronomy,
Nottingham, NG7 2RD, UK\\ 
$^{5}$~Astronomy Department, University of Michigan, Ann Arbor, MI
48109, USA\\ 
$^{6}$~Physics Department, University of California, Davis, CA 95616, USA\\ 
$^{7}$~Institute of Geophysics and Planetary Physics, Lawrence
Livermore National Laboratory, Livermore,\\ 
CA 94551, USA\\
$^{8}$~Institut de Ci\`{e}ncies del Cosmos (ICCUB-IEEC), Departmento de
F\'{\i}sica, Mart\'{\i} i Franqu\'{e}s 1, 08034\\ 
Barcelona, Spain\\ 
$^{9}$~Institute of Cosmology and Gravitation, Dennis Sciama Building,
Burnaby Road, Portsmouth, PO1 3FX, UK\\ 
$^{10}$~Helsinki Institute of Physics, P.O. Box 64, FIN-00014 University of Helsinki, Finland\\
$^{11}$~Astrophysics Research Institute, Liverpool John Moores
University, Twelve Quays House, Egerton Wharf,\\ 
Birkenhead, CH41 1LD, UK\\ 
$^{12}$~Centro de Astrof\'{\i}sica da Universidade do Porto, Rua das
Estrelas, 4150-762, Porto, Portugal\\
$^{13}$~Departamento de F\'{\i}sica e Astronomia, Faculdade de
Ci\^{e}ncias, Universidade do Porto, Rua do Campo\\
Alegre, 687, 4169-007 Porto, Portugal\\
$^{14}$~Department of Physics, Institute for Computational Cosmology, Durham University, South Road, Durham,\\
DH1 3LE, UK\\
$^{15}$~Jodrell Bank Centre for Astrophysics, School of Physics and Astronomy, The University of Manchester,\\ Manchester, M13 9PL, UK \\
$^{16}$~The Oskar Klein Centre for Cosmoparticle Physics, Department
of Physics, Stockholm University, AlbaNova,\\
SE-106 91 Stockholm, Sweden\\
$^{17}$~SUPA, Institute for Astronomy, University of Edinburgh, Royal
Observatory, Edinburgh, EH9 3HJ, UK\\ 
$^{18}$~Astrophysics, Cosmology and Gravity Centre (ACGC), Deptarment of Mathematics and Applied Mathematics, \\
University of Cape Town, Rondebosch 7701, South Africa\\
$^{19}$~ESO, Alonso de Cordova 3107, Vitacura, Santiago, Chile\\
$^{20}$~Astronomy Department, Yale University, New Haven, CT, 06520, USA\\
$^{21}$~Universit\'e de Nice Sophia-Antipolis, CNRS, Observatoire de la C\^ote d'Azur, UMR 6202 CASSIOP\'EE, \\
BP 4229, F-06304 Nice Cedex 4, France\\
$^{22}$~Laborat\'orio Interinstitucional de e-Astronomia - LIneA, Rua Gal. Jos\'e Cristino 77, Rio de Janeiro, \\
RJ - 20921-400, Brazil\\
$^{23}$~Observat\'orio Nacional, R. Gal. Jos\'e Cristino 77, BR Rio de Janeiro, RJ 20921-400, Brazil
}

\date{Accepted 2011 ??.  
      Received 2011 ??; 
      in original form 2011 ??}

\pagerange{\pageref{firstpage}--\pageref{lastpage}}
\pubyear{2011}

\begin{document}

\maketitle

\label{firstpage}

\begin{abstract}

The \textit{XMM} Cluster Survey (XCS) is a serendipitous search for galaxy clusters using all publicly available data in the \textit{XMM--Newton} Science Archive. Its main aims are to measure cosmological parameters and trace the evolution of X-ray scaling relations. In this paper we present the first data release from the \textit{XMM} Cluster Survey (XCS-DR1). This consists of \nABSa\, optically confirmed, serendipitously detected, X-ray clusters. Of these clusters, \nABSg\, are new to the literature and \nABSh\, are new X-ray discoveries. We present \nABSb\, clusters with a redshift estimate ($\nABSd<z<\nABSe$), including \nABSi\, clusters with spectroscopic redshifts. In addition, we have measured X-ray temperatures ($T_{\rm X}$) for \nABSc\, clusters ($\nABSfi < T_{\rm X} < \nABSf$ keV). We highlight seven interesting subsamples of XCS-DR1 clusters: (i) \nABSj\, clusters at high redshift ($z>\nABSk$, including a new spectroscopically-confirmed cluster at $z=1.01$); (ii) \nABSl\, clusters with high $T_{\rm X}$ ($>\nABSm$ keV); (iii) \nABSn\, clusters/groups with low $T_{\rm X}$ ($<\nABSo$ keV); (iv) \nABSp\, clusters with measured $T_{\rm X}$ values in the SDSS `Stripe 82' co-add region; (v) \nABSq\, clusters with measured $T_{\rm X}$ values in the Dark Energy Survey region; (vi) \nABSr\, clusters detected with sufficient counts to permit mass measurements (under the assumption of hydrostatic equilibrium); (vii) \nABSs\, clusters that can be used for applications such as the derivation of cosmological parameters and the measurement of cluster scaling relations. The X-ray analysis methodology used to construct and analyse the XCS-DR1 cluster sample has been presented in a companion paper, \cite{LD10}.


\end{abstract}


\section{Introduction}
\label{introduction}

\begin{figure*}
\includegraphics[width=9.5cm]{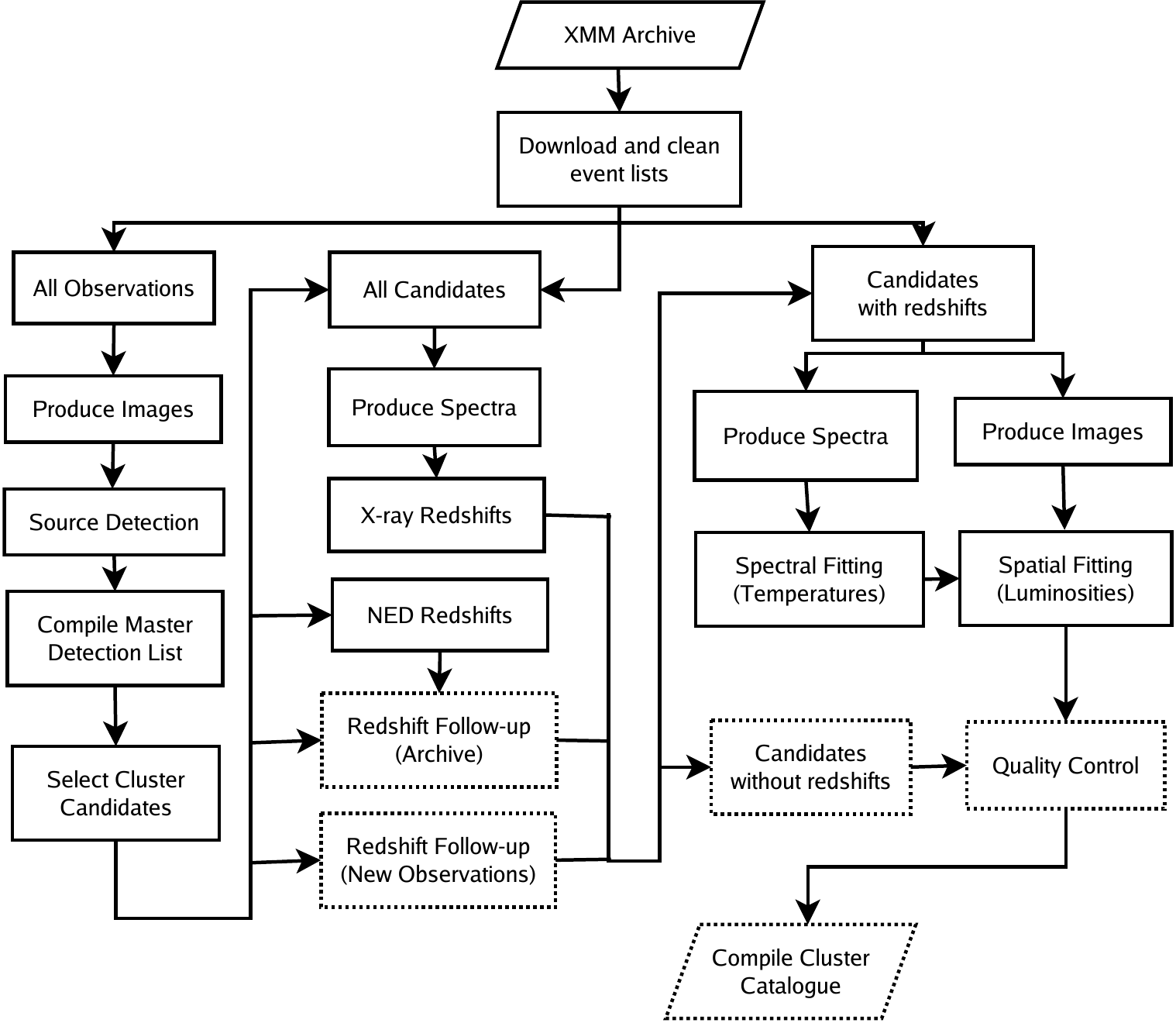}
\caption{Figure taken from LD10: \textit{Flowchart showing an overview of the XCS analysis methodology. This 
illustrates the sequence by which data from the \emph{XMM} archive is used
to create a catalogue of galaxy clusters.} The components indicated with dashed outlines are described in this manuscript, the remainder are described in LD10.
\label{ELDFig1}}
\end{figure*}

Clusters of galaxies provide an opportunity to explore the underlying cosmological model and the processes governing structure formation (see \citealt{Voit:2005Rv, Allen-2011} for reviews) and so several large-area surveys for clusters are currently underway.  In this paper we present the \textit{XMM} Cluster Survey (XCS), a search for serendipitous galaxy clusters in archival \textit{XMM--Newton}\footnote{http://xmm.esac.esa.int} (\textit{XMM} hereafter) data, using the signature of X-ray extent. The original XCS concept was described in \citet{Romer:2001}. The main goals of the survey are to (\textit{i}) measure cosmological parameters, (\textit{ii}) measure the evolution of the X-ray luminosity--temperature scaling relation ($L_{\rm X} - T_{\rm X}$ relation, hereafter), (\textit{iii}) study galaxy properties in clusters to high redshift, and (\textit{iv}) provide the community with a high quality, homogeneously selected, X-ray cluster sample. XCS highlights to date include the detection, and subsequent multi-wavelength follow-up, of a $z = 1.46$ cluster (XMMXCS J2215.9$-$1738; \citealt{Stanford:2006, Hilton-2007, Hilton-2009, Hilton-2010}), studies of galaxy evolution in high-redshift clusters \citep{collins-2009-458,Stott:2010}, and forecasts of the performance of XCS for cosmological parameter estimation and cluster scaling relations (\citealt{Sahlen-2009}, S09 hereafter). In a companion paper, (\citealt{LD10}, henceforth LD10), we describe the XCS X-ray data analysis strategy, including the XCS Automated Pipeline Algorithm ({\sc Xapa} hereafter). In this paper we describe the corresponding optical data analysis strategy and present the first XCS data release (XCS-DR1 hereafter). A schematic of the XCS methodology is reproduced (from LD10) in Fig.~\ref{ELDFig1}. The components indicated with solid outlines were discussed in LD10. Those with dashed outlines are discussed herein: Redshift Follow-up (New Observations) in Section~\ref{z-new}; Redshift Follow-up (Archive) in Section~\ref{z-archive}; Quality Control in Section~\ref{Quality Control}; and Compile Cluster Catalogue in Section~\ref{The catalogue}. Summaries and discussions are presented in Section~\ref{Discussion}. Brief conclusions are made in Section~\ref{conclusions}.

In this paper we have relied heavily on the red-sequence, or colour--magnitude relation (CMR), technique to derive photometric redshifts using one colour ($r-z$) CCD imaging \citep{Ostrander-1998,GY00,Lopez-Cruz-2004}. This technique takes advantage of the fact that cluster cores are populated with passively-evolving elliptical galaxies that dominate the bright end of the luminosity function. The mass--metallicity relation of these ellipticals, when expressed in colour--magnitude space, has only a small intrinsic scatter \citep{Sandage-1978, Bower-1992, Kodama-1997, Stanford-1998, Lopez-Cruz-2004} and has become known as the E/S0 ridgeline or the cluster red-sequence. The red-sequence has been found to be remarkably homogeneous between clusters at the same redshift and has been detected to $z > 1$ \citep{Lidman-2008, mei-2009, Hilton-2009, Papovich-2010}, meaning it can be used as a tool for cluster detection out to high redshifts (e.g. \citealt{GY00, gladders-2004, Wilson-2009, Muzzin-2009, Demarco-2010, Papovich-2010}). The red-sequence can also be used to measure cluster redshifts because, by using appropriately placed filters, one can track the migration of the 4000\AA\, break feature in the spectrum of passive ellipticals. It is this redshift application that we make use of in XCS. 

We note that, in the following, we have used the following terms in an XCS-specific manner;  count, ObsID, candidate,  candidate$^{300}$, and cluster$^{300}$.  `Count' is used as a shortening of the phrase `background-subtracted (0.5-2.0 keV) photon count as determined by \textsc{Xapa}'. As explained in LD10, these count values have not been corrected for photons falling outside the \textsc{Xapa} defined aperture (that is done during an additional spatial fitting step once the cluster redshifts are known). The count values pertain to the number of photons gathered from a single ObsID, where `ObsID' is used herein to refer to each of the complex sets of \emph{XMM} exposures and calibration files that comprise the \nINTRa\, \emph{XMM} observations processed so far by XCS. If a candidate was detected in multiple ObsIDs, then the highest recorded count is used. `Candidate' is used with reference to the LD10 definition of an XCS cluster candidate, i.e. a {\sc Xapa}-detected \textit{XMM} source, detected with 50 or more counts, that has been classified -- without warning flags -- as being more extended than the instrument point spread function (PSF). Moreover, candidates must not lie in the Galactic plane or near the Magellanic Clouds. Candidates must have also passed the target filters, i.e. are genuine serendipitous detections (as far as we can tell using automated methods). To date, we have selected a total of \nINTRb\, candidates (LD10). A subset of \nINTRc\,, the `candidates$^{300}$', are of particular importance, as these were detected with 300 or more counts (see below). Similarly, `clusters$^{300}$' are candidates$^{300}$ that have been optically confirmed as clusters.

The significance of the 300 count threshold mentioned above is twofold. First, and most importantly -- because we require temperature measurements for most of our key scientific goals -- we have determined (LD10) that we can derive temperatures with acceptable errors for $T_{\rm X}>2$ keV clusters to this count limit (although we note that it is still possible to measure $T_{\rm X}$ values with fewer counts, especially for cool clusters/groups, and there are many such examples in XCS-DR1). Second, we have demonstrated using selection function simulations (LD10), that {\sc Xapa} will detect most ($> 70$ per cent) of the clusters$^{300}$ that lie within the field of view of an ObsID. 

Throughout, we assume a concordance cosmology  ($\Omega_{m} =0.3$, $\Omega_{\Lambda} = 0.7$, and $H_{0} = 70$ km s$^{-1}$ Mpc$^{-1}$) and error bounds are quoted by their $1\sigma$ limits. XCS reduced X-ray images and optical images (colour-composite and greyscale) of the XCS-DR1 clusters mentioned in the text can be viewed at http://xcs-home.org/datareleases (see Section~\ref{The catalogue}).

\section{Redshift follow-up (new observations)}
\label{z-new}

We have carried out several observing campaigns in order to measure redshifts for XCS clusters. We describe our photometric follow-up in Section~\ref{NXS}, the derivation of redshifts from that photometric follow-up in Section~\ref{redsequence algorithm}, and our spectroscopic follow-up in Section~\ref{spec zs}.

\subsection{The NOAO-\textit{XMM} Cluster Survey}
\label{NXS}

The NOAO--\textit{XMM} Cluster Survey, or NXS, was a two-band imaging survey that gathered data across the Northern and Southern Celestial hemispheres to $r \simeq 24$ over 38 nights.  It was carried out at the National Optical Astronomy Observatory (NOAO) 4-m facilities at Kitt Peak National Observatory (KPNO) and Cerro Tololo Inter-American Observatory (CTIO) during six observing campaigns between November 2005 and April 2008. Slightly more time (by two nights) was allocated to the Southern hemisphere due to larger optical archival coverage in the North.  During the NXS, both the KPNO and CTIO 4-m telescopes were equipped with wide-field MOSAIC CCD cameras. The KPNO MOSAIC I and CTIO MOSAIC II cameras consist of a $4\times2$ array of $2048\times4096$ pixel CCDs. These CCDs are separated by gaps of 35 pixels between columns and 50 pixels between rows. Both cameras are controlled by four \textsc{Arcon} CCD controllers that read out eight amplifiers for MOSAIC I (one per chip) and 16 amplifiers for MOSAIC II (two per chip). The similarity of the two MOSAIC instruments has allowed the NXS to produce a homogeneous data set across the sky. The MOSAIC I and II pixel scale of 0.26\,arcsec/pixel and 0.27\,arcsec/pixel respectively, provides a total imaging area of 0.36\,deg$^{2}$ and 0.38\,deg$^{2}$ on the sky. By comparison, the \textit{XMM} field of view is circular with a diameter of 30 arcmin. Thus, each NXS image encompasses one \textit{XMM} image. Since each ObsID typically contains multiple candidates, we opted to centre the MOSAIC camera near the aim-point of the ObsID, rather than on a specific candidate.  With several thousand candidates to choose from, and only a limited number of nights at our disposal, emphasis was placed on imaging candidates$^{300}$ when possible.

\begin{figure*}
\begin{center}
{
\includegraphics[scale=0.21]{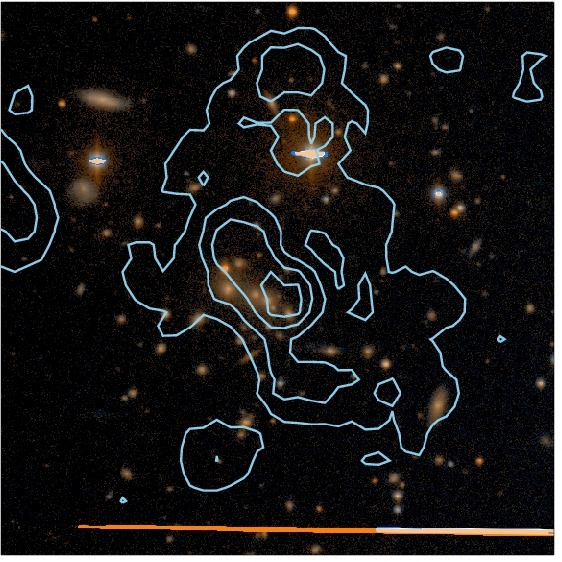}
}
\subfigure
{
 \includegraphics[scale=0.21]{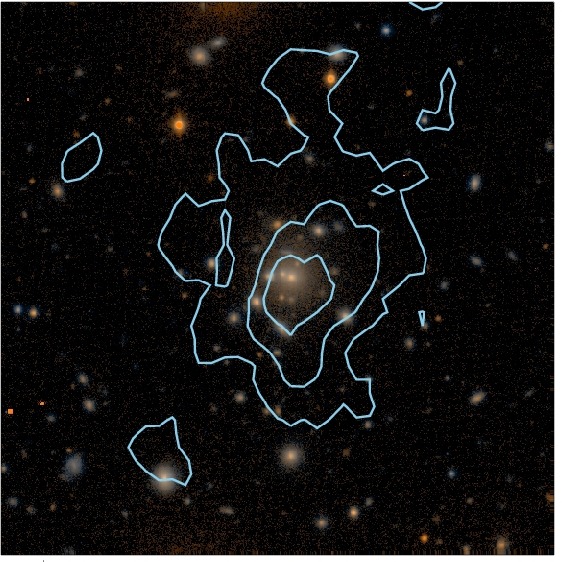}
}
\subfigure
{
\includegraphics[scale=0.21]{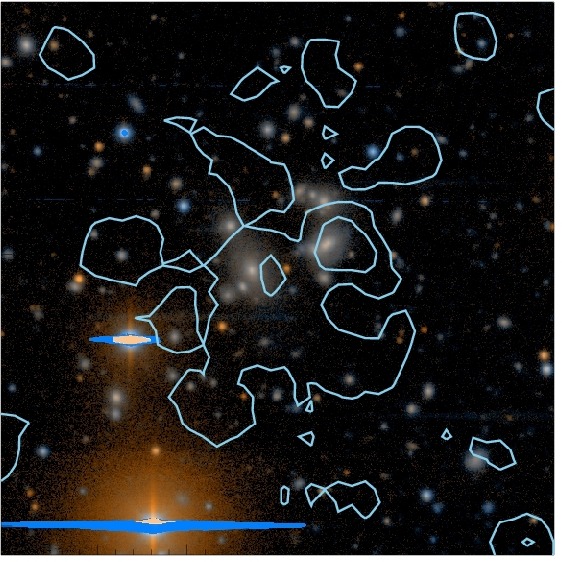} 
}
\subfigure
{
\includegraphics[scale=0.21]{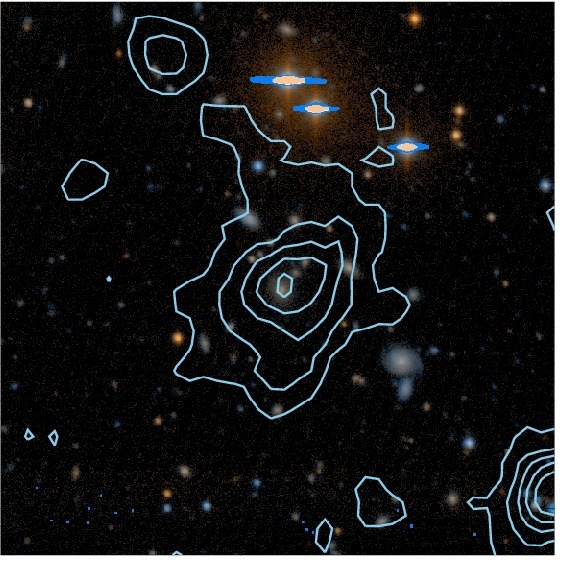}
}
\\ 
{
\includegraphics[scale=0.21]{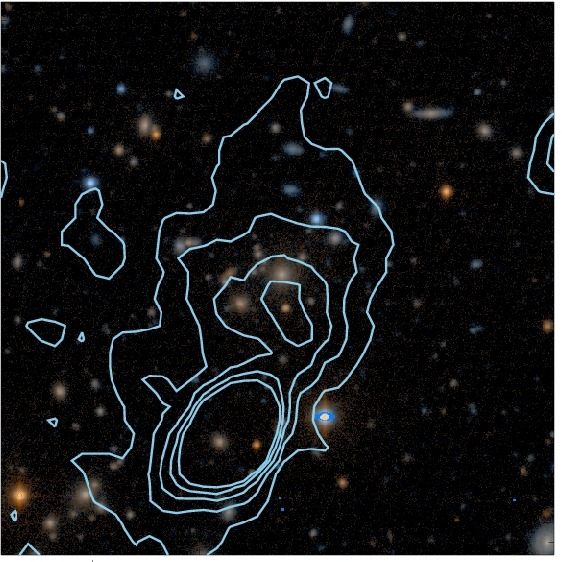}
}
\subfigure
{
\includegraphics[scale=0.21]{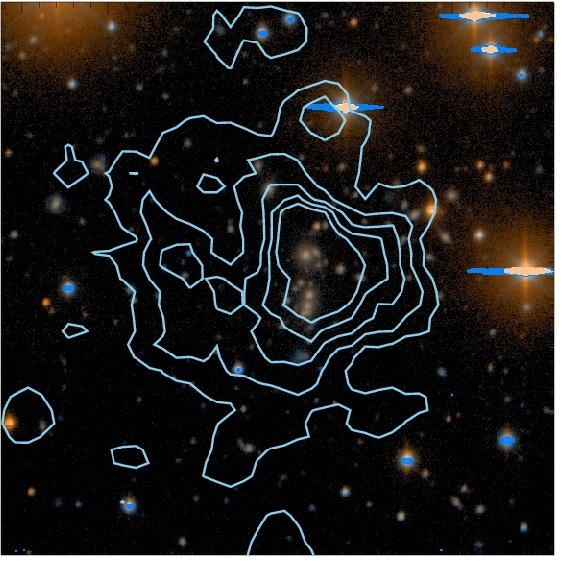}
}
\subfigure
{
\includegraphics[scale=0.21]{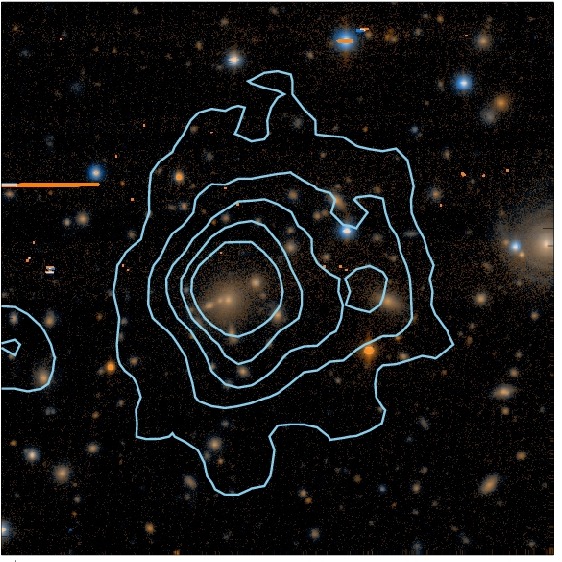}
}
\subfigure
{
\includegraphics[scale=0.21]{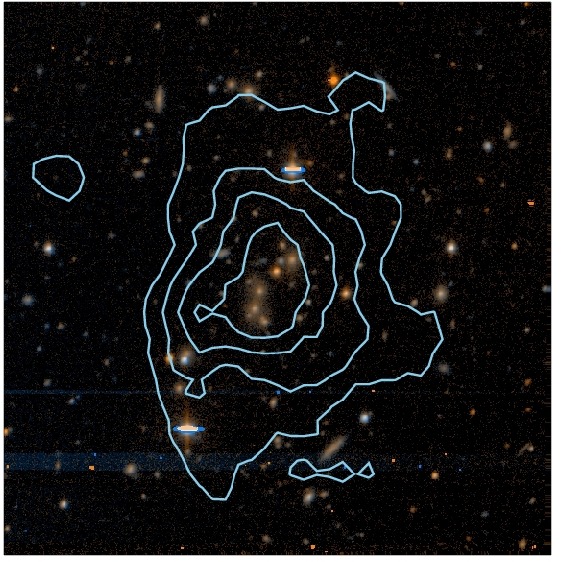}
} 
\\
{
\includegraphics[scale=0.21]{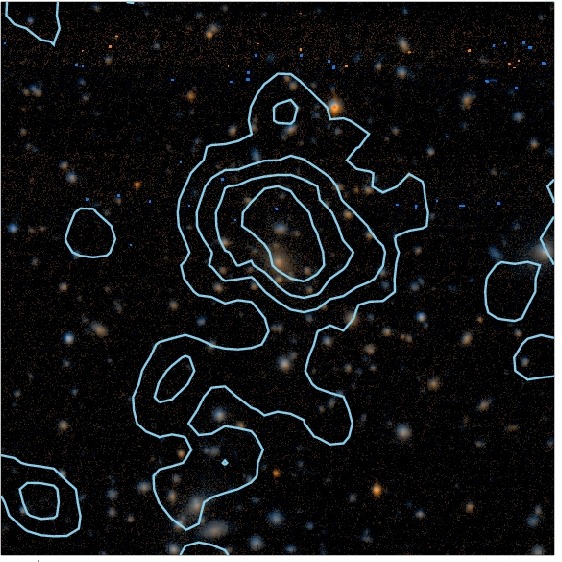}
}
\subfigure
{
\includegraphics[scale=0.21]{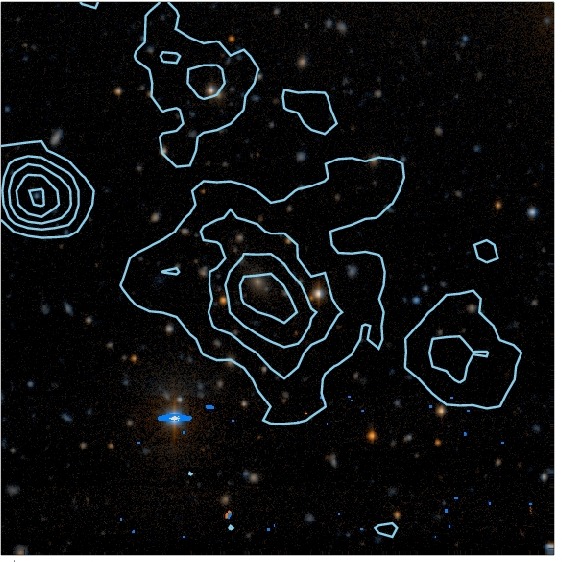}
}
\subfigure
{
\includegraphics[scale=0.21]{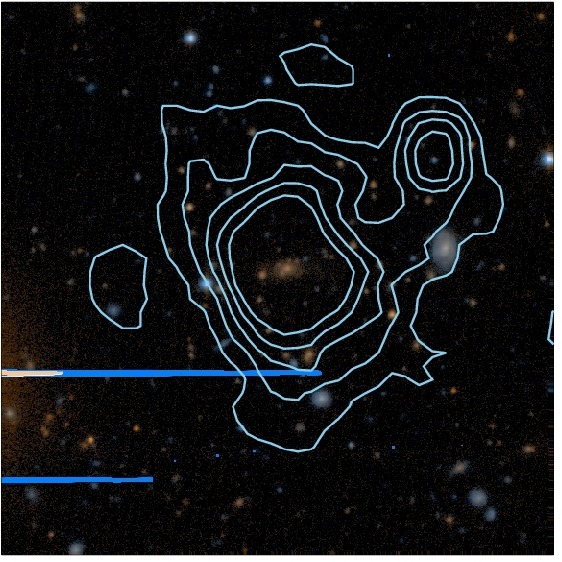}
}
\subfigure
{
\includegraphics[scale=0.21]{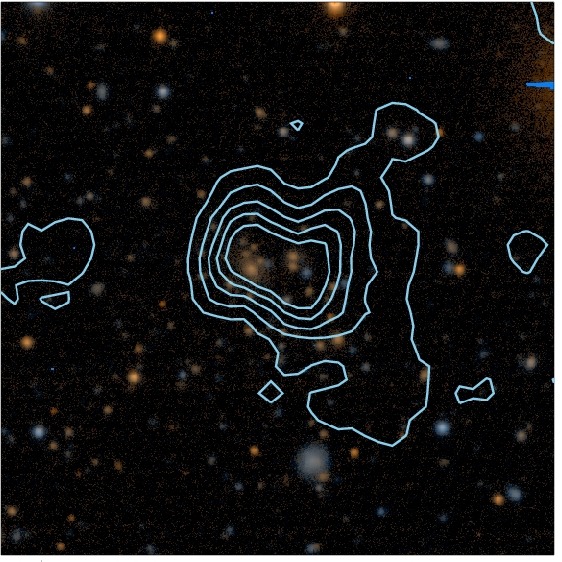}
}
\\
{
\includegraphics[scale=0.21]{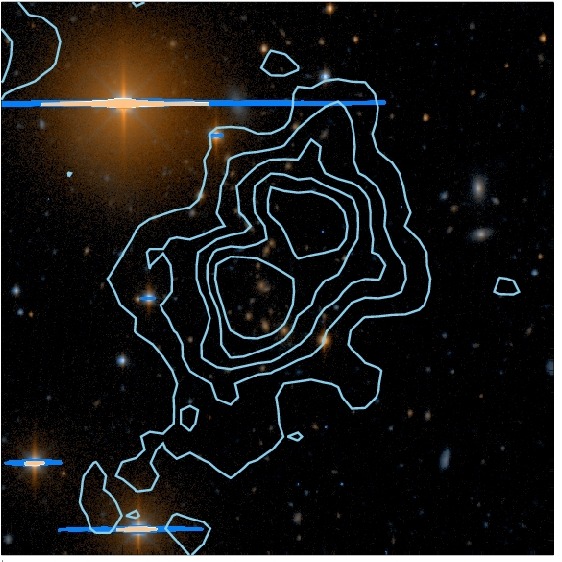}
}
\subfigure
{
\includegraphics[scale=0.21]{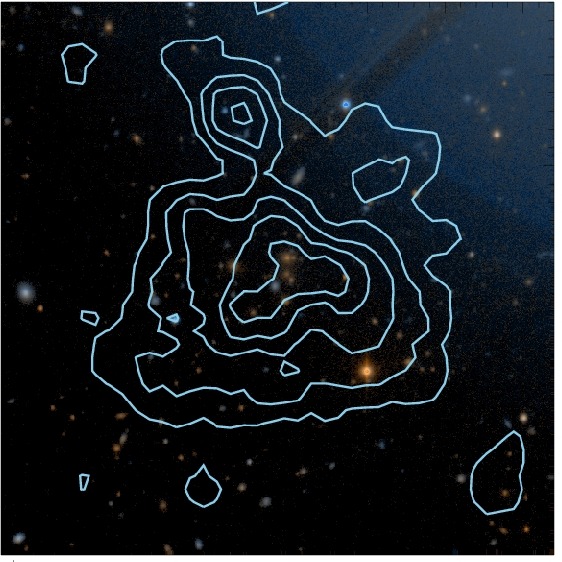}
}
\subfigure
{
\includegraphics[scale=0.21]{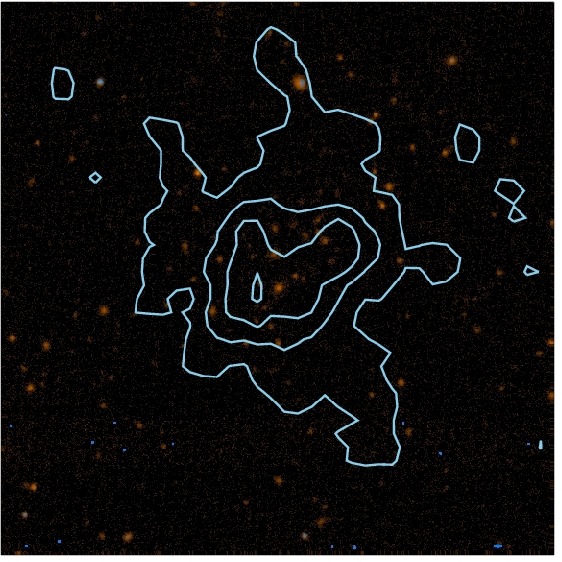}
}
\subfigure
{
\includegraphics[scale=0.21]{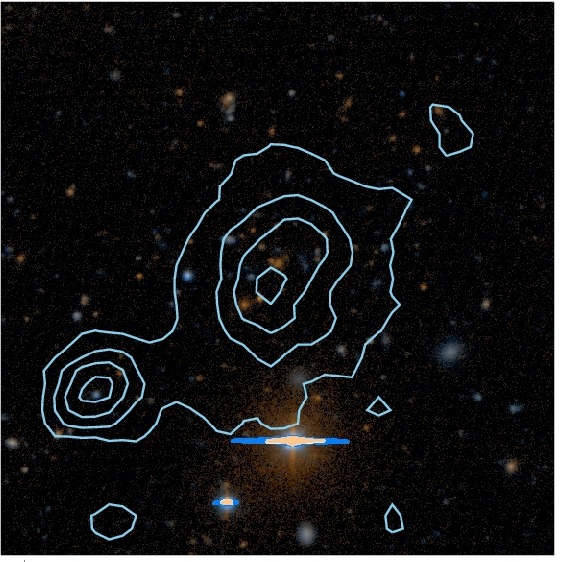}
}
\caption{A selection of optically confirmed XCS clusters as imaged by the NOAO-XMM Cluster Survey (NXS) and classified as \textit{gold} in \textit{Zoo}$^{\rm NXS}$. These clusters have corresponding redshifts and X-ray temperature measurements, and none of them have been previously catalogued in the literature.  False colour-composite images are $3\times3$\,arcmin with X-ray contours overlaid in blue. From left to right and top to bottom, the compilation displays the clusters: XMMXCS J130649.9$-$233128.5 at $z=0.21$; XMMXCS J232221.3$+$193855.0 at $z=0.23$; XMMXCS J205405.9$-$154736.5 at $z=0.27$; XMMXCS J223852.3$-$202612.2 at $z=0.35$; XMMXCS J011140.3$-$453908.0 at $z=0.367$; XMMXCS J075427.8$+$220950.9 at $z=0.40$; XMMXCS J232124.6$+$194514.8 at $z=0.40$; XMMXCS J063945.9$+$821847.3 at $z=0.41$; XMMXCS J003439.4$-$120715.8 at $z=0.44$; XMMXCS J011624.2$+$325717.0 at $z=0.45$; XMMXCS J092545.7$+$305856.9 at $z=0.52$; XMMXCS J212748.7$-$450151.9 at $z=0.56$; XMMXCS J011023.8$+$330544.1 at $z=0.60$;  XMMXCS J011632.1$+$330325.0 at $z=0.64$; XMMXCS J100115.5$+$250611.5 at $z=0.763$; and XMMXCS J025006.4$-$310400.8 at $z=0.91$.}
\label{fig:nxscompilation}
\end{center}
\end{figure*}

The primary aim of the NXS was to efficiently provide photometric redshifts for candidates to $z\simeq 1$. Therefore, the Sloan Digital Sky Survey (SDSS\footnote{www.sdss.org}, \citealt{York-2000}) $r$-band and $z$-band filters were chosen for the survey, because these straddle the 4000\AA\, break over the approximate redshift range $0.3<z<0.6$. This enhances the magnitude contrast of elliptical red-sequence galaxies detected in both bands at $z\sim 0.5$, and enables the estimation of red-sequence redshifts to $z>1$ \citep{GY00}. MOSAIC observations were made in a typical sequence of $2\times600$\,s $r$-band and $3\times750$\,s $z$-band exposures. Hereafter the set of NXS observations towards an ObsID will be referred to as an NXS field, where the NXS fieldID (see Table~\ref{t_nxs}) is set to the respective ObsID name. Exposures were offset by 30\,arcsec in R.A. and Dec. to eliminate MOSAIC chip gaps and aid the removal of cosmic rays and satellite trails in the final stacked images. If the original sequence of exposures was not taken under photometric conditions, then additional, short exposures were obtained (when possible) under photometric conditions at a later date, for calibration purposes. Over the course of the NXS project, \nZNEWa\, NXS fields, containing a total of \nZNEWb\, candidates, were observed. All of the raw data are publicly available at the NOAO Archive by searching the NOAO Portal\footnote{Information and a Data Handbook which describes how to access NOAO archival data and use the NOAO Portal is available at this URL: http://www.noao.edu/sdm/help.php} \citep{Miller-2007} for the Program ID: 2005B-0045. The total uncompressed data volume taken as part of the NXS survey is just over 500GB. This includes 1589 science exposures and another $\sim$2000 calibration images. A summary of the NXS observations is presented in Table~\ref{t_nxs}. Examples of NXS images of  XCS-DR1 clusters are shown in Fig.~\ref{fig:nxscompilation}.


\begin{table*}
   \caption{A summary of reduced observations taken by the NOAO-XMM Cluster Survey. $r$ and $z$ refer to exposures taken in the SDSS $r$ and $z$-band filters, respectively. A full version of Table \ref{t_nxs} is provided in electronic format in the online version of the article. These tables are ordered by increasing NXS FieldID number.}

   \label{t_nxs}
   \begin{tabular}{cccccccc}
     \hline
     \hline
     NXS FieldID & R.A. (J2000)  & Dec. (J2000) & No. Exposures & Integration Time & Seeing  & Depth  & Run \\
     & & & $r/z$ & $r/z$ (s) & $r/z$ (arcsec) & $r/z$ (mag) & \\
\hline
0002940101 &  13:07:04.7 & $-$23:38:51.3 & 2/3 &  1200$/$1500 &  1.2$/$1.0 & 24.1$/$23.4 &  4  \\
0010620101 &  05:15:45.0 & $+$01:03:14.6 & 2/2 &  1200$/$1000 &  1.1$/$1.9 & 24.4$/$23.8 &  1  \\ 
0012440301 &  22:05:04.3 & $-$01:54:19.1 & 2/3 &  1200$/$1500 &  1.4$/$1.4 & 24.5$/$23.2 &  4  \\ 
0021540101 &  15:06:27.4 & $+$01:37:55.2 & 2/3 &  1200$/$1500 &  1.4$/$1.1 & 25.3$/$23.6 &  4  \\ 
0025540301 &  08:38:25.5 & $+$25:43:40.0 & 2/3 &  1200$/$1500 &  2.0$/$1.5 & 25.0$/$23.7 &  3  \\ 
0025541601 &  01:24:40.8 & $+$03:46:30.7 & 2/3 &  1200$/$1500 &  1.0$/$1.1 & 25.3$/$22.5 &  1  \\ 
0029340101 &  06:41:43.3 & $+$82:14:31.4 & 2/2 &  1200$/$1000 &  1.0$/$0.9 & 25.1$/$23.9 &  1  \\ 
0032141201 &  13:05:11.8 & $-$10:19:22.0 & 2/3 &  1200$/$1500 &  1.1$/$1.0 & 24.6$/$23.5 &  4  \\ 
0037980301 &  02:25:25.7 & $-$03:50:59.2 & 2/3 &  1200$/$1500 &  1.3$/$1.0 & 25.6$/$23.0 &  5  \\ 
0037981601 &  02:23:14.3 & $-$02:48:56.3 & 2/3 &  1200$/$1500 &  1.8$/$1.8 & 25.0$/$23.8 &  2  \\ 
\hline
\end{tabular}
\end{table*}

\subsubsection{NXS data reduction}

Images were reduced using the \texttt{MSCRED} package \citep{Valdes:1998} written for the \texttt{IRAF} environment. \texttt{MSCRED} is specifically written to reduce data taken by the NOAO MOSAIC I and MOSAIC II cameras. We briefly summarize the reduction procedures below.

After correction for cross-talk between amplifiers, and overscan trimming, the images were bias and dome flat-field corrected. Next, artifacts were corrected by generating a fringe frame and pupil-ghost frame from science images and subtracting these templates from each individual science image (MOSAIC I and II $z$-band images both suffer from interference fringing, and MOSAIC I $z$-band images also suffer from a pupil ghost). A further flat-field correction was then applied using a sky-flat. Usually this was generated by combining suitable science images taken under similar conditions, but for the March 2008 observing campaign (due to the low number of NXS fields observed), it was necessary to make use of a sky-flat generated by the NOAO MOSAIC reduction pipeline\footnote{The pipeline implements a Master sky-flat generated from every science frame taken at the CTIO 4-m telescope and produces pre-stacked, sky-flattened, WCS (World Coordinate System) corrected images.}. Cosmic rays, bad pixels and bleed trails were automatically identified and added to bad-pixel masks. Satellite trails were identified and masked by eye\footnote{Satellite trails were masked using the script sat-b-gon.pl written by Matthew Hunt, available from www.ifa.hawaii.edu.}. An astrometric solution for each image was generated from the USNO-A2.0 catalogue using the automated task \texttt{msccmatch}. This solution was then used to rebin the image to a constant pixel scale to compensate for distortions across the MOSAIC field of view. The large-scale sky gradient was then removed from each image. Individual images of a particular field were then stacked by matching the background sky levels and excluding bad pixels. 

Source detection and photometric measurements were performed on the stacked images using \texttt{SExtractor} \citep{BertinArnouts:1996} operated in dual-image mode. In this mode, source positions and apertures were identified in $z$-band images and then the photometry was performed in both $r$ and $z$ bands simultaneously to produce matched object catalogues. To facilitate dual-image mode, the $r$ and $z$ band images were registered to pixel-level accuracy to allow matched aperture photometry across both bands. To avoid introducing erroneous colour estimates in the resulting galaxy catalogues, regions near bright stars were excluded in the NXS images prior to source detection. Such regions were identified in NXS images by performing an initial run of \texttt{SExtractor} using a high detection threshold to locate large extended sources ($>$ 3000 connected pixels) that also contained saturated pixels. The corresponding regions were then masked in the NXS exposure maps. The final object catalogues were then produced, using a second run of \texttt{SExtractor}, utilising these updated NXS exposure maps. We adopt Kron magnitudes (\texttt{MAGAUTO}) to estimate galaxy magnitudes and isophotal magnitudes (\texttt{MAGISO}) to calculate galaxy colours. 

Photometric calibration was achieved predominantly through the use of NXS fields that happened to lie within the survey regions of the SDSS Sixth Data Release (DR6, \citealt{Adelman-McCarthy:2008}), because SDSS sources have photometric calibration in both $r$ and $z$ bands accurate to within 3 per cent. Where this was not possible, observations were made of regions that contained either Southern Standard stars \citep{Smith:2002} or Landolt stars \citep{Landolt-1992} measured in the SDSS photometric system. In addition, two NXS fields imaged during the second observing campaign were also used to calibrate subsequent runs. 

For NXS fields with SDSS DR6 overlap, standard star catalogues were generated by extracting \textsc{model} magnitudes from the SDSS DR6 \textsc{PhotoObj} table for stars containing standard flags for clean photometry in the $r$ and $z$ band, respectively. The positions of these SDSS stars were then cross-matched with sources in the NXS object catalogues, using 1\,arcsec matching radii, to produce a matched catalogue of stars with both instrumental and corresponding standard magnitudes. Similarly, Southern Standard star positions, as well as the positions of calibrated objects in the two designated NXS calibration fields, were cross-matched with the NXS object catalogues using a 1\,arcsec matching radius. In the case of Landolt stars, these were identified and photometered using the aperture photometry tool in \texttt{GAIA}. 

Using the \texttt{IRAF} task \texttt{FITPARAMS}, the resulting catalogues of instrumental and corresponding standard magnitudes were compared in order to fit for a single zeropoint for each night, or partial night, that was deemed to be photometric. If multiple NXS fields containing calibration stars were observed over the course of a night, then the matched catalogues of instrumental and standard stars of each field were combined to fit for a single zeropoint. These updated zeropoints were then applied to the appropriate NXS object catalogues. Galactic extinction corrections were subsequently applied to NXS object magnitudes based on the dust maps and software of \citet{Schlegel:1998}.

Star--galaxy separation was determined for each NXS image using the method presented in \citet{Metcalfe-1991}, based on identifying the locus of stars in the concentration--magnitude plane. A concentration parameter was computed using aperture magnitudes measured within four and twelve pixels ($\sim1$ and $\sim3$ arcsec) in diameter. Star--galaxy separation was performed in the $r$-band and results in a clear separation of stars from galaxies at magnitudes typically brighter than $r\sim22$. At fainter magnitudes, we classified all objects to be galaxies, regardless of their concentration parameter.

The \nZNEWc\, NXS fields taken under photometric conditions have a typical seeing of 1.39 and 1.23\, arcsec in the $r$ and $z$-band, respectively. For an additional \nZNEWd\, NXS fields, it was possible to calibrate them \textit{a posteriori} using short integrations taken on subsequent photometric nights.  Observations of another \nZNEWf\, NXS fields were taken under non-photometric conditions, but the images were still of sufficient quality that they could be used for the optical identification work described in Section~\ref{ClusterZoo}. The mean depth of the calibrated NXS fields, as given by the 5$\sigma$ point source detection limit, are $r=25.0$ and $z=23.8$.  Based on the \citet{Bruzual-2003} population synthesis models and the assumption that the bulk of the signal in detecting a cluster comes from the galaxies brighter than about 0.5 magnitudes below $L_{\star}$ \citep{GY00}, these limits should be sufficient to detect clusters, and measure CMR-redshifts, to $z\simeq1$. In total, \nZNEWe\, candidates are contained within photometrically-calibrated NXS fields. 

\subsection{Photometric redshifts}
\label{redsequence algorithm}

We have applied the CMR-redshift technique to single-colour ($r-z$) photometric images of candidates. The photometry has either come from the NXS project (Section~\ref{NXS}) or from the SDSS (Section~\ref{Application to the SDSS DR7}). Our redshift algorithm is based on that presented in \citet{GY00}, in that it identifies overdensities of galaxies exhibiting a red-sequence and assigns a redshift based on the red-sequence colour. However it differs from \citet{GY00} in that it assumes the cluster centre is defined by the X-ray centroid of the corresponding XCS candidate (rather than by the centroid of a galaxy overdensity). 

For each candidate, potential cluster galaxies are extracted from within twice the X-ray extent as measured by the {\sc Xapa} algorithm. The colour distribution of these galaxies is then compared to that of an assumed field galaxy sample (normalised to the cluster area) to identify potential overdensities of red-sequence galaxies. For NXS, a local field sample was generated separately for each NXS field. This involved masking out all extended X-ray sources in that field using a fixed circular aperture of radius 0.15\,deg. We note that this mask aperture was always larger than that used to extract cluster galaxies. 

We then constructed a matched-filter to detect red-sequences via a maximum-likelihood fit. According to convention (e.g. \citealt{Postman-1996}; \citealt{Koester:2007}), we refer to the likelihood of there being a cluster red-sequence at a particular redshift, and with a certain number of galaxy members, as the \textit{ridgeline likelihood}. We chose to maximise our likelihood using the Cash statistic (\citealt{Cash-1979}, adapted to the form shown in Eq.~\ref{likelihood}), because the number of extracted red-sequence galaxies is often low compared to the local field sample. This likelihood is evaluated for each potential cluster galaxy, $x$, and summed over the candidate as a whole, as follows:

\begin{equation}
\mathcal{L}=\displaystyle\sum_{x=0}^{x=D}[\ln(b(z)+\Lambda_{N}M(z))] - D\label{Eqn:likelihood},
\label{likelihood}
\end{equation}
where $\mathcal{L}$ is the negative log likelihood; $z$ is the cluster redshift; $b(z)$ is the background distribution, given by the colour distribution of the local field sample; $\Lambda_{N}$ is a measure of cluster richness and corresponds to the total number of cluster galaxies above the background distribution; $M(z)$ is our red-sequence cluster model; and $D$ is the total number of galaxies within twice the X-ray extent. The red-sequence cluster model we adopt is a Gaussian distribution in colour \citep{Koester:2007}: 

\begin{equation}
M(z)=\frac{1}{\sqrt{2\pi}\sigma}{\rm exp} \frac{(x_{r-z}-RS_{\rm col})^{2}}{2\sigma^{2}}\label{Eqn:clusterModel};
\end{equation}

\begin{equation}
\sigma=\sqrt{\sigma^{2}_{r-z}+RS_{\rm width}^{2}}\label{GaussianWidth},
\end{equation}
where  $\sigma^2$ is the variance of the cluster model red-sequence; $x_{r-z}$ is the colour of a sampled galaxy; $RS_{\rm col}$ is the assumed red-sequence colour at the cluster redshift being evaluated; $\sigma_{r-z}$ is the uncertainty in the colour $x_{r-z}$; and $RS_{\rm width}$ is the intrinsic width of the red sequence, assumed to be 0.05 in colour \citep{Lopez-Cruz-2004} and constant with redshift.

The maximum ridgeline likelihood is found by considering a grid of red-sequence colours at redshifts $0.1 \leq z \leq 1.0$ and richness $0\leq \Lambda_{N} \leq 50$ in discrete steps of $\Delta z =0.01$ and $\Delta \Lambda_{N}=1$ respectively. Each model redshift is converted into a red-sequence colour using a theoretical map of red-sequence colour relations with redshift. This map is based on the slope of the composite red-sequence of 73 clusters at $z \sim 0.1$ detected by the SDSS-C4 survey \citep{miller-2005}, which is then evolved with redshift using the \citet{Bruzual-2003} population synthesis models assuming a single-burst Salpeter IMF with a formation redshift of $z_{f}=2.5$ \citep{gladders-2004}. 

In addition, a simultaneous estimate of the cluster richness ($\Lambda_{N}$) is produced together with an estimate of the cluster redshift. We note that the $\Lambda_{N}$ richness can be considered to be a lower limit to the true richness. This is because the adopted search radius (of twice the {\sc Xapa} X-ray extent) is typically less than R$_{200}$ (an approximation to the virial radius, defined as the radius at which the overdensity has fallen to 200 times that of the critical density). Moveover, the entire R$_{200}$ region is not always contained within an NXS field (the fields are centred on the ObsID aim point rather than on a specific candidate).

The best-fitting redshift, or CMR-redshift, will have an associated error that depends on a range of factors including the true redshift, the sensitivity of the image, the accuracy of the photometric calibration, the quality of the local field sample, and the appropriateness of the red-sequence model. The error on an individual CMR-redshift can only be determined once spectroscopic follow-up has taken place, but the typical error on a CMR-redshift can be externally determined via comparisons with measured spectroscopic redshifts (see Section~\ref{photoz error}). That said, we do calculate $\chi^{2}$ error estimates for the individual redshift measurements and these $\chi^{2}$ errors are used as an indication of the quality of the individual fits. 

We deemed a CMR-redshift fit to be unreliable if the $\chi^{2}$ error was too high ($\sigma_z > 0.1$), or if the richness was too low ($\Lambda_{N} < 5$), or if the NXS images were taken under non-photometric conditions. Excluding these unreliable fits, a total of \nZNEWg\,  CMR-redshift measurements were made using NXS data. We note that the drop in number between the \nZNEWe\, candidates contained within photometrically calibrated NXS fields (see above) to the \nZNEWg\, with reliable CMR-redshifts, is primarily due to the $\Lambda_{N} < 5$ cut. We assess the accuracy of NXS CMR-redshifts in Section~\ref{photoz error}.

\subsection{Spectroscopic redshifts}
\label{spec zs}

\begin{table*}
\caption{Spectroscopic follow-up by XCS team members. The $N(z)$ column lists the number of concordant redshifts obtained for each cluster. For clusters with only one spectroscopic redshift, the $z$ column gives the redshift of the suspected BCG; otherwise the $z$ column gives the mean galaxy redshift. The rightmost column lists the date(s) of observation, the observing programme number (for ESO or Gemini), or references, as appropriate. Uncertainties on the spectroscopic redshift values are not presented but are assumed to be at the level of the cluster velocity dispersion, i.e. $\sigma_v < 2000$ km$\,$s$^{-1}$. (Note that many more spectroscopic cluster redshifts are presented in XCS-DR1, but those were obtained from archives or from the literature.)}
\label{t_specz}
\begin{tabular}{|c|c|c|c|c|c|}
\hline
\hline
XCS ID  & $z$ & $N(z)$ & Telescope/Instrument & Comments\\
\hline
XMMXCS J003548.2$-$432232.4        & 0.633  & 12    & GMOS/Gemini                           & GS2010B-Q-46\\
XMMXCS J011140.3$-$453908.0        & 0.367  & 11    & NTT/EFOSC2                            & 28, 30-31 Jul 2008 (Program ID: 081.A-0843(A))\\
XMMXCS J012400.0$+$035110.8        & 0.883  & 7     & Keck/DEIMOS                           & 21 Sep 2006\\
XMMXCS J015241.1$-$133855.9        & 0.825  & 10    & Keck/DEIMOS                           & 2 Sep 2005, 21 Sep 2006\\
XMMXCS J023346.0$-$085048.5        & 0.25   & 1     & NTT/EMMI                              & 15 Sep 2006 (Program ID: 077.A-0437(A))\\
XMMXCS J025006.4$-$310400.8        & 0.908  & 6     & Gemini/GMOS                           & GS2010B-Q-46\\
XMMXCS J030145.5$+$000335.8        & 0.694  & 3     & Gemini/GMOS                           & GS-2010B-Q-46\\
XMMXCS J030317.4$+$001238.4        & 0.594  & 1     & NTT/EMMI                              & 15 Sep 2006 (Program ID: 077.A-0437(A))\\
XMMXCS J032553.3$-$061719.9        & 0.322  & 2     & NTT/EMMI                              & 9 Dec 2007 (Program ID: 080.A-0024(C))\\
XMMXCS J035417.0$-$001006.6        & 0.214  & 2     & NTT/EMMI                              & 9 Dec 2007 (Program ID: 080.A-0024(C))\\
XMMXCS J041944.6$+$143904.5        & 0.196  & 2     & NTT/EMMI                              & 17 Oct 2006 (Program ID: 078.A-0325(C))\\
XMMXCS J045506.3$-$532343.8        & 0.410  & 1     & NTT/EMMI                              & 13 Dec 2006 (Program ID: 078.A-0325(A))\\
XMMXCS J051610.0$+$010954.0        & 0.318  & 2     & NTT/EMMI                              & 7 Dec 2007 (Program ID: 080.A-0024(C))\\
XMMXCS J080612.6$+$152309.0        & 0.41   & 1     & WHT/ISIS                              & 1-3 Dec 2007 (Program ID: P53)\\
XMMXCS J091821.9$+$211446.0        & 1.007  & 16    & Gemini/GMOS                           & GN-2010B-Q-65\\
XMMXCS J095105.7$+$391742.9        & 0.47   & 1     & WHT/ISIS                              & 1-3 Dec 2007 (Program ID: P53)\\
XMMXCS J095940.8$+$023111.3        & 0.720  & 14    & Gemini/GMOS                           & GS2010B-Q-46\\
XMMXCS J100115.3$+$250612.4        & 0.763  & 12    & Gemini/GMOS	                    & GN-2010B-Q-65\\
XMMXCS J100201.7$+$021332.8        & 0.832  & 6     & Gemini/GMOS                           & GS2009B-Q-80\\
XMMXCS J102136.9$+$125643.2        & 0.325  & 1     & NTT/EMMI                              & 14 Dec 2006 (Program ID: 078.A-0325(C))\\
XMMXCS J104422.2$+$213025.2        & 0.515  & 7     & Gemini/GMOS                           & GN-2010B-Q-65\\
XMMXCS J105040.6$+$573741.4        & 0.689  & 12    & Gemini/GMOS                           & GN-2010B-Q-65\\
XMMXCS J111645.5$+$180047.7        & 0.662  & 7     & Gemini/GMOS                           & GN2005B-Q-56\\
XMMXCS J111726.0$+$074327.7        & 0.482  & 15    & Gemini/GMOS                           & GS2010B-Q-46\\
XMMXCS J112349.3$+$052956.8	   & 0.652  & 11    & Gemini/GMOS                           & GS-2010B-Q-46\\
XMMXCS J130601.4$+$180145.9        & 0.93   & 3     & Keck/LRIS                             & 10 Feb 2005\\
XMMXCS J150652.9$+$014424.8        & 0.653  & 2     & NTT/EFOSC2                            & 29 Jul 2008 (Program ID: 081.A-0843(A))\\
XMMXCS J153643.9$-$141024.2        & 0.40   & 2     & NTT/EFOSC2                            & 30 Jul 2008 (Program ID: 081.A-0843(A))\\
XMMXCS J200703.1$-$443757.6        & 0.202  & 1     & NTT/EFOSC2                            & 31 Jul 2008 (Program ID: 081.A-0843(A))\\
XMMXCS J204134.7$-$350901.2        & 0.425  & 1     & NTT/EFOSC2                            & 30 Jul 2008 (Program ID: 081.A-0843(A))\\
XMMXCS J212807.6$-$445417.3        & 0.538  & 4     & NTT/EFOSC2                            & 27-28, 30-31 Jul 2008 (Program ID: 081.A-0843(A))\\
XMMXCS J215221.0$-$273022.6        & 0.826  & 9	    & Gemini/GMOS                           & GS-2010B-Q-46\\
XMMXCS J221559.6$-$173816.2        & 1.457  & 31    & Various                               & See \citet{Stanford:2006,Hilton-2007,Hilton-2009,Hilton-2010}\\
XMMXCS J231852.3$-$423147.6        & 0.114  & 1     & NTT/EMMI                              & 10 Dec 2007 (Program ID: 080.A-0024(C))\\
XMMXCS J235708.6$-$241449.2        & 0.588  & 10    & NTT/EFOSC2                            & 27, 29-31 Jul, 1 Aug 2008 (Program ID: 081.A-0843(A))\\
\hline
\end{tabular}
\end{table*}

Table~\ref{t_specz} lists the mean spectroscopic cluster redshifts obtained by members of the XCS team for \nZNEWh\, candidates. Of the objects in the table, only the spectroscopic redshift of XMMXCS J2215.9$-$1738 ($z=1.46$) has previously been published \citep{Stanford:2006, Hilton-2007, Hilton-2009, Hilton-2010}.  At the start of our spectroscopic  programme, candidates were selected for follow-up either to fill R.A. gaps during the follow-up of the Sloan Digital Sky Survey-II Supernova Survey \citep{Ostman-2011, Frieman-2008}, or because they were judged by eye to be potential $z>1$ clusters (and hence suitable for Keck or Gemini follow-up). However, as the project matured, the target selection was informed by X-ray redshift estimates (see LD10), thus allowing us to design a follow-up programme that both sampled the $L_{\rm X}-T_{\rm X}(z)$ relation and allowed us to determine the accuracy of the CMR-redshifts (Section~\ref{photoz error}).

\begin{figure*}
\includegraphics[scale=0.3]{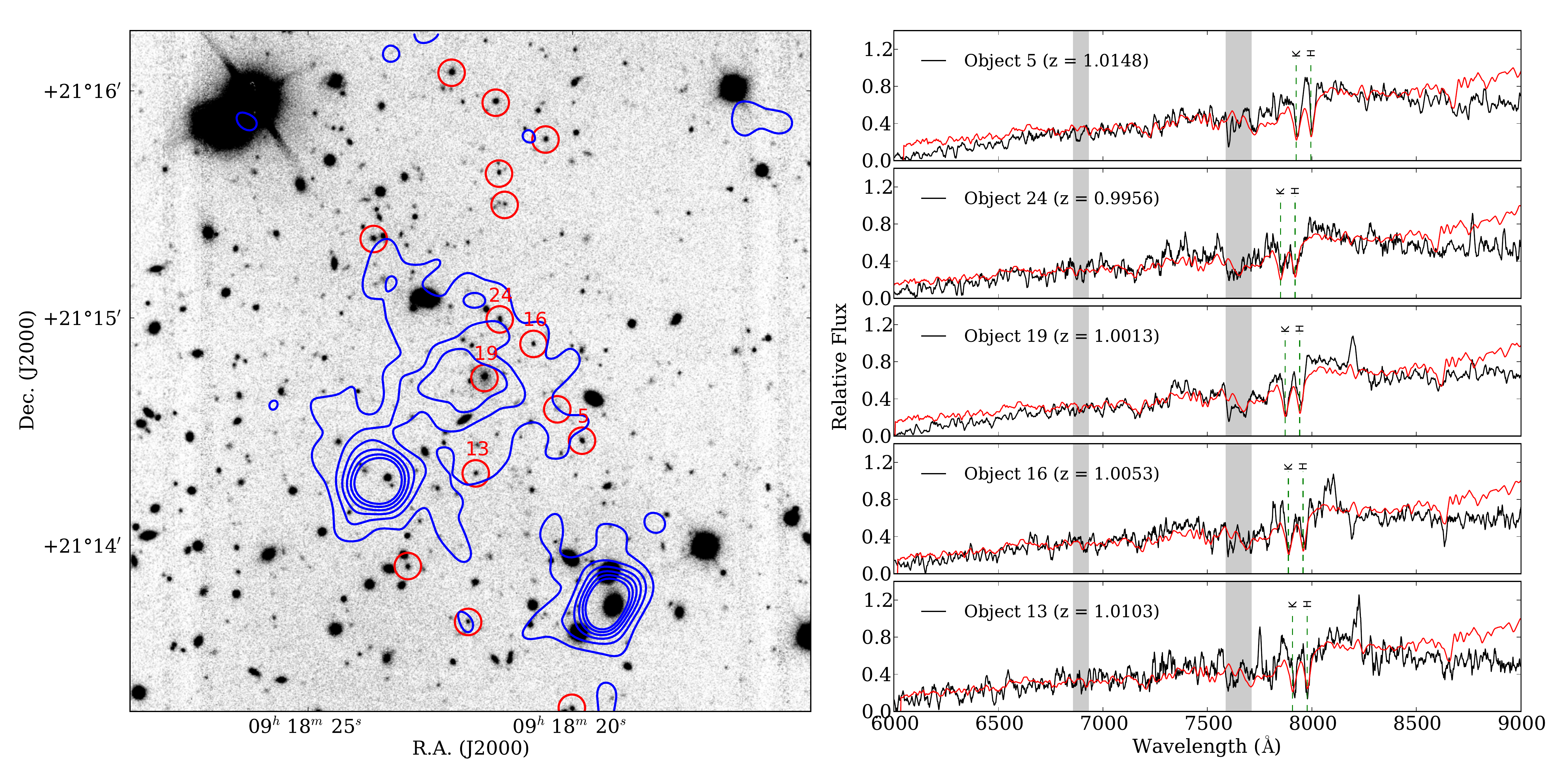}
\caption{The $z=1.01$ cluster XMMXCS J091821.9$+$211446.0. The left hand panel shows a $3\times3$\,arcmin Gemini GMOS i-band image, with X-ray contours overlaid (blue) and spectroscopically identified cluster members circled in red. The right-hand panel shows the GMOS spectra (black lines; not flux calibrated) of a selection of members highlighted in the image. A redshifted LRG spectral template (red line) is shown for each galaxy.}
\label{J0918}
\end{figure*}

The spectroscopic observations were performed over several years at a number of different observatories with a variety of instruments, as summarised in Table~\ref{t_specz}. Data taken with the DEep Imaging and Multi-Object Spectrograph \citep[DEIMOS;][]{Faber-2003} at the Keck observatory were processed with version 1.1.4 of \texttt{spec2d}, the pipeline developed for the DEEP2 galaxy redshift survey \citep{Davis_2003}. All Gemini Multi-Object Spectrograph \citep[GMOS;][]{Hook_2003} observations were obtained in nod-and-shuffle mode, and reduced in a manner similar to that described in \citet{Hilton-2010}. An example cluster spectroscopically confirmed in one of the GMOS observing programmes is shown in Fig.~\ref{J0918}. The ESO Multi-Mode Instrument \citep[EMMI;][]{Dekker_1986} and ESO Faint Object Spectrograph and Camera \citep[EFOSC2;][]{Buzzoni_1984, Snodgrass_2008} at the NTT were used to obtain long-slit spectroscopy of likely (as judged by eye) Brightest Cluster Galaxies (BCGs). Multi-object spectroscopic observations were also obtained for some clusters using EFOSC2. All of the data obtained at ESO were reduced using \texttt{IRAF} in the standard manner. 

Redshifts were measured either from visually identified spectral features or using the cross-correlation method implemented in the \texttt{rvsao IRAF} package \citep{KurtzMink_1998}. Table~\ref{t_specz} lists the number of secure concordant redshifts obtained for each cluster. A number of clusters listed in Table~\ref{t_specz} have only one redshift measurement; in these cases, the quoted redshift is that of the likely BCG.

We highlight here two clusters from Table~\ref{t_specz}. First, a new (to the literature) $z>1$ cluster, optically confirmed in the NXS (Section~\ref{NXS}), with multi-object spectroscopic confirmation (XMMXCS J091821.9$+$211446.0, $z=1.01$, Fig.~\ref{J0918}). Second, a new (to the literature) cluster, XMMXCS J015241.1$-$133855.9 at $z=0.83$, within a projected distance of $\sim$8.7\,Mpc from the well-studied merger system XMMXCS J015242.2$-$135746.8 (or WARP J0152.7$-$1357, \citealt{Ebeling00, Romer00, Demarco-2005, Girardi-2005, Maughan-2006}).

\section{Redshift follow-up (archive)}
\label{z-archive}

In addition to our own redshift follow-up work, we have been able to extract a large number of redshifts (both spectroscopic and photometric) for our candidates using data archives and from the literature.  We describe the extraction of redshifts from the SDSS Seventh Data Release (DR7; \citealt{Abazajian:2009}) in Sections~\ref{Application to the SDSS DR7} (photometric) and \ref{LRG Speczs} (spectroscopic), and the extraction of redshifts from the literature in Section~\ref{archive-z-lit}. We note that no XCS-determined X-ray redshifts are included in XCS-DR1. These redshifts have been shown to be reliable (at the $\Delta_z<0.1$ level) in 75 per cent of cases (LD10) and so, in principle, we could have used them for XCS-DR1. However, in practice, there were only \nZARCa\, overlaps between the sample of candidates with X-ray redshifts and the XCS-DR1 list, and all of these have other redshift determinations of higher quality.

\subsection{Redshifts from SDSS (photometric)}
\label{Application to the SDSS DR7}

The red-sequence redshift algorithm described above (Section~\ref{redsequence algorithm}) was also applied to each of the candidates that fall in the SDSS DR7 footprint\footnote{During the preparation of this manuscript, the SDSS Eighth Data Release was made public (DR8; \citealt{Aihara-2011}). Importantly, this covers more area than SDSS DR7, so we will be exploiting SDSS DR8 for XCS follow-up in future publications.}. Included in SDSS DR7 is a 270\,deg$^2$ co-added stripe, known as Stripe 82, and referred to as S82 hereafter, that reaches a depth $\sim$2 magnitudes fainter than the regular survey.  We have used both datasets to determine CMR-redshifts from SDSS DR7.  At the time of writing (June 2011), \nZARCb\, candidates lie within the SDSS DR7 footprint, of which \nZARCc\, lie within the S82 footprint.

Galaxy samples were extracted from the \textsc{Galaxy View}, which contains photometric information for all \textsc{primary} objects imaged by SDSS and subsequently classified as galaxies. We use the SDSS measurement \textsc{ModelMag} to provide galaxy magnitudes and calculate colours for each galaxy, and apply the Galactic extinction corrections supplied by the SDSS based on the dust maps by \citet{Schlegel:1998}. We specify that all galaxies must contain the standard flags for clean photometry in both the $r$ and $z$ bands. In this manner, potential cluster galaxy samples were generated by retrieving de-reddened  model $r$ and $z$ magnitudes for all galaxies falling within twice the {\sc Xapa} extent of each candidate contained within the SDSS DR7 and S82 footprints, respectively. 

As the SDSS is a large, homogeneous survey, a universal field sample could be constructed (rather than the field-by-field approach adopted for NXS, Section~\ref{redsequence algorithm}). For this, a random sample of 50 (20) ObsIDs within the SDSS DR7 (S82) footprint were selected as the basis for a field sample. Pointings with incomplete SDSS coverage, or those containing image defects or saturated objects, were not used. De-reddened  model $r$-band and $z$-band magnitudes were retrieved for all galaxies with clean photometry across the regions covered by each of the 50 (20) ObsIDs. 

To minimize contamination of the field sample by galaxies within clusters, areas that overlapped either with candidates or with known clusters (identified using NED) were masked from the field sample. In the case of XCS candidates, we do not \textit{a priori} know their redshift or temperature, so a fixed radius of 0.15 degrees was used (as was the case for NXS). In the case of NED clusters, the redshifts are usually known, but the temperatures typically are not. So, to be conservative, we used an R$_{200}$ radius for the masking that assumes a cluster temperature of $T_{\rm X}=4$ keV (less than 30 per cent of XCS clusters are hotter than this). The R$_{200}$ values were calculated according to the method outlined in Section 3.2 of S09. This process yielded a single combined field sample containing 52,660 (207,693) galaxies covering a combined total area of 18.10\,deg$^2$ (5.87\,deg$^2$) derived from SDSS DR7 (S82).

Similar to the approach taken with NXS CMR-redshifts (Section~\ref{redsequence algorithm}), we deemed a SDSS DR7 (S82) CMR-redshift fit to be unreliable if the $\chi^{2}$ error was too high ($\sigma_z > 0.1$), or if the richness was too low ($\Lambda_{N} < 5$). After excluding these unreliable fits, and candidates with less than 100 counts (DR7 only; see Section~\ref{ClusterZoo}), a total of \nZARCd\, and \nZARCe\, CMR-redshift measurements were made using DR7 and S82 data, respectively.  We assess the accuracy of DR7 and S82 CMR-redshifts in Section~\ref{photoz error}.

\subsection{Redshifts from SDSS (spectroscopic)}
\label{LRG Speczs}

Luminous Red Galaxies, or LRGs, have been targeted by SDSS for spectroscopic follow-up using colour and magnitude cuts designed to select luminous ($L>3L_*$), intrinsically red, elliptical galaxies \citep{Eisenstein-2001}. As LRGs predominantly reside in the central regions of dense cluster environments, we can make the assumption that an identified LRG coincident with an X-ray emitting cluster, is part of that cluster. The spectroscopic redshift of this LRG (or group of LRGs) can then be adopted as the cluster redshift. 

Because the 4000\AA\, break migrates with redshift, two colour cuts are necessary to select LRGs within SDSS imaging: we use the low-redshift ($z < 0.45$) colour cuts of \citet{Eisenstein-2001} and the high-redshift ($0.45 < z < 0.7$) cuts of \citet{Padmanabhan-2005} and \citet{Collister-2007}. From the resulting combined sample, an LRG (and its spectroscopic redshift) is assigned to an XCS candidate if it lies within 175\,kpc of the X-ray centroid (assuming the redshift of the LRG). This matching radius was chosen because it was both free from high-levels of contamination (the ratio of real-to-false matches was found to be 18 per cent when comparing the assigned LRG redshift to corresponding cluster redshifts in the 400d catalogue; \citealt{2007ApJS..172..561B}), and because it was consistent with the results from \citet{LinMohr-2004} who found that 70 per cent of BCGs are located within 5 per cent of the cluster virial radius (R$_{200}$) from the X-ray centroid. There are instances however where multiple LRGs are assigned to a given XCS candidate. Some of these will be groups of LRGs belonging the same cluster halo. These are identified by scanning in redshift intervals of $\Delta z =0.05$ and counting the number of (assumed) Gaussian colour error distributions overlapping in segments of $\Delta z=0.1$. A cluster redshift is then assigned from a given group of LRGs using the following hierarchy: if one distinct group of LRGs is found, then the weighted mean redshift and weighted error of that group is assigned to the XCS candidate; if more than one group of LRGs is found, the group with the larger number of LRGs is chosen; if the number of LRGs within two groups are identical, then the group with the redshift closest to the CMR-redshift determined from the deepest imaging data (Sections~\ref{redsequence algorithm} and \ref{Application to the SDSS DR7}) is chosen. 

It was found, using eye-ball inspection, that when low-redshift ($z<0.08$) LRGs were associated with XCS candidates, the matches were typically erroneous. This is because the 175\,kpc search radius subtends a large angle on the sky at low LRG redshifts. Therefore, only LRG redshifts at $z \geq 0.08$ were typically used for XCS-DR1. However, we did make exceptions if the candidate had a measured CMR-redshift pegged at the algorithm's minimum value  ($z=0.1$). In this instance, the candidate was judged to be at low redshift and so could be safely associated with LRG redshifts below $z=0.08$. There are \nZARCf\, such cases in XCS-DR1: XMMXCS J010720.2$+$141604.2, XMMXCS J015315.0$+$010214.2, XMMXCS J115112.0$+$550655.5, XMMXCS J134326.9$+$554648.3, and XMMXCS J163015.6$+$243423.2. In summary, \nZARCg\, candidates were associated with spectroscopic derived from SDSS LRGs.

\subsection{Redshifts from the literature}
\label{archive-z-lit}

All candidates have been cross matched using a simple automated NED query to determine whether they have been catalogued by an earlier cluster survey (Section~\ref{The catalogue}). In addition, a more complex NED query has been used to determine which of the candidates can be associated with a published redshift. This search involves an iterative analysis of the \textit{XMM} data, and the technical aspects have been described in LD10. To date \nZARCh\, literature redshifts, or $z_{\rm lit}$'s, have been extracted from NED using this process. The automated nature of the $z_{\rm lit}$ collection means that not all of the extracted redshifts are correct. Therefore, for XCS-DR1 we have taken a conservative approach of only using literature redshifts if $z_{\rm lit} \geq 0.08$. After applying this cut\footnote{In principle, as was the case for the LRG redshifts (Section~\ref{LRG Speczs}), we would have been prepared to assign $z_{\rm lit}<0.08$ values to clusters if the measured CMR-redshift pegged at the algorithm's minimum value  ($z=0.1$). However, in practice there were no such candidates.}, \nZARCm\, $z_{\rm lit}$ values remain. To these, we have added by hand \nZARCo\, redshifts that were not in NED at the time when the automated $z_{\rm lit}$ extraction was performed (\nZARCn\, taken from a recent data release by the XMM--LSS survey by \citealt{Adami_10}, three redshifts taken from a parallel study, \citet{Harrison11}, H11 hereafter, see Section~\ref{best z}, two redshifts taken from \citealt{Suhada-2011}, a single redshift taken from \citealt{Lamer-2008}, and a single redshift taken from \citealt{2005Msngr.120...33B}). In addition, we updated \nZARCq\, redshifts in NED with improved values (\nZARCp\, XMM--LSS redshifts taken from \citealt{Adami_10}, the redshift for the cluster RX J105346.6+573517 taken from \citealt{Hashimoto-2005}, and the redshift for the cluster XMMXCS J2215.9$-$1738 taken from \citealt{Stanford:2006}; see Table~\ref{t_specz}). 

The NED-based $z_{\rm lit}$ collection method cannot discern automatically whether individual redshifts were spectroscopic or photometric. However, this information is important to XCS, both to assess the reliability of derived quantities (especially X-ray luminosities) and to determine the typical error on XCS photometric redshifts (Section~\ref{photoz error}). Therefore, we have made a manual check of the respective publication(s) for each of the \nZARCr\, XCS-DR1 clusters with associated $z_{\rm lit}$ values (\nZARCs\, coming from the automatic NED search, the remainder coming from the sources described above).

\section{Quality control}
\label{Quality Control}

A quality control step is necessary for XCS-DR1 because candidates are selected in a fully automated fashion (Fig.~\ref{ELDFig1}).  Whilst automation is important to XCS -- for both efficiency and to maintain statistical robustness -- it can result in contamination of the candidate list by: \textit{(i)} extended non-cluster X-ray sources (e.g. low-redshift galaxies); \textit{(ii)} non-extended X-ray sources (e.g. blended point sources); and \textit{(iii)} clusters that were the intended target of the respective ObsID (or physically associated with it). Therefore, some quality control must be applied before releasing a confirmed cluster catalogue based on a given input candidate list. This has been carried out for XCS-DR1 using one or more of the following: an {\sc XCS-Zoo} (Section~\ref{ClusterZoo}); information from the literature (Section~\ref{Litz confirmation}); our own spectroscopy (Section~\ref{Litz confirmation}); and checks of the ObsID headers (Section~\ref{targetcheck}).

\subsection{Candidate identification using {\sc XCS-Zoo}}
\label{ClusterZoo}

Both the name and the methodology of {\sc XCS-Zoo} were inspired by the SDSS Galaxy Zoo project \citep{Lintott-2008}. The Galaxy Zoo project took advantage of community input to morphologically classify SDSS galaxies over the web. The {\sc XCS-Zoo} project is similar, in that it draws on a team of volunteers -- either members of XCS or astronomers at affiliated universities -- to classify XCS cluster candidates, and this classification is done using eye-ball inspection via a web interface. However, {\sc XCS-Zoo}  is on a much smaller scale than Galaxy Zoo. Moreover, unlike the hundreds of thousands of Galaxy Zoo volunteers, all 23 {\sc XCS-Zoo} participants are co-authors of this paper.  

The {\sc XCS-Zoo} allowed us to establish, by consensus, whether a candidate had an obvious optical cluster counterpart. Candidates were included in {\sc XCS-Zoo} if optical\footnote{In principle, useful information related to candidate identification could be derived from a wide range of observations, including radio and infra-red, but to date (June 2011) we have only used optical data.} CCD imaging was available from the NXS (Section~\ref{NXS}) or SDSS DR7 (both the regular survey and S82).  A separate {\sc XCS-Zoo} was undertaken for each of the three imaging surveys, and hereafter we refer to these as \textit{Zoo}$^{\rm NXS}$, \textit{Zoo}$^{\rm DR7}$, and \textit{Zoo}$^{\rm S82}$, respectively. Each candidate was classified at least five times per \textit{Zoo}, even if they were covered by multiple imaging surveys. The number of candidates that could potentially have been classified by \textit{Zoo}$^{\rm DR7}$ was much larger (\nZARCb\,) than the other two \textit{Zoos} (\nQCONTa\, in total), and so we set a minimum X-ray count ($>100$) threshold for \textit{Zoo}$^{\rm DR7}$. This reduced the number of candidates included in \textit{Zoo}$^{\rm DR7}$ to a more manageable \nQCONTb. 

The inspected candidates were classified into one of the following categories of cluster: \textit{gold}; \textit{silver}; and \textit{bronze}. A fourth category (\textit{other}) was used for any remaining candidates (Section~\ref{Zoo-other}). The {\sc XCS-Zoo} categorisation of each source was based upon the following information: a series of X-ray image cutouts ($3\times3$, $6\times6$, $12\times12$ \,arcmin), highlighting X-ray contours and the region enclosed by the {\sc Xapa} X-ray extent; a corresponding series of colour-composite optical images (with and without X-ray contours overdrawn); and an image highlighting the location of the candidate within the ObsID.  

\begin{figure*}
\begin{center}
\subfigure
{          
\includegraphics[scale=0.4]{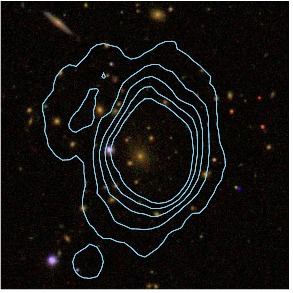}
} 
\subfigure
{
\includegraphics[scale=0.4]{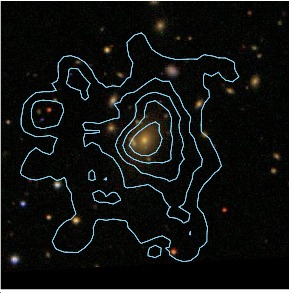}
}
\subfigure
{
\includegraphics[scale=0.4]{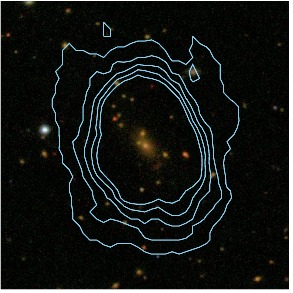}
}
\subfigure
{
\includegraphics[scale=0.4]{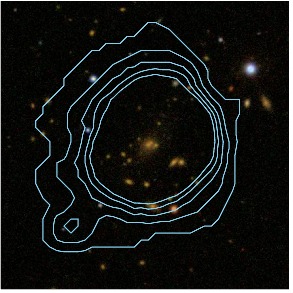} 
}
\\
\subfigure
{
\includegraphics[scale=0.215]{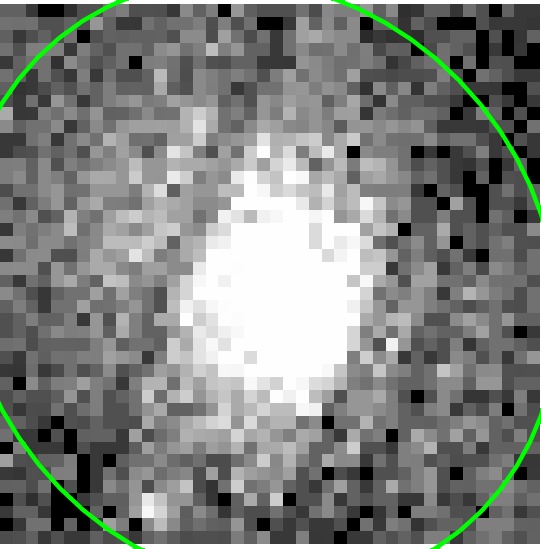}
}
\subfigure
{
\includegraphics[scale=0.215]{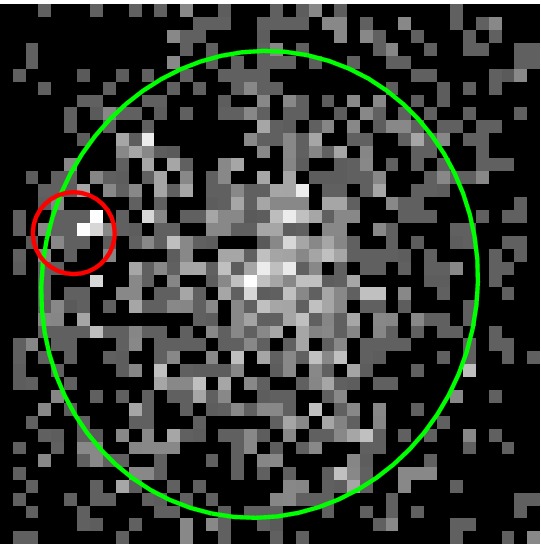}
}
\subfigure
{
\includegraphics[scale=0.215]{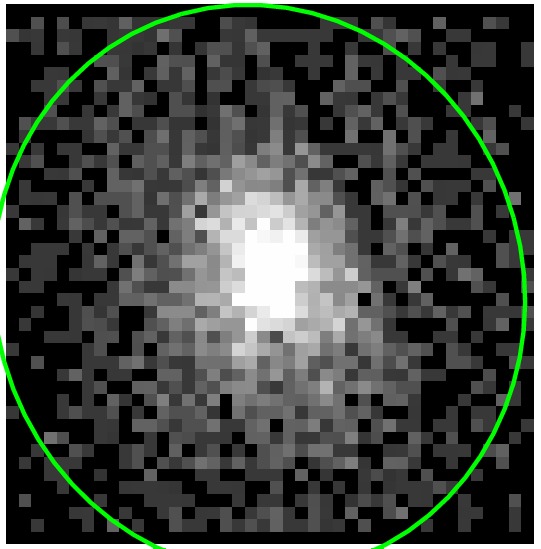}
}
\subfigure
{
\includegraphics[scale=0.215]{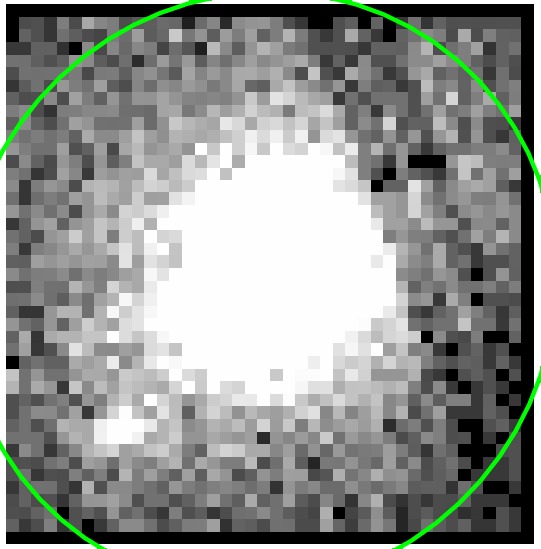}
}
\caption{The first four clusters presented in Table \ref{BigTable} classified as \textit{gold} in  \textit{Zoo}$^{\rm DR7}$ (Section~\ref{ClusterZoo}). False colour-composite images are $3\times3$\,arcmin with X-ray contours overlaid in blue. Corresponding X-ray images are shown below each optical image (lighter regions show areas of increased X-ray flux).  The shape of the {\sc Xapa}-detected extended (point) source ellipses are highlighted in green (red). From left to right, the clusters are:  XMMXCS J001737.4$-$005235.4 at $z=0.21$; XMMXCS J010858.7$+$132557.7  at $z=0.15$; XMMXCS J083454.8$+$553420.9 at $z=0.24$; and XMMXCS J092018.9$+$370617.7 at $z=0.21$.}
\label{fig:SDSSGold}
\end{center}
\end{figure*}

\begin{figure*}
\begin{center}
\subfigure
{          
\includegraphics[scale=0.4]{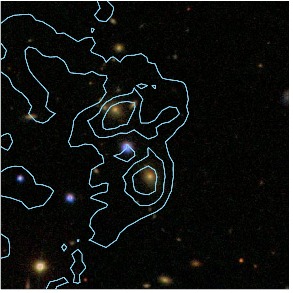}
}
\subfigure
{ 
\includegraphics[scale=0.4]{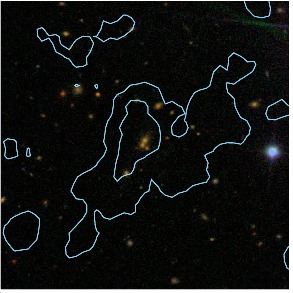}
}
\subfigure
{  
\includegraphics[scale=0.4]{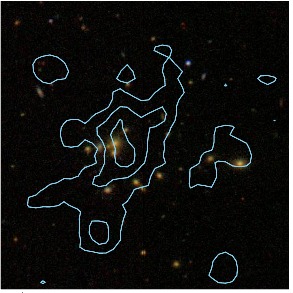}
}
\subfigure
{ 
\includegraphics[scale=0.4]{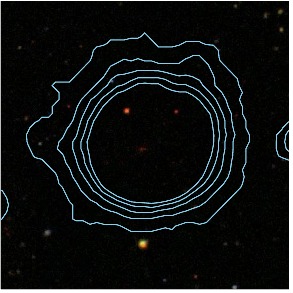} 
}
\\               
\subfigure
{ 
\includegraphics[scale=0.215]{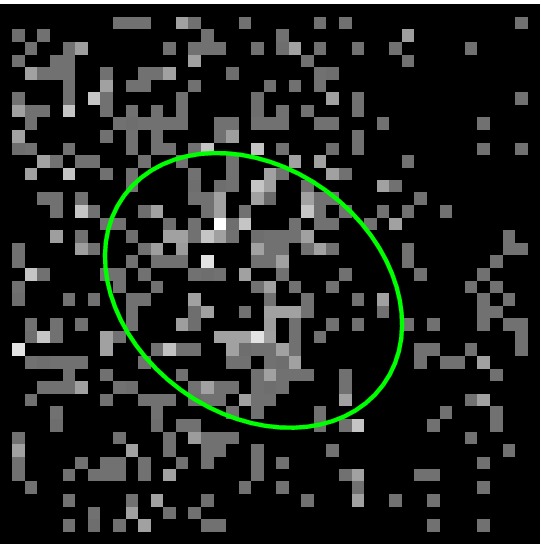}
}
\subfigure
{  
\includegraphics[scale=0.215]{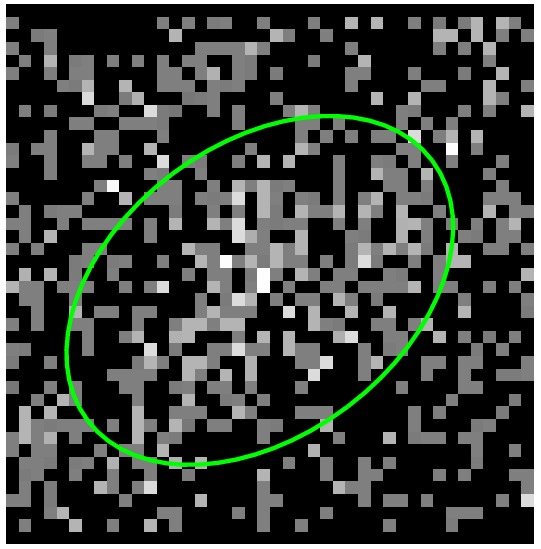}
}
\subfigure
{
\includegraphics[scale=0.215]{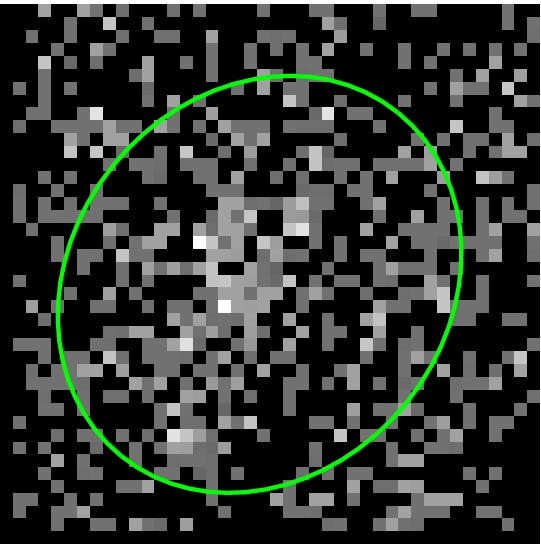}
}
\subfigure
{         
\includegraphics[scale=0.215]{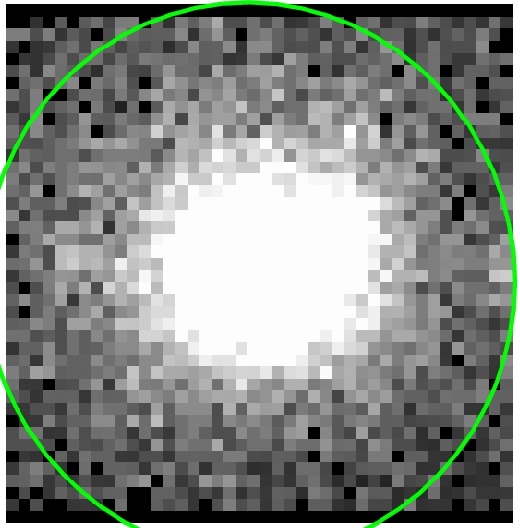}
}
\caption{Four XCS-DR1 clusters that have been classified as \textit{silver} in  \textit{Zoo}$^{\rm DR7}$ (Section~\ref{ClusterZoo}). False colour-composite images are $3\times3$\,arcmin with X-ray contours overlaid in blue. Corresponding X-ray images are shown below each optical image (lighter regions show areas of increased X-ray flux).  The shape of the {\sc Xapa}-detected extended source ellipses are highlighted in green. The right-most cluster is an example where the classification (as \textit{silver}) was based predominantly on the X-ray data, the other three are examples where the classification (as \textit{silver}) was based predominantly on the galaxy overdensity (these three represent the first \textit{silver} entries in  Table \ref{BigTable}).  From left to right, the clusters are: XMMXCS J004231.6$+$005119.9 at $z=0.15$; XMMXCS J004252.6$+$004303.1 at $z=0.27$; XMMXCS J004333.7$+$010109.6 at $z=0.20$; and XMMXCS J122658.1$+$333250.9 at $z=0.89$.}
\label{fig:SDSSSilver}
\end{center}
\end{figure*}

\begin{figure*}
\begin{center}
\subfigure
{
\includegraphics[scale=0.4]{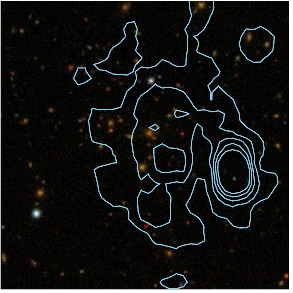}
}
\subfigure
{
\includegraphics[scale=0.4]{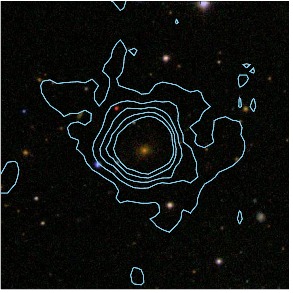}
}
\subfigure
{
\includegraphics[scale=0.4]{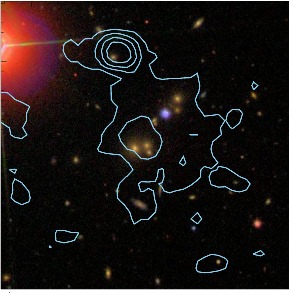}
}
\subfigure
{        
\includegraphics[scale=0.4]{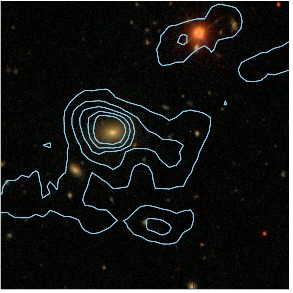}
} 
\\
\subfigure
{
\includegraphics[scale=0.215]{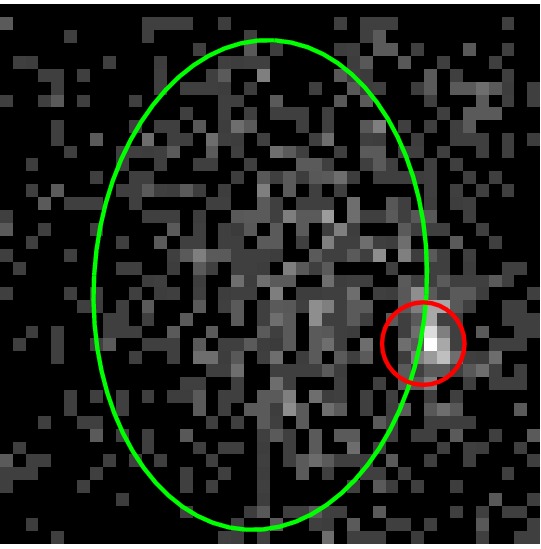}
}
\subfigure
{
\includegraphics[scale=0.215]{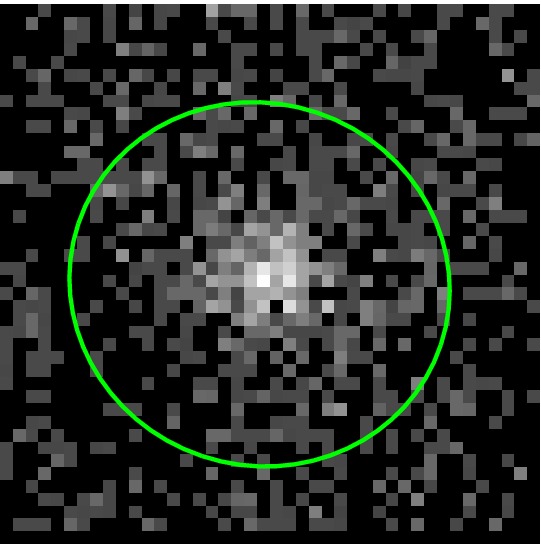}
}
\subfigure
{
\includegraphics[scale=0.215]{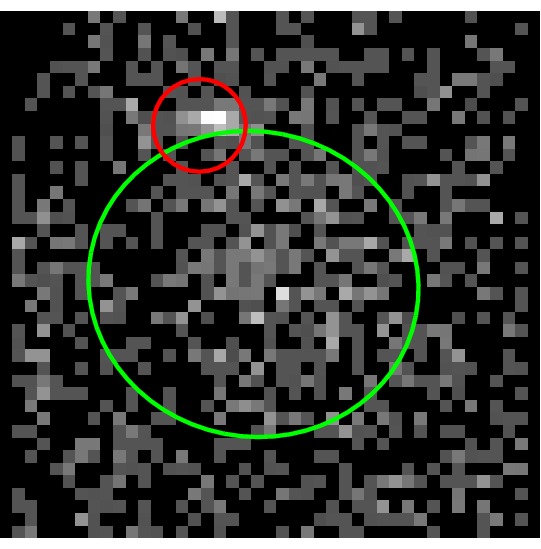}
}
\subfigure
{
\includegraphics[scale=0.215]{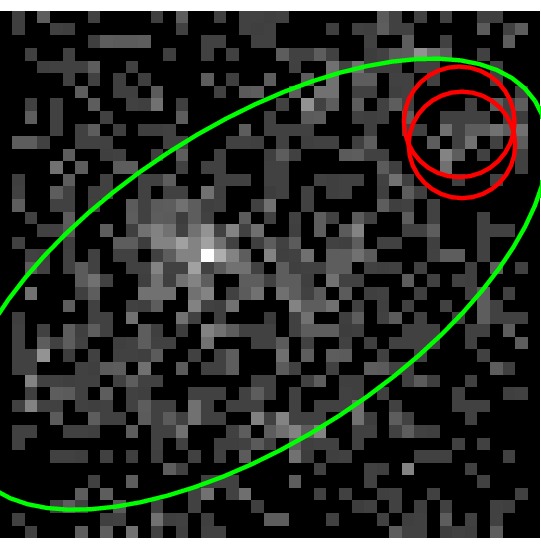}
}
\caption{The first four clusters in Table \ref{BigTable} classified as \textit{bronze} in \textit{Zoo}$^{\rm DR7}$ (Section~\ref{ClusterZoo}), all four have been optically confirmed using information in the literature (Section~\ref{Litz confirmation}). False colour-composite images are $3\times3$\,arcmin with X-ray contours overlaid in blue. Corresponding X-ray images are shown below each optical image (lighter regions show areas of increased X-ray flux).  The shape of the {\sc Xapa}-detected extended (point) source ellipses are  highlighted in green (red). From left to right, the clusters are: XMMXCS J092111.0$+$302758.2 at $z=0.43$;  XMMXCS J095951.4$+$014052.1 at $z=0.37$; XMMXCS J101056.3$+$555711.5 at $z=0.17$; and XMMXCS J103100.1$+$305134.9 at $z=0.14$.}
\label{fig:SDSSBronze}
\end{center}
\end{figure*}

To be assigned a classification of \textit{gold}, a candidate must have an unambiguous overdensity of galaxies coincident (i.e. within the extent of the {\sc Xapa} defined source ellipse) with an unambiguous\footnote{All XCS candidates are extended in a statistical sense, but only the high signal-to-noise sources stand out to the human eye as being unambiguously extended and without blend contamination.} extended X-ray source (Fig.~\ref{fig:SDSSGold}). Candidates classified as \textit{silver} must have either an unambiguous overdensity of galaxies associated with an acceptable extended X-ray source, or an unambiguous extended X-ray source associated with a suspected galaxy overdensity and/or BCG (Fig.~\ref{fig:SDSSSilver}). Candidates classified as \textit{bronze} were judged likely to be clusters, but could not be confirmed as such using only the information available in {\sc XCS-Zoo} (Fig.~\ref{fig:SDSSBronze}). 

Each category was allocated an integer (from 1 to 4), with 4 for \textit{gold}  through to 1 for  \textit{other}. The average value (after rounding down) was adopted for a particular candidate, based on the five (or more) classifications available per {\sc XCS-Zoo}. If a candidate was included in more than one {\sc XCS-Zoo}, and had gained different average categorisations, then the category with the highest numerical score was adopted.

\begin{figure*}
\begin{center}
\subfigure
{        
\includegraphics[scale=0.4]{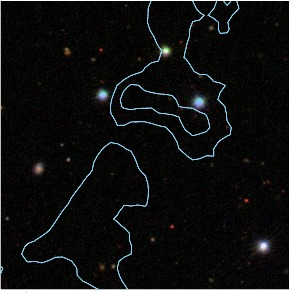}
}
\subfigure
{ 
\includegraphics[scale=0.4]{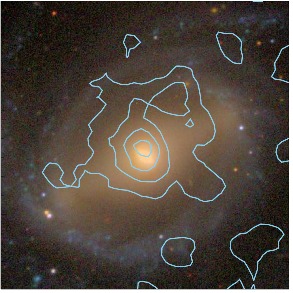}
}
\subfigure
{ 
\includegraphics[scale=0.4]{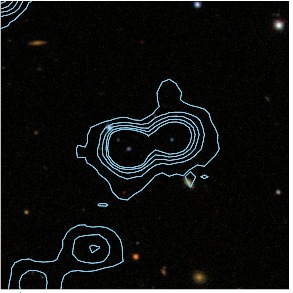}
}
\subfigure
{
\includegraphics[scale=0.4]{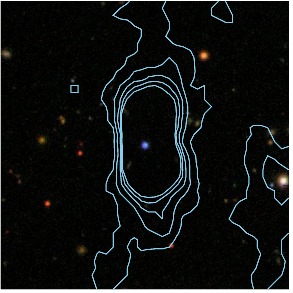}
}
\\      
\subfigure
{ 
\includegraphics[scale=0.215]{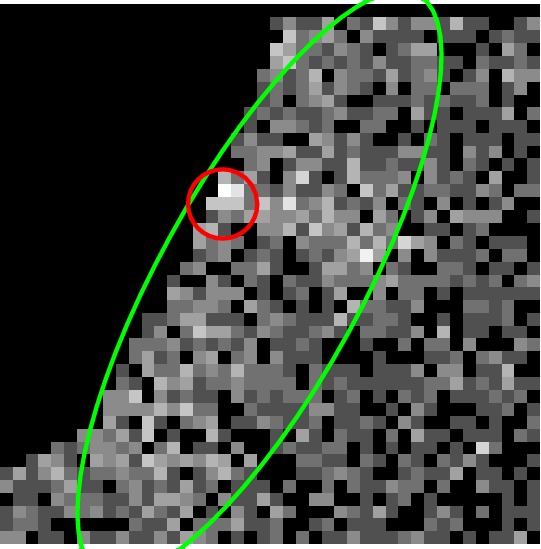}
}
\subfigure
{
\includegraphics[scale=0.215]{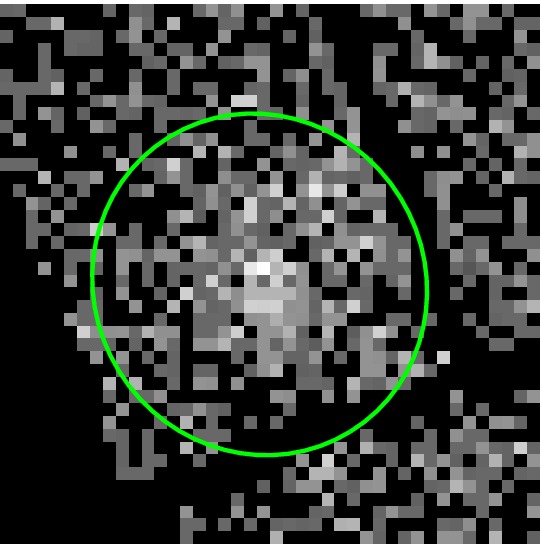}
}
\subfigure
{
\includegraphics[scale=0.215]{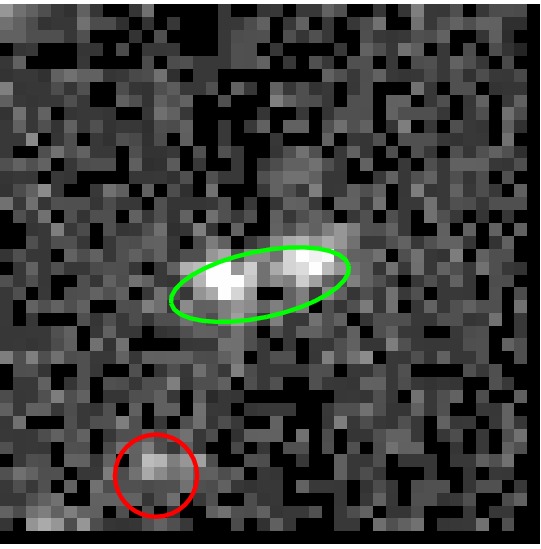}
}
\subfigure
{ 
\includegraphics[scale=0.215]{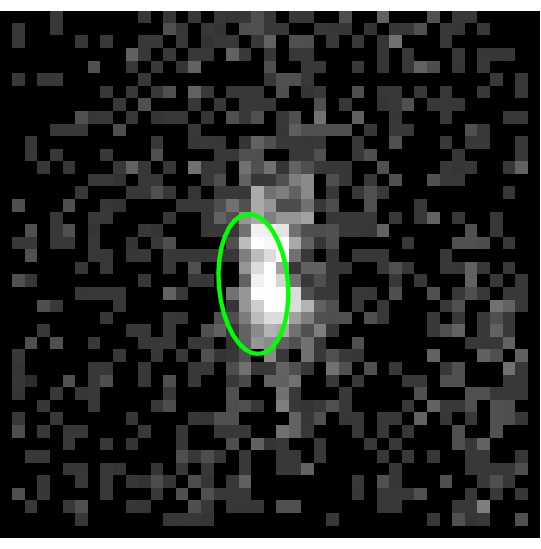}
}
\caption{A selection of XCS sources classified as \textit{other} in \textit{Zoo}$^{\rm DR7}$ (Section~\ref{ClusterZoo}). None of these objects are included in XCS-DR1$^{\dagger}$. False colour-composite images are $3\times3$\,arcmin with X-ray contours overlaid in blue. Corresponding X-ray images are shown below each optical image (lighter regions show areas of increased X-ray flux). The shape of the {\sc Xapa}-detected extended (point) source ellipse is highlighted in green (red). Reasons for a classification as \textit{other} include artifacts at the edge of ObsID masks (far left);  extended X-ray sources not associated with a galaxy cluster, such as a low-redshift galaxy (mid left); cases where neighbouring X-ray point sources have been blended by {\sc Xapa} into an erroneous extended source (mid right); and finally, cases of point sources misclassified as extended (because the point spread function model at the edge of the XMM field-of-view is inadequate; far right).}
\label{fig:SDSSOther}
\end{center}
\end{figure*}

\begin{figure*}
\begin{center}
\subfigure
{
\includegraphics[scale=0.4]{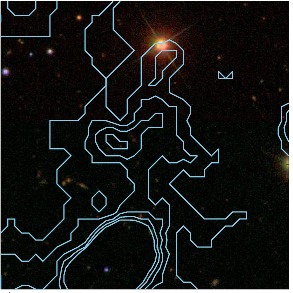}
}
\subfigure
{       
\includegraphics[scale=0.4]{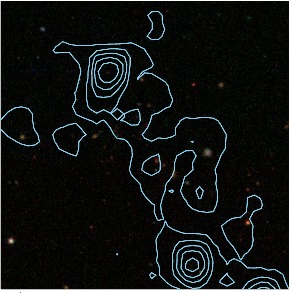}
}
\subfigure
{       
\includegraphics[scale=0.4]{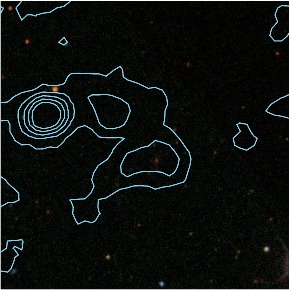}
}
\subfigure
{
\includegraphics[scale=0.4]{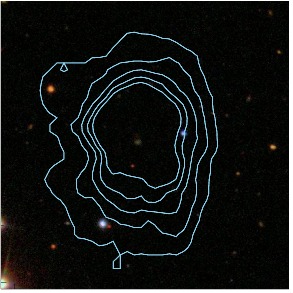}
} 
\\
\subfigure
{
\includegraphics[scale=0.40]{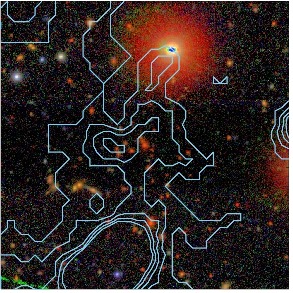}
}
\subfigure
{ 
\includegraphics[scale=0.40]{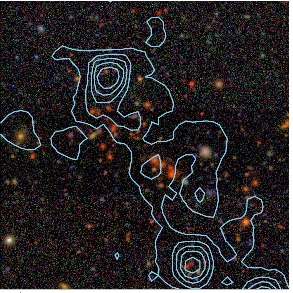}
}
\subfigure
{
\includegraphics[scale=0.21]{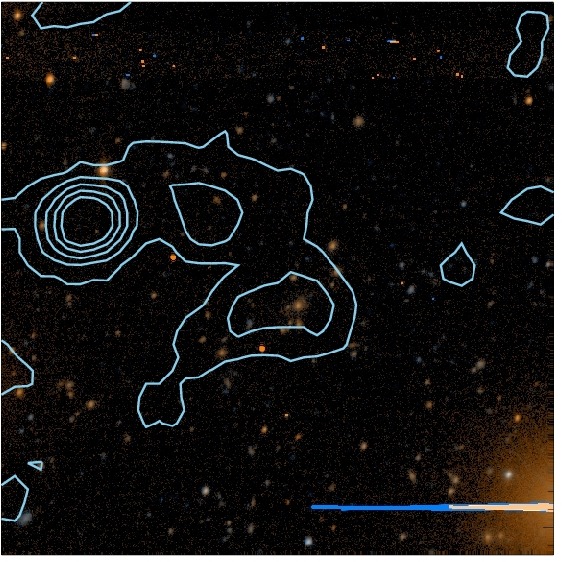}
}
\subfigure
{
\includegraphics[scale=0.21]{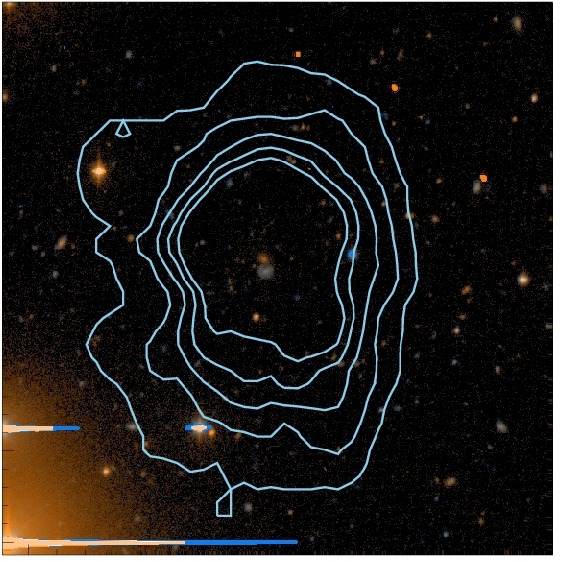}
}
\caption{Four examples of XCS-DR1 clusters in classified as \textit{bronze} in  \textit{Zoo}$^{\rm DR7}$ (Section~\ref{ClusterZoo}) that were subsequently classified as \textit{gold} or \textit{silver} in  \textit{Zoo}$^{\rm NXS}$ or  \textit{Zoo}$^{\rm S82}$. False colour-composite images are $3\times3$\,arcmin with X-ray contours overlaid in blue. Images from SDSS DR7 are shown above the corresponding deeper image (Stripe 82 on far and mid left, NXS on far and mid right). From left to right the clusters are: XMMXCS J030205.1$-$000003.6 at $z=0.65$; XMMXCS J030317.4$+$001238.4 at $z=0.59$; XMMXCS J083115.0$+$523453.9 at $z=0.52$ and XMMXCS J083025.9$+$524128.4 at $z=0.99$.}
\label{fig:SDSSBronze2Gold}
\end{center}
\end{figure*}

Excluding duplicates, the number of candidates classified as \textit{gold}, \textit{silver}, \textit{bronze}  and \textit{other} via {\sc XCS-Zoo} was \nQCONTc, \nQCONTd, \nQCONTe, and \nQCONTf, respectively. Including duplicates, \nQCONTg, \nQCONTb, and \nZARCc\, candidates were classified by \textit{Zoo}$^{\rm NXS}$, \textit{Zoo}$^{\rm DR7}$, and \textit{Zoo}$^{\rm S82}$, respectively. For the purposes of XCS-DR1, we have decided to include all candidates with \textit{gold} and \textit{silver} classifications, because we judge those to have been confirmed as clusters. By contrast, only a subset of those with \textit{bronze} classifications are included in XCS-DR1 because, based on {\sc XCS-Zoo} alone, we cannot be sure they are clusters (even if they have measured CMR-redshifts). Therefore, only  the \nQCONTh\, \textit{bronze} candidates that have been confirmed as being clusters by some other (to {\sc XCS-Zoo}) method, are included (Section~\ref{Litz confirmation}). Once deeper optical imaging, and/or multi-object spectroscopy, is available, we expect that many of the \nQCONTe\, candidates in the \textit{bronze} category will be confirmed as clusters. This has already been demonstrated in \nQCONTi\, cases where candidates that were categorised as \textit{bronze} in \textit{Zoo}$^{\rm DR7}$ were \textit{silver} or \textit{gold} in the \textit{Zoo}$^{\rm NXS}$ or \textit{Zoo}$^{\rm S82}$  (Fig.~\ref{fig:SDSSBronze2Gold}). In summary, excluding duplicates, \nQCONTj, \nQCONTk, and \nQCONTh\, candidates\footnote{The slight decrease compared to the numbers mentioned in the paragraph above is a result of the removal of some clusters that were either ObsID targets or associated with ObsID targets, Section~\ref{targetcheck}.} classified as \textit{gold}, \textit{silver}, and \textit{bronze} (and none of those  classified as \textit{other}, Section~\ref{Zoo-other}) appear in XCS-DR1 as confirmed clusters. 

In principle, we would like to include all the remaining clusters that fall within the NXS, DR7 and S82 footprints in future data releases. These comprise of 311 \textit{bronze} clusters and \nQCONTf\, \textit{other} objects. In practice, this is too many to follow-up individually, so we have decided to concentrate our efforts on the candidates$^{300}$. Applying the count threshold reduces the numbers of candidates requiring follow-up by roughly two thirds. Moreover, we have found (see Section~\ref{Zoo-other}) that 75 per cent of the \textit{other} candidates$^{300}$ do not require additional follow-up, but can rather be removed immediately (as contaminants) without impacting the completeness of a final cluster catalogue. Thus only \nQCONTl\, \textit{other}, in addition to the  95 \textit{bronze}, candidates$^{300}$ require additional follow-up. This process has recently begun based on imaging campaigns at the WHT and SALT telescopes\footnote{www.ing.iac.es, www.salt.ac.za}. The identities, and redshifts, of the candidates with fewer counts are likely to remain unknown until more sensitive large-area imaging surveys are publicly available (e.g. from the Dark Energy Survey, LSST, or Pan-Starrs4)\footnote{darkenergysurvey.org; lsst.org; pan-starrs.ifa.hawaii.edu}.


\subsubsection{Candidates not classified as clusters}
\label{Zoo-other}

The {\sc XCS-Zoo} exercise was primarily designed to pick out the obvious clusters in the candidate list; these clusters can be used in the short term for a variety of scientific applications (Section~\ref{disc: subsamples}) and in the longer term can be used to inform improvements to both the optical and X-ray methodology used by XCS. Therefore, anything that wasn't obviously a cluster ended up in the \textit{other} category. On the completion of  {\sc XCS-Zoo}, we reviewed all the candidates$^{300}$ in the \textit{other} category and found they could be sub-divided into the following classes:

\begin{enumerate}

\item
Masking or reduction issues ($\simeq 50$ per cent): Before running the {\sc Xapa} software on a given ObsID, the XCS generated image is examined by eye. Any sub-regions unsuitable for cluster searching are saved into a mask file, and some files are removed from the pipeline entirely. The removed files include those with atypically high backgrounds (this can occur if one of the \textit{XMM} cameras was behaving abnormally during the exposure). The masked regions include those covered by large extended objects, such as low-redshift clusters, or those with out-of-time bleed trails (see LD10).  The purpose of these eye-ball checks is to correct the XCS survey area for regions where serendipitous clusters could not have been found. However, {\sc XCS-Zoo} has shown that several high background files had not been excluded. Moreover, some of the image masks were not large enough and, as a result, {\sc Xapa} was either mistaking discontinuities at the mask edges as `sources' (e.g. Fig.~\ref{fig:SDSSOther}, far left panel), or detecting multiple portions of a large cluster as separate sources (because the largest {\sc Xapa} wavelet was too small to encompass the whole object). Both these problems can be solved by improving the checks of reduced images before they are passed to {\sc Xapa}. In future the checks will be made independently by at least two experienced XCS members (rather than relying on student volunteers, as was done previously). We are confident, therefore, that future generations of the candidate list will not be similarly contaminated by masking or reduction issues.


\item
Require additional follow-up: a) identity unknown ($\simeq 13$ per cent): In these cases, the identity of the candidate will not be established until more data is available. If any of these candidates are distant clusters, then they will be revealed using deeper optical (or IR) imaging (as was the case for the examples shown in Fig.~\ref{fig:SDSSBronze2Gold}). However, if any of them are blends or other artefacts (see items $v$ and $vi$), then additional X-ray imaging might be required, e.g. using the \textit{Chandra} X-ray observatory\footnote{http://chandra.harvard.edu}, because it has much higher spatial resolution than \textit{XMM}. 

\item
Require additional follow-up: b) clusters ($\simeq 12$ per cent): The {\sc XCS-Zoo} categories were set by rounding down the average value. So it was inevitable that some candidates judged likely to be clusters by some classifiers would end up in the \textit{other} category, rather than \textit{bronze}. We note that in one case (XMMXCS J074528.1$+$280011.3\symbolfootnote[2]{Images of this object are not available from http://xcs-home.org/datareleases because it is not part of XCS-DR1. Please contact the authors for more information.}), the candidate could have been classed as \textit{silver} because it is an `unambiguous extended X-ray source associated with a suspected galaxy overdensity'. However, the overdensity was only revealed after extra (to {\sc XCS-Zoo}) manipulation of the SDSS data; the location of the X-ray source falls under a bright star diffraction spike in the SDSS image. This cluster was detected with 1690 counts and so would easily yield a $T_{\rm X}$ value, if the redshift were known.

\item
Non-cluster X-ray source: a) extended but not a cluster ($\simeq 11$ per cent): The {\sc Xapa} software is designed to pick up extended objects, rather than clusters specifically, so contamination of the candidate list by non-cluster extended sources is to be expected. Fortunately, clusters are the only type of extended X-ray source outside of the Solar-System\footnote{Jupiter did originally appear in our list of \textit{other} candidates and, because it does not have a fixed location on the sky, did not have an SDSS counterpart. Even so, it was still identifiable, as not being a cluster, on the basis of its peculiar X-ray profile.} that are bright enough to be detected at high redshift, so any other types of extended sources, such as low-redshift galaxies, supernovae remnants or star-formation regions (e.g. Fig.~\ref{fig:SDSSOther}, mid-left panel), are straightforward to identify using {\sc XCS-Zoo}. We are improving our automated NED checks in order to remove more of this type of contaminating object from the candidate lists in future.

\item
Non-cluster X-ray source: b) blend ($\simeq 8$ per cent): Despite using multi-scale wavelet detections, {\sc Xapa} sometimes confuses emission from two or more neighbouring point sources as being the extended emission from a single object. Several obvious cases of blended emission were identified using {\sc XCS-Zoo}, either from the X-ray data directly, and/or with reference to optical images (e.g. Fig.~\ref{fig:SDSSOther}, mid-right panel). Blends will continue to affect our candidate lists in future, due to the limited spatial resolution of \textit{XMM}, and the most effective way to remove them will be to continue to use an exercise like {\sc XCS-Zoo}.

\item
Non-cluster X-ray source: c) bow-tie-shaped point source ($\simeq 6$ per cent):  {\sc Xapa} uses an \textit{XMM}-supplied circularly-symmetric PSF model to distinguish between point-like and extended sources. It is well-known that this model fails to describe the bow-tie-shaped nature of point source images at large off-axis angles. Such sources can be erroneously classified as extended by {\sc Xapa} and several examples were identified using {\sc XCS-Zoo} (e.g. Fig.~\ref{fig:SDSSOther}, far right panel). It is possible that an improved PSF model would help prevent these objects contaminating future candidate lists. If not, they can continue to be excluded at the {\sc XCS-Zoo} stage.

\end{enumerate}

\subsection{Candidate identification using the literature or multi-object spectroscopy}
\label{Litz confirmation}

In addition to using {\sc XCS-Zoo} (Section~\ref{ClusterZoo}), we have used information in the literature and our own multi-object spectroscopy to confirm candidates as clusters. To this end, we have examined candidates with associated redshifts that were either not part of {\sc XCS-Zoo} at all, or that were classified by it as \textit{bronze}. For those with redshifts from our spectroscopic follow-up campaign (Section~\ref{spec zs}), we have judged them to be confirmed as clusters if there are multiple concordant galaxy redshifts. There are \nQCONTm\, such cases. In addition, we confirmed one cluster, XMMXCS J231852.3$-$423147.6,  despite it having a spectroscopic redshift based on only one galaxy, because it was associated with an obvious galaxy overdensity in the Digitised Sky Survey (DSS\footnote{http://archive.stsci.edu/dss/}). For those with redshifts from the literature ($z_{\rm lit}$; Section~\ref{archive-z-lit}), we have used published material, in combination with the DSS to confirm \nQCONTn\, candidates as clusters (see below for explanation and FIg.~\ref{fig:SDSSBronze} for examples). XCS-DR1 clusters confirmed in either of these ways can be recognised because they either carry no indication of their {\sc XCS-Zoo} classification (because they were not part of it), or are flagged as \textit{b} for \textit{bronze} (see Section~\ref{table columns}).

For the $z_{\rm lit}$ candidates, we used the following criteria as evidence for confirmation: \textit{(i)} association with a galaxy overdensity that is obvious to the eye in the DSS (e.g. XMMXCS J000141.4$-$154031.3, $z=0.12$, and XMMXCS J005603.0$-$373248.0, $z=0.17$);  and/or  \textit{(ii)} association with a galaxy overdensity that is obvious to the eye in published optical/IR images (e.g. XMMXCS J000141.4$-$154031.3, $z=0.12$, and XMMXCS J005603.0$-$373248.0, $z=0.17$); and/or \textit{(iii)} $z_{\rm lit}$ values based on multi-object spectroscopy (e.g. XMMXCS J010422.4$-$063004.5, $z=0.95$, and XMMXCS J022738.5$-$031801.3, $z=0.84$);  and/or \textit{(iv)} membership in an optical/IR cluster catalogue that was constructed using an objective galaxy-based technique (e.g. XMMXCS J022618.3$-$040000.1, $z=0.20$, and XMMXCS J100053.2$+$022831.6, $z=0.3$). Moreover, the \textit{XMM} image of the candidate should be consistent with an extended X-ray source without blend contamination. We have been deliberately conservative with these `literature confirmations'. As a consequence, several candidates with associated $z_{\rm lit}$ values were not included in XCS-DR1 because they could not be confirmed using available resources, for example: XMMXCS J105251.8$+$573156.0$^{\dagger}$ ($z=0.58$).

\subsection{Candidates that were not serendipitous detections}
\label{targetcheck}

It is vital to the statistical integrity of XCS-derived cluster samples, that all the clusters are detected serendipitously by \textit{XMM}.  Therefore, filters are applied before the candidate list is generated to remove non-serendipitous or `target' clusters (LD10). The filters are generally very effective, even when the telescope was positioned so that the target clusters were detected away from the aim-point. Target filtering works both when the respective ObsID is classified in the \textit{XMM} database (and ObsID header) as having a cluster target, and when it does not. In the latter case, target clusters are identified by cross-checking the PI-supplied target names against NED. The NED cross-check can recognise most cluster names, but not when those names include atypical representations of cluster coordinates (see below for examples), so we can expect a small number of `target' clusters to contaminate the candidate list. Therefore, during {\sc XCS-Zoo}, we flagged up any candidates that might possibly be targets (based on their extent relative to their location in the ObsID) as a precautionary measure. We checked the ObsID headers for all of those so flagged individually. We also checked the headers for any reamining confirmed clusters with {\sc Xapa} centroids that were separated by 3 arcmin or less from the ObsID aim-point.

Of those candidates that were checked, the majority were confirmed to be valid members of the candidate list, e.g. XMMXCS J100029.2$+$024137.4 ($z=0.35$; \citealt{2007ApJS..172..182F}), which was detected near the \textit{XMM} aim-point, but was still a genuine serendipitous detection because that ObsID was part of a blind survey towards the COSMOS field \citep{Scoville-2007}. Other examples include candidates that were indeed the target of the ObsID viewed during {\sc XCS-Zoo}, but were also detected serendipitously in at least one other ObsID. In these instances, the ObsID with the target cluster was chosen by {\sc Xapa} to represent the candidate because it contained the most detected counts. Examples of such candidates include XMMXCS J130832.6$+$534214.2 ($z=0.33$) and XMMXCS J052215.4$-$362513.7 ($z=0.47$). 

Only seven candidates flagged as being potential target clusters by {\sc XCS-Zoo} turned out to be so when individually checked. These have not been included in XCS-DR1. In these seven cases, the PI-supplied target name did not match any of those listed for that cluster in NED. Examples include XMMXCS J092021.2$+$303005.7$^{\dagger}$, which is associated in NED the with literature cluster NSC J092017$+$303027 (\citealt{2003AJ....125.2064G}; $z=0.29$), but which had a target name of DLS09201$+$3029. Another example is XMMXCS J131914.6$-$005911.6$^{\dagger}$, which is associated in NED with the literature cluster SDSS CE J199.807541$-$00.985108 (\citealt{2002AJ....123.1807G}; $z=0.09$), but which had a target name of 2PI0.084J1319.3$-$005. 

It is also important to avoid including clusters in XCS samples that, despite not being the ObsID target, are physically associated with it; in these cases there is a better than random chance of the cluster entering the XCS survey volume. Therefore, a NED-based filter is run before the candidate list is drawn up to identify such cases (see LD10). However, this only works if the candidate's redshift is available in NED, and in many cases it is not. Therefore, we ran a similar filter on an initial XCS-DR1 list to highlight clusters with similar redshifts to their respective ObsID target (whether that target is a cluster or not). We found one such case, XMMXCS J083057.0$+$655059.2$^{\dagger}$, the cluster redshift was $z=0.21$ and the target redshift was  $z=0.18$, and so this cluster was removed from XCS-DR1.

\section{The XCS-DR1 cluster catalogue}
\label{The catalogue}

The first XCS data release (XCS-DR1) is presented in Tables~\ref{BigTable} and \ref{BigTable2}. It consists of \nABSa\, candidates that we have optically confirmed as being serendipitously detected X-ray clusters (Section~\ref{Quality Control}), \nCCATa\, of which were detected with more than 300 counts. The contents of each of the columns in Tables~\ref{BigTable} and \ref{BigTable2}, and the associated webpage, are explained in Section~\ref{table columns}. We describe the selection of redshifts for the clusters in Sections~\ref{best z}. In Section~\ref{photoz error}, we discuss the errors on the CMR-redshifts derived from both our own observations and from archival data. In Section~\ref{altname}, we describe the selection of alternative names for the clusters.

\begin{table*}
\caption{The XCS-DR1 Cluster Catalogue: Part I, redshifts and X-ray temperatures. A full version of Table \ref{BigTable} is provided in electronic format in the online version of the article. Descriptions of column entries and superscripts are provided in Section~\ref{table columns}.}
\label{BigTable}
\begin{tabular}{lrllrlr}
\hline
\hline
XCS ID    & Counts    & $z$    &  $z$-source    & $T_{X}$   & Alternative name & References \\
          &           &        &                & keV       &                  & [Name, $z$]\\
(1)       & (2)       & (3)    &  (4)           &  (5)      & (6)              & (7)\\
\hline
XMMXCS J000013.9$-$251052.1	& 878	& 0.08	& Lit$^{3}$$^{g}$$^{*}$  & 1.8$_{-0.2}^{+0.4}$ & APMCC 948	& [1,1]  \\
XMMXCS J000029.8$-$251211.4	& 652	& 0.15	& NXS$^{s}$$^{*}$  & 0.81$_{-0.05}^{+0.04}$ & 	&  \\
XMMXCS J000103.8$-$250353.6	& 362	& 0.91	& NXS$^{s}$  & 	& 	&   \\
XMMXCS J000141.4$-$154031.3	& 1135	& 0.12	& Lit$^{3}$  & 1.8$_{-0.1}^{+0.3}$ & RXC J0001.6$-$1540	& [2,2]  \\
XMMXCS J000626.2$+$195944.2	& 118	& 0.46	& NXS$^{g}$  & 	& 	&   \\
XMMXCS J001116.1$+$005211.3	& 155	& 0.36	& S82$^{s}$  & 0.7$_{-0.1}^{+0.1}$ & 	&  \\
XMMXCS J001328.5$-$272319.0	& 484	&	& NXS$^{1}$$^{s}$	&	& 	&  \\
XMMXCS J001345.2$-$271654.8	& 164	&	& NXS$^{1}$$^{s}$	&	& 	&  \\
XMMXCS J001639.1$-$010211.5	& 403	& 0.17	& S82$^{s}$$^{*}$  & 1.7$_{-0.4}^{+1.5}$ & MaxBCG J004.16184$-$01.03538	& [3,--]  \\
\hline
\end{tabular}
\end{table*}

\begin{table*}
\caption{The XCS-DR1 Cluster Catalogue: Part II, X-ray luminosities. A full version of Table \ref{BigTable2} is provided in electronic format in the online version of the article. Descriptions of column entries and superscripts fare provided in Section~\ref{table columns}.}
\label{BigTable2}
\begin{tabular}{clclclc|c|c|c|c}
\hline
\hline
XCS ID       &  $L_{500}$                  	& $R_{500}$	& $L_{200}$                  	& $R_{200}$	&$\beta$ &$r_c$		& Model\\
  & $10^{44}$ erg s$^{-1}$	&kpc  & $10^{44}$ erg s$^{-1}$	&kpc			&	&kpc					& used\\
(1)     & (2) & (3)      &  (4)      & (5)	& (6)	& (7)	& (8)		\\
\hline
XMMXCS J000013.9$-$251052.1 & 0.066$_{-0.009}^{+0.017}$ & 566$_{-28}^{+73}$ & 0.096$_{-0.015}^{+0.032}$ & 858$_{-43}^{+111}$ & 0.350$_{-0.021}^{+0.009}$ & 0.100$_{-0.016}^{+0.013}$ & 2 \\
XMMXCS J000029.8$-$251211.4 & 0.060$_{-0.021}^{+0.041}$ & 343$_{-11}^{+10}$ & 0.075$_{-0.036}^{+0.131}$ & 520$_{-16}^{+15}$ & 0.438$_{-0.088}^{+0.512}$ & 0.100$_{-0.000}^{+1.900}$ & 3 \\
XMMXCS J000103.8$-$250353.6 &  &  &  &  &  &  &  \\
XMMXCS J000141.4$-$154031.3 & 0.177 & 561$_{-16}^{+55}$ & 0.167 & 561$_{-16}^{+55}$ & 0.666 & 0.589 & 0 \\
XMMXCS J000626.2$+$195944.2 &  &  &  &  &  &  &  \\
XMMXCS J001116.1$+$005211.3 & 0.035$_{-0.027}^{+0.016}$ & 259$_{-13}^{+23}$ & 0.073$_{-0.059}^{+0.032}$ & 393$_{-20}^{+35}$ & 0.350$_{-0.046}^{+0.049}$ & 2.000$_{-0.335}^{+0.356}$ & 3 \\
XMMXCS J001328.5$-$272319.0 &  &  &  &  &  &  &  \\
XMMXCS J001345.2$-$271654.8 &  &  &  &  &  &  &  \\
XMMXCS J001639.1$-$010211.5 & 0.084$_{-0.068}^{+0.148}$ & 506$_{-81}^{+225}$ & 0.108$_{-0.092}^{+0.295}$ & 768$_{-123}^{+342}$ & 0.733$_{-0.148}^{+0.155}$ & 2.000$_{-1.015}^{+1.683}$ & 2 \\
\hline
\end{tabular}
\end{table*}

\subsection{The XCS-DR1 data table and webpage}
\label{table columns}

The columns in Table~\ref{BigTable} contain the following information:

\begin{description}

\item[(1)] 
The XCS name. Contained within the name are the positional coordinates (R.A. and Dec. in J2000) of the {\sc Xapa} determined X-ray centroid. 

\item[(2)] 
The number of counts (0.5-2.0 keV) detected from each cluster (see Section~\ref{introduction} for a definition of `counts').

\item[(3)] 
The adopted cluster redshift (Section~\ref{best z}).  

\item[(4)] 
The source of the redshift and, where applicable, the {\sc XCS-Zoo} classification. Superscripts $g$, $s$, and $b$ denote {\sc XCS-Zoo} classifications of \textit{gold}, \textit{silver} and \textit{bronze} respectively. If a cluster redshift is not presented in Column three, then Column four indicates which {\sc XCS-Zoo} was used to provide the optical confirmation. In these cases, a superscript 1 refers to instances where the NXS images used for the optical confirmation were taken under non-photometric conditions, and a superscript 2 denotes cases where the CMR-redshift was considered to be unreliable (Section~\ref{redsequence algorithm}). Superscripts 3, 4 and 5 denote literature redshifts derived from spectroscopic, photometric and X-ray data respectively. The symbol `*' denotes clusters that form part of a preliminary statistical subsample (Section~\ref{subsamp: statsam}).

\item[(5)] 
The measured X-ray temperature for each cluster, and the $1\sigma$ errors. The temperature fits are redshift dependent and those presented here  assume that the value listed in Column three is correct. 

\item[(6)] 
The adopted alternative cluster name taken from the literature (Section~\ref{altname}).

\item[(7)] 
The reference for the alternative name given in Column six, and, where applicable, a reference for the redshift in Column three.

\end{description}

The columns in Table~\ref{BigTable2} contain the following information:

\begin{description}

\item[(1)]  The XCS name (i.e. as column 1 in Table~\ref{BigTable}).
\item[(2)]  The bolometric ($0.05-100$\,keV band) luminosity in units of 10$^{44}$ erg s$^{-1}$ within a radius of $R_{500}$, and the 1$\sigma$ errors. Only clusters that have measured $T_{\rm X}$ values have $L_{\rm X}$ information (see LD10).
\item[(3)] The $R_{500}$ value used to derive the luminosity in Column 2.
\item[(4)] The bolometric ($0.05-100$\,keV band) luminosity in units of 10$^{44}$ erg s$^{-1}$ within a radius of $R_{200}$, and the 1$\sigma$ errors. Only clusters that have measured $T_{\rm X}$ values have $L_{\rm X}$ information (see LD10).
\item[(5)] The $R_{200}$ value used to derive the luminosity in Column 4.
\item[(6)] The power-law slope of the fitted $\beta$-profile. 
\item[(7)] The core radius in units of kpc. 
\item[(8)] The spatial model used to derive the luminosity. These models are defined in LD10, but  summarised here for completeness. They are all based on spherical $\beta$-profile model \citep{cavaliere76a}. In Model 0, both $\beta=2/3$ and $r_c$ are fixed (the later being estimated using the measured $T_{\rm X}$), and it is, therefore, not appropriate to include errors with the derived Lx values. Model 1 has $\beta$ fixed at the canonical value of 2/3, but allows $r_c$ to vary. Model 2 also allows $\beta$  to vary. Model 3 is similar to 2, but includes a central cusp (to replicate an AGN or a cool core). 

\end{description}

Similar information to that in Table~\ref{BigTable} appears on the XCS-DR1 webpage table (accessible from http://xcs-home.org/datareleases). The webpage table can be ordered by right ascension, redshift, and temperature. The key advantage of the webpage table, over Table~\ref{BigTable}, is that each XCS name connects to a separate page that contains X-ray and optical, greyscale and colour-composite, images. There are $3\times3$\,arcmin \textit{XMM} and optical cut-outs, plus the full field-of-view of the respective ObsID. The X-ray images can be viewed with or without the {\sc Xapa} defined source outlines, and the optical images can be viewed with or without \textit{XMM} surface brightness contours. The optical cut-outs are taken from NXS, SDSS DR7 or SDSS S82, where available. In the event of more than one of these being available, the deepest image is presented. When none are available (i.e. when the candidate was confirmed as being a cluster using either the literature or our own spectroscopy, Section~\ref{Litz confirmation}), the DSS image is shown.

A machine-readable version of XCS-DR1, that includes the information in both Table~\ref{BigTable} and Table~\ref{BigTable2}, is available from the same URL.

\subsection{Selection of the cluster redshift}
\label{best z}

A particular cluster may be associated with multiple redshift estimates, but each cluster is presented with only a single redshift in Column 3 of Table~\ref{BigTable}. That redshift is chosen based on the quality of the measurement, with spectroscopic redshifts almost always favoured over photometric redshifts, as we describe below. In total, \nABSb\, redshifts appear in XCS-DR1, as summarised in Table~\ref{bestZsummary}. We note that the right most column of Table~\ref{bestZsummary} sums to four more than \nABSb. This is because the three H11 entries are also included in the `Lit (added/changed by hand)' row. We have also double counted XMMXCS J221559.6$-$173816.2 because we used its literature redshift, but since that redshift based on our own spectroscopy, it also appears in the `Spec' row.

If a cluster has more than one spectroscopic redshift, then the typical hierarchy is as follows: those obtained by XCS team members are favoured over those from SDSS LRGs, which in turn are favoured over those taken from the literature (with $z_{\rm lit}$ values based on optical spectroscopy bring prioritised over those from X-ray spectroscopy). The redshift source for these clusters are indicated in Column 4 as `Spec', `LRG' and `Lit$^3$' respectively (Lit$^{5}$ is used in the case of X-ray literature redshifts). For the purposes of XCS-DR1, the uncertainty on the optical spectroscopic redshifts is assumed to be insignificant (being at the level of the cluster velocity dispersion, i.e. $\sigma_v < 2000$ km$\,$s$^{-1}$). The errors on the two X-ray redshifts used in XCS-DR1 are $\pm$0.005 and $\pm$0.03 respectively for XMMXCS J004624.5$+$420429.5 (or RX J0046.4+4204 at $z_x=0.3$; \citealt{2006ApJ...641..756K}), and XMMXCS J083025.9$+$524128.4 (or 2XMM J083026$+$524133 at $z_x=0.99$; \citealt{Lamer-2008}).

The adopted hierarchy of spectroscopic redshifts is based on the assumption that spectra obtained closest to the {\sc Xapa} centroid are most likely to be of cluster members. Since most of the literature redshifts were obtained for clusters selected using optical methods, and the optical centroid can differ from the {\sc Xapa} one, it seemed prudent to use our own (or LRG) spectra rather than published values (even when the published value was based on more galaxies). We have also chosen to adopt a spectroscopic redshift from H11 over our own LRG redshift for the cluster XMMXCS J030659.8$+$000824.9. The H11 study, a search for `fossil' systems in the \textit{XMM} archive, involves a redshift allocation process based on SDSS spectroscopy. The H11 method differs from that described in Section~\ref{LRG Speczs} in that it is based on all available galaxy spectra, rather than just those for LRGs. There is significant overlap (in the common region, i.e. the SDSS footprint) between the objects in the H11 study and the candidate list used for this paper (although, in H11, non-serendipitous -- or \textit{XMM}-target -- sources were also included). We have therefore cross-checked the XCS-DR1 redshift assigned by the default hierarchy against the H11 determined values using \nCCATb\, XCS-DR1 clusters. The redshifts match very closely with \nCCATc\, exceptions. We have examined these and determined that in \nCCATd\, instance (XMMXCS J030659.8$+$000824.9), the H11 value was more reliable than our default choice, because the redshift was based on galaxies closer to the {\sc Xapa} centroid (in the other, XMMXCS J133605.0+514531.2, it was clear from the DR7 image that our default redshift choice, $z=0.53$ , was more appropriate than that of H11, $z_{H11}=0.234$). We note, as previously mentioned in Section~\ref{archive-z-lit}, that we have also used H11 redshifts for two confirmed XCS-DR1 clusters that had no other available redshift information. All three of the H11 determined redshifts listed in Table~\ref{BigTable} are denoted as `H11' in column 4.

If no spectroscopic redshift was available, but a CMR-redshift was, then this will be listed in column 3, with its origin indicated as `NXS', `DR7' or `S82' in column 4. The uncertainty associated with these CMR-redshifts is $\sigma_z=0.08$, $\sigma_z=0.03$ and $\sigma_z=0.03$ respectively (see Section~\ref{photoz error}).  When more than one CMR-redshift is available for a particular cluster, then the value chosen is governed by the relative quality of imaging, so that those taken from S82 are favoured over those from NXS, which are in turn favoured over those from SDSS DR7. The CMR-redshift algorithm is not able to determine redshifts below $z=0.1$ (see Section~\ref{redsequence algorithm}). Thus, CMR-redshifts with values of exactly $z=0.1$ are taken as upper limits and presented in Table~\ref{BigTable} as $z\leq0.1$. In one instance (XMMXCS J064423.6$+$822626.5), we suspect that the CMR-redshift ($z_{\rm CMR} = 0.84$) is a catastrophic failure, based on the appearance of the NXS image (the cluster falls over a MOSAIC chip gap; Section~\ref{NXS}) and we have not included a redshift for it XCS-DR1. 

Finally, when no other redshift information is available, but a non-spectroscopic literature redshift is, then that is used in XCS-DR1. One exception to this hierarchy was the use of a photometric literature redshift ($z=0.29$) for XMMXCS J090101.5$+$600606.2. That redshift came from from the MaxBCG catalogue \citep{Koester:2007} and was chosen over the corresponding SDSS DR7 CMR-redshift ($z=0.23$) because that system had been included in some initial testing of a multi-colour-based method to determine CMR-redshifts (Section~\ref{disc: follow-up-Archive}), and that method placed the cluster closer to the MaxBCG value.

\begin{table*}
\caption{A summary of the redshifts used in XCS-DR1.}
   \label{bestZsummary}
   \begin{tabular}{llll}
     \hline
     \hline
     Redshift Source 			& Candidates 	& Overlap with	& Used in \\
                               		&               & DR1 clusters	&  XCS-DR1 \\ \hline
Lit (auto NED query)			&\nZARCh	&\nCCATq	&\nCCATi\\
LRG						&\nZARCg	&\nCCATr		&\nCCATj\\
CMR-NXS					&\nZNEWg	&\nCCATs	&\nCCATk\\
CMR-DR7					&\nZARCd	&\nCCATt		&\nCCATl\\
Spec						&\nZNEWh	&\nZNEWh	&\nCCATm\\
CMR-S82					&\nZARCe	&\nCCATu	&\nCCATn\\
Lit (added/changed by hand)	&\nCCATo	&\nCCATo	&\nCCATo\\
H11 method				&\textit{n.a.}		&\nCCATb	&\nCCATp\\ 
\hline
\end{tabular}
\end{table*}

\subsection{Photometric redshift accuracy}
\label{photoz error}

We have derived CMR-redshifts from NXS and SDSS DR7 and S82 data for \nZNEWg, \nZARCd, and \nZARCe\, confirmed clusters respectively (including duplicates), see Sections~\ref{redsequence algorithm}, \ref{Application to the SDSS DR7} and Table~\ref{bestZsummary}. We have been able to extract optical spectroscopic redshifts for a fraction of these candidates using either our own observations, the SDSS archive, or the literature (Sections~\ref{spec zs}, \ref{LRG Speczs}, and \ref{archive-z-lit} respectively). These spectroscopic redshifts have allowed us to determine the typical accuracy, and the catastrophic failure rate, of the CMR-redshifts presented in XCS-DR1.

\begin{figure}
\includegraphics[width=84mm]{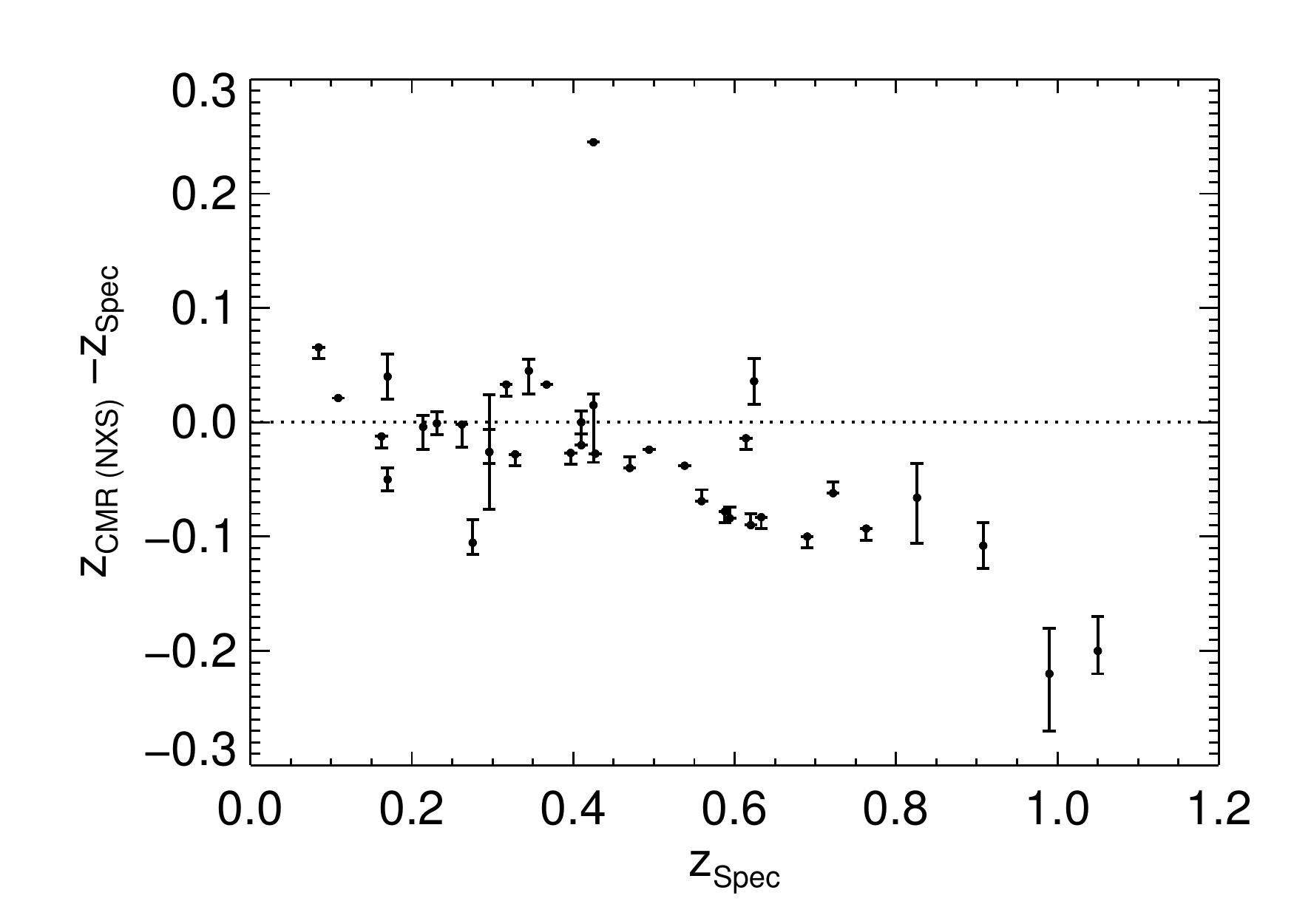}
\caption{A comparison between photometric CMR-redshifts obtained from NXS imaging (Section~\ref{redsequence algorithm}) and corresponding spectroscopic redshifts (Sections~\ref{spec zs}, \ref{LRG Speczs} and \ref{archive-z-lit}). The comparison uses \nZCOMPa\, clusters optically confirmed as \textit{gold} or \textit{silver} by \textit{Zoo}$^{\rm NXS}$ (Section~\ref{ClusterZoo}). Note that one of these \nZCOMPa\, is not shown in the figure (XMMXCS J221559.6$-$173816.2), because it has an anomonously large redshift offset (see Table~\ref{t_outliers}). The bars indicate the statistical 1$\sigma$ limits on each CMR-redshift (Section~\ref{redsequence algorithm}). Only those clusters with CMR-redshifts obtained from photometrically calibrated data with a minimum of 5 galaxies and with statistical uncertainties of $\sigma_z<0.1$ have been used in the comparison. The dotted line shows the one-to-one relation.}
\label{NXS_speczcomp}
\end{figure}

\begin{figure}
\includegraphics[width=84mm]{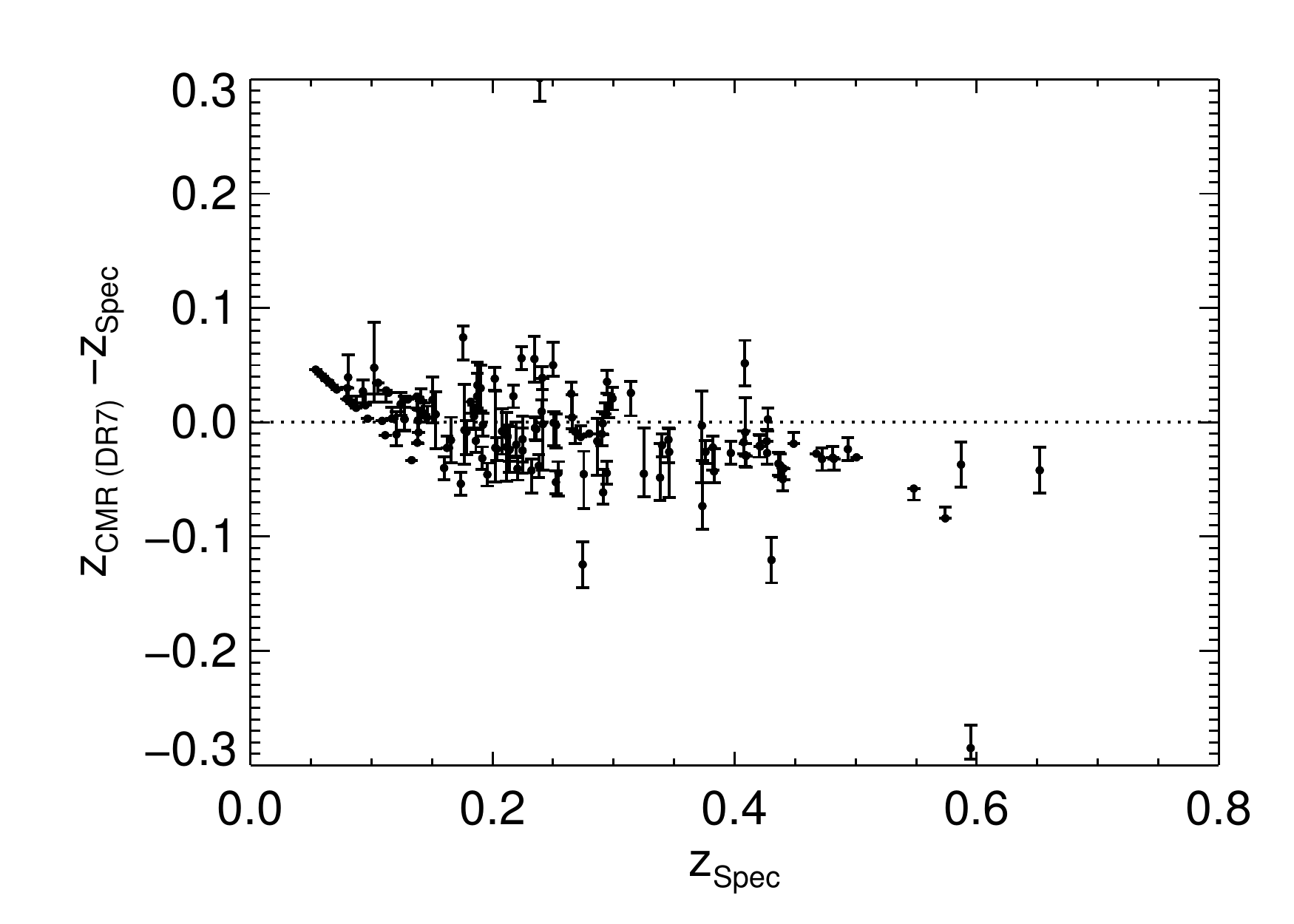}
\caption{A comparison between photometric CMR-redshifts obtained from SDSS DR7 imaging (Section~\ref{Application to the SDSS DR7}) and corresponding spectroscopic redshifts (Sections~\ref{spec zs}, \ref{LRG Speczs} and \ref{archive-z-lit}). The comparison uses \nZCOMPb\, clusters optically confirmed as \textit{gold} or \textit{silver} by \textit{Zoo}$^{\rm DR7}$ (Section~\ref{ClusterZoo}). The bars indicate the statistical 1$\sigma$ limits on each CMR-redshift (Section~\ref{redsequence algorithm}). Only those clusters with CMR-redshifts obtained from a minimum of 5 galaxies and with statistical uncertainties of $\sigma_z<0.1$ have been used in the comparison. The dotted line shows the one-to-one relation.}
\label{SDSS_speczcomp}
\end{figure}

\begin{figure}
\includegraphics[width=84mm]{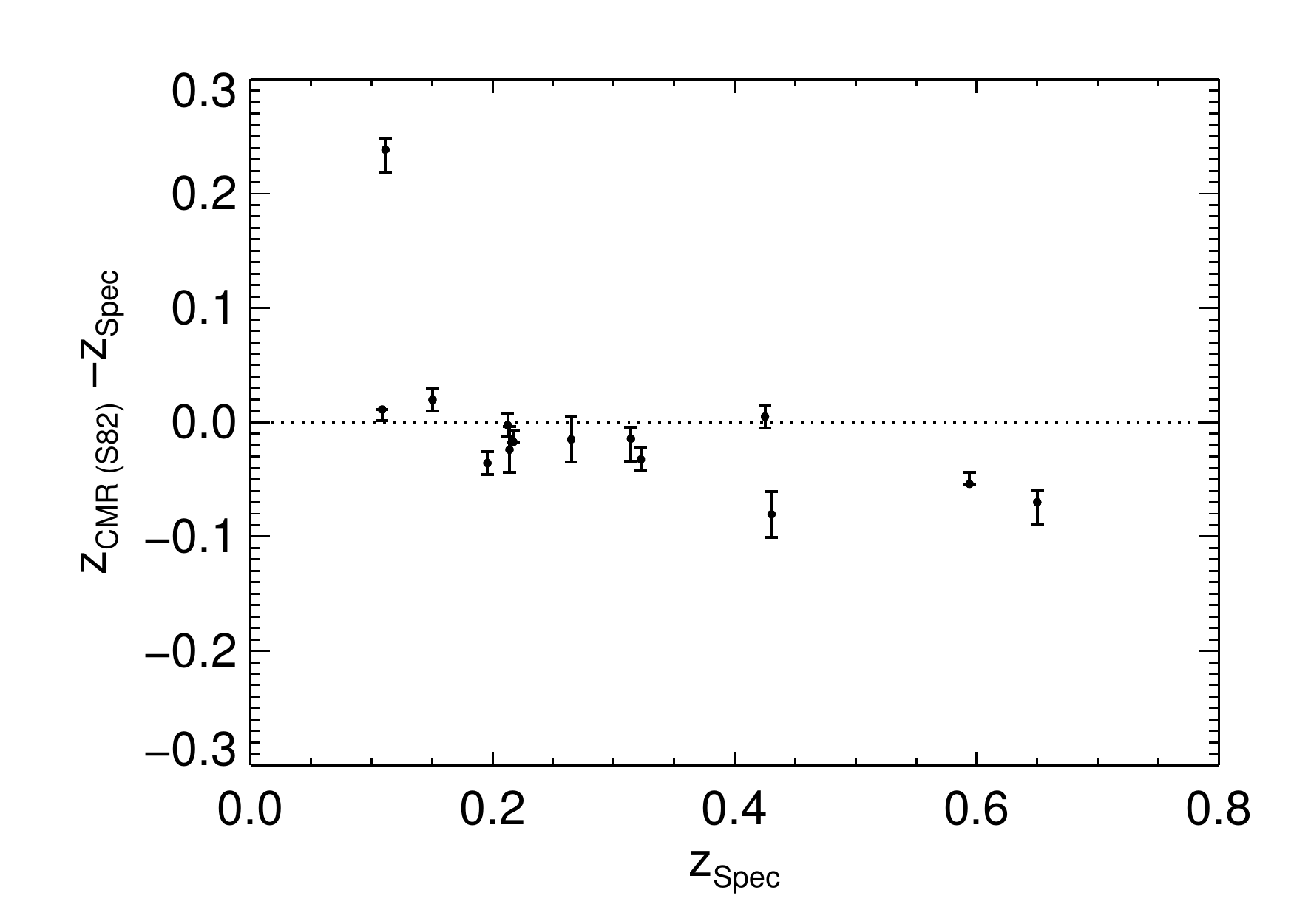}
\caption{A comparison between photometric CMR-redshifts obtained from the SDSS S82 (Section~\ref{Application to the SDSS DR7}) and corresponding spectroscopic redshifts (Sections~\ref{spec zs}, \ref{LRG Speczs} and \ref{archive-z-lit}). The comparison uses \nZCOMPc\, clusters optically confirmed as \textit{gold} or \textit{silver} by \textit{Zoo}$^{\rm S82}$ (Section~\ref{ClusterZoo}). The bars indicate the statistical 1$\sigma$ limits on each CMR-redshift (Section~\ref{redsequence algorithm}). Only those clusters with CMR-redshifts obtained from a minimum of 5 galaxies and with statistical uncertainties of $\sigma_z<0.1$ have been used in the comparison. The dotted line shows the one-to-one relation.}
\label{Stripe82_speczcomp}
\end{figure}

Figures~\ref{NXS_speczcomp}, \ref{SDSS_speczcomp} and \ref{Stripe82_speczcomp} show the results for NXS, SDSS DR7 and S82 respectively, under the assumption that the spectroscopic redshift is the true value. We note that only \textit{gold} and \textit{silver} clusters from XCS-DR1 were included in this comparison. The typical CMR-redshift uncertainty has thus been determined, from a 3$\sigma$ clipped mean, to be $\sigma_z=0.08$, $\sigma_z=0.03$ and $\sigma_z=0.03$ for NXS, SDSS DR7 and S82 respectively. These redshift uncertainties, are similar to those obtained by other authors using the single colour technique.  For example, \citet{gladders-2004b} estimate red-sequence redshift errors of $\sigma_z =0.05$ between $0.2<z<1$ using a single colour. 

We note that a mean offset of $\Delta z=0.03$, $\Delta z=0.01$ and $\Delta z=0.02$ was found for NXS, SDSS DR7 and S82 respectively, plus there is a trend to systematically underestimate CMR-redshifts above $z_{\rm spec} \simeq 0.5$ (Fig.~\ref{NXS_speczcomp}). Given that these offsets are smaller than the statistical errors, and there are very few spectroscopic redshifts available beyond $z \simeq 0.5$ with which to explore the redshift trend, we have not to adjust the CMR-redshifts to compensate for them. 

\begin{table*}
\caption{XCS clusters with  CMR-redshifts that were more than $3\sigma$ from their respective spectroscopic redshift. The XCS name is given in Column 1. The CMR-redshift, and the imaging survey from which it was derived, are listed in Columns 2 and 3. The measured spectroscopic redshift and its source are given in Columns 4 and 5.}
\label{t_outliers}
\begin{tabular}{|l|l|l|l|l|}
\hline
\hline
XCS ID  & $z_{CMR}$ & imaging & $z_{\rm spec}$ & $z_{\rm spec}$ \\
 & value  & survey & & source \\ 
(1)       & (2)       & (3)    &  (4)           &  (5)\\
\hline
XMMXCS J030644.2$-$000112.7        & 0.35  & S82  & 0.11    & LRG (Section~\ref{LRG Speczs})                         \\
XMMXCS J033556.2$+$003214.7        & 0.31  & DR7  & 0.43    & LRG  (Section~\ref{LRG Speczs})    \\
XMMXCS J124100.8$+$325959.9        & 0.15  & DR7  & 0.27    & LRG    (Section~\ref{LRG Speczs})                    \\
XMMXCS J133514.1$+$374905.8        & 0.31  & DR7  & 0.60    & Literature  (Section~\ref{archive-z-lit}) \\ 
XMMXCS J163341.0$+$571420.1        & 0.54  & DR7  & 0.24    &  Literature  (Section~\ref{archive-z-lit})  \\ 
XMMXCS J204134.7$-$350901.2        & 0.67  & NXS  & 0.43    & XCS      (Section~\ref{spec zs})     \\
XMMXCS J221559.6$-$173816.2        & 0.42  & NXS  & 1.46    & XCS    (Section~\ref{spec zs})                        \\ 
\hline
\end{tabular}
\end{table*}

All of the 3$\sigma$ redshift outliers are listed in Table \ref{t_outliers}. These outliers either represent catastrophic failures of the CMR technique or indicate incidences where the adopted spectroscopic redshift was wrong. In the following cases the CMR-redshift method has broken down:  XMMXCS J133514.1$+$374905.8 and XMMXCS J221559.6$-$173816.2 (both clusters are at redshifts beyond the grasp of their respective imaging surveys). In the case of XMMXCS J163341.0$+$571420.1, the candidate lies right at the edge of the SDSS footprint, so that only a fraction of the extraction region contains catalogued SDSS galaxies; we will refine our CMR-redshift techniques to filter out such objects in future. In the remaining cases  (XMMXCS J030644.2$-$000112.7, XMMXCS J033556.2$+$003214.7, XMMXCS J124100.8$+$325959.9, and XMMXCS J204134.7$-$350901.2), it is difficult to say if the fault lies with the CMR-redshift method or with the adopted spectroscopic redshift, as the $z_{\rm spec}$ values are based on a single galaxy. More spectroscopy would be needed to confirm these redshifts, and hence improve our estimate of the CMR-redshift catastrophic failure rate.  However, assuming all entries in Table~\ref{t_outliers} to be CMR failures, the failure rate is then $\simeq$5, $\simeq$3\, and $\simeq$7 per cent for NXS, SDSS DR7, and S82, respectively. 


\subsection{Selection of alternative names}
\label{altname}

Many of the XCS-DR1 clusters have been catalogued before by previous authors. In order to give due credit to this earlier work, we have matched XCS-DR1 clusters to catalogued clusters using an automated NED query. This query was run separately to that used to extract literature redshifts (Section~\ref{archive-z-lit}) and involved a simple search for any NED object classified as a galaxy cluster within a fixed radius of the {\sc Xapa}-defined centroid. The radius used was the mean of the major and minor axes of the {\sc Xapa} defined source ellipse. In the event of several NED matches within this radius, the default top choice listed by NED (ordered by seperation, on the date the query was made) was used. We concede that this approach is not ideal, in that it might not select the historical name of a cluster (e.g. one taken from the Abell catalogue, \citealt{1958ApJS....3..211A}), if another catalogue (e.g. MaxBCG,  \citealt{Koester-2007b}) has an entry with an optical centroid closer to the {\sc Xapa} position. We also note that if the selected redshift (Section~\ref{best z}) of the XCS-DR1 cluster does not come from a literature source, then the redshift in XCS-DR1 might differ from the NED redshift for the previously catalogued cluster. In addition to the automated NED query, some alternative names have been added by hand from papers too recent to have been in NED at the time the search was carried out (see Section~\ref{archive-z-lit}). In total, \nCCATe\, of the XCS-DR1 clusters have been matched with alternative names

\section{Summary and discussion}
\label{Discussion}

In Fig.~\ref{ELDFig1} we introduced the steps involved in the development of XCS-DR1. In this paper we have focussed on the steps involving redshift follow-up, quality control and cluster catalogue compilation. In this section we summarise those aspects and highlight areas for future development. 

\subsection{Redshift follow-up (new observations)}
\label{disc: follow-up-NEW}

In Section~\ref{z-new} we described new observations obtained to confirm XCS candidates as clusters and measure cluster redshifts. Section~\ref{NXS} focussed on the methodology of the NXS (NOAO--\textit{XMM} Cluster Survey) imaging programme and Section~\ref{redsequence algorithm} presented the CMR-redshift method used to estimate redshifts from NXS galaxy catalogues. Over the course of 38 nights, a total of \nZNEWb\, candidates were targeted by NXS using the MOSAIC cameras at the KPNO and CTIO 4-m telescopes. Excluding unreliable fits, a total of \nZNEWg\, CMR-redshift measurements were made using NXS data, \nCCATk\, of which appear in XCS-DR1. The accuracy and catastrophic failure rate of the NXS CMR-redshifts were estimated to be $\sigma_z=0.08$ and $\simeq$5 per cent. 

The original NXS goals were \textit{(i)} to image, and derive CMR-redshifts for, as many XCS candidates as possible, whilst \textit{(ii)} also providing a useful data set for the X-ray community, i.e. deep imaging in two bands over a large number of \textit{XMM} fields. With regard to our progress towards those goals, we are satisfied with the procedures used for NXS target selection, data reduction and photometric calibration.  That said, it may be possible to improve the CMR-redshift accuracy, and reduce the catastrophic failure rate, by investigating different (to \citealt{Bruzual-2003}) population synthesis models, to see if  the redshift drift seen in Fig.~\ref{NXS_speczcomp} can be reduced. It might also be possible to improve the quality of the field galaxy sample used for the NXS CMR-redshifts, by using similarly sensitive archival data. 


Looking ahead, plan secure additional CCD imaging on 4-meter class telescopes. This will allow us to improve on, and extend, the current NXS results in three ways. First, we would like to calibrate the \nZNEWf\, NXS-fields that currently lack photometric calibration; this would allow additional CMR-redshifts to be extracted from NXS. Second, there are hundreds of candidates that have yet to be included in an {\sc XCS-Zoo} identification exercise to the depth of \textit{Zoo}$^{\rm NXS}$ and  \textit{Zoo}$^{\rm S82}$, including \nDISCa\, candidates$^{300}$ that have not been included in any {\sc XCS-Zoo}, and a further \nDISCb\, candidates$^{300}$ that were classified as \textit{bronze} in \textit{Zoo}$^{\rm DR7}$.  As mentioned earlier (Section~\ref{ClusterZoo}), we have already started the imaging follow-up of these objects. However we note that these new images will not be as useful to the X-ray community, as those obtained during NXS, because they will not necessarily provide imaging across entire \textit{XMM} field (opting to target specific candidates instead). Third, we could improve the accuracy of the existing CMR-redshifts using additional observations through other filters. Other authors have shown that multi-colour photometric redshifts are more accurate than our single-colour ones. For example, \citet{Song-2011} and \citet{High-2010} achieve typical uncertainties of 2 per cent in $\Delta_{z}/(1+z)$, to $z<0.5$ and $z<1$ respectively, using multi-colour data (both studies use the mean colour of the red-sequence as the cluster redshift estimator). Two other multi-colour cluster finders,  MaxBCG and GMBCG (\citealt{Koester-2007b} and \citealt{Hao-2010}) achieve uncertainties of $\sigma_z \simeq 0.01$ ($z<0.3$) and $\sigma_z = 0.015$ ($z<0.55$) respectively (both studies use the cluster red-sequence for cluster finding, but adopt the photometric redshift of the identified BCG as the cluster redshift). The addition of extra filters will mostly benefit the high-redshift end, but bluer filters would be of benefit at the low-redshift end also (i.e. allowing us to bridge the 4000\AA\, break at $z<0.3$). 

Section~\ref{spec zs} presented the results to date (June 2011) from optical spectroscopy performed by XCS team members; \nDISCd\, new (and one previously published) spectroscopic cluster redshifts (Table~\ref{t_specz}). We highlighted a new (to the literature) $z>1$ cluster with multi-object spectroscopic confirmation (XMMXCS J091821.9$+$211446.0, $z=1.01$, Fig.~\ref{J0918}), and a new (to the literature) cluster, XMMXCS J015241.1$-$133855.9 at $z=0.83$, that is most likely associated with the well-studied merger system XMMXCS J015242.2$-$135746.8 (or WARP J0152.7$-$1357). The new spectroscopic redshifts have been invaluable with regard to the calibration of the CMR-redshifts and the derivation of X-ray temperatures. We intend to continue the spectroscopic follow-up of XCS candidates for several more years, with the emphasis on multi-object spectroscopy where possible.

\subsection{Redshift follow-up (archive)}
\label{disc: follow-up-Archive}

In Section~\ref{z-archive}, we described how we used data in public archives, and the literature, to collect more redshifts for our candidates (see Table~\ref{bestZsummary} for a summary). In Section~\ref{Application to the SDSS DR7}, we described how we applied the CMR-redshift technique designed for NXS to SDSS DR7 and S82 data. Excluding unreliable fits, a total of \nZARCd\, and \nZARCe\, CMR-redshift measurements were made using regular SDSS DR7 and S82 data respectively, \nCCATl\, and \nCCATn\, of which appear in XCS-DR1. The accuracy (and catastrophic failure rates) of the SDSS CMR-redshifts were estimated to be $\sigma_z=0.03$ ($\simeq$3 per cent) and $\sigma_z=0.03$ ($\simeq$7 per cent) for SDSS DR7 and S82 respectively. With regard to the future improvement of SDSS CMR-redshifts, many of the statements made above, with regard to NXS, also apply here. However, in the case of SDSS, we already have the option of using more than one colour, so we are now investigating the use of multi-colour data to derive CMR-redshifts from SDSS DR8. We also note that SDSS DR8 covers more area than DR7, so should yield additional (to XCS-DR1) redshifts.

In Section~\ref{LRG Speczs}, we described how spectroscopic redshifts were extracted from the SDSS archive under the assumption that LRGs reside in the centres of X-ray clusters. A total of \nZNEWi\, spectroscopic redshifts were determined from SDSS LRGs, \nCCATj\, of which appear in XCS-DR1. An additional three spectroscopic redshifts were adopted for XCS-DR1 clusters from the H11 XCS study. We look forward to the public release of additional SDSS spectroscopy from the SDSS-III BOSS project\footnote{www.sdss3.org}, because this will allow us to extract more spectroscopic redshifts for XCS clusters.

In Section~\ref{archive-z-lit}, we described how redshifts were extracted from the literature (mostly via automated queries to NED) and matched to XCS candidates. Whilst the use of $z_{\rm lit}$'s does cut down on the quantity of new optical follow-up required, it does have associated risks (i.e. that the selected redshift is not appropriate). Therefore, when assigning redshifts to clusters (if more than one redshift source was available), we tended not to use the NED derived value. A total of \nZARCh\, $z_{\rm lit}$ values were collected from NED, of which \nCCATi\, are presented in XCS-DR1. An additional \nCCATo\, redshifts from literature sources are included in XCS-DR1, bringing the total to \nZARCi\, (\nZARCk\, of these being spectroscopic in nature).

\subsection{Quality control}
\label{disc: quality control}
 
In Section~\ref{Quality Control}, we described the procedures used to confirm the identity of candidates as serendipitously detected clusters. In Section~\ref{ClusterZoo} we described an exercise, {\sc XCS-Zoo}, that used the consensus opinion of at least five (of 23) volunteers to classify candidates. Those candidates classified as \textit{gold} and \textit{silver} were judged to have been `confirmed' as clusters and appear in XCS-DR1.  Those classified as \textit{bronze} do not appear, unless additional information was available. In summary,  \nDISCe\, candidates were `confirmed' as being clusters (i.e. were classified as \textit{gold} or \textit{silver}) using  {\sc XCS-Zoo}.

The remaining category,  \textit{other},  was used for candidates that did not fall into the  \textit{gold}, \textit{silver} or \textit{bronze} categories. Just over half of all candidates classified by {\sc XCS-Zoo} were placed in this category. Subsequently to {\sc XCS-Zoo}, we classified the \textit{other} candidates$^{300}$, finding that $\simeq 25$ per cent required more optical and/or X-ray follow-up before they could be identified (Section~\ref{Zoo-other}). The other $\simeq 75$ per cent could be removed from the candidate list without introducing incompleteness in the final XCS cluster sample. We described how we will be able to reduce, by a half, the number of contaminating objects entering candidate$^{300}$ list in future (by improving the checks of reduced \textit{XMM} images before {\sc Xapa} is run).

Overall, we feel that {\sc XCS-Zoo}, was a very worthwhile exercise. It has allowed us to efficiently identify several hundreds of X-ray clusters and has highlighted areas where the XCS candidate selection needs to be improved. We are planning to run {\sc XCS-Zoo} again using additional (to NXS, S82 and DR7) datasets such as SDSS DR8, DSS, CFHTLS, VISTA-Video and UKIDSS-DXS\footnote{www.cfht.hawaii.edu/Science/CFHLS; www.eso.org/vista; www.ukidss.org}. For the next generation of {\sc XCS-Zoo} we will include all candidates (i.e. not impose the $>100$ count threshold on \textit{Zoo}$^{\rm DR8}$, as it was in \textit{Zoo}$^{\rm DR7}$).

In Section~\ref{Litz confirmation} we described both how we interrogated the literature in order to confirm an additional \nQCONTn\, clusters and how we used our own spectroscopic follow-up to confirm \nQCONTm\, more. This process allowed us to confirm \nQCONTh\, \textit{bronze} clusters from {\sc XCS-Zoo} (Fig.~\ref{fig:SDSSBronze}), and several of those clusters have been included in the preliminary statistical subsample (Section~\ref{subsamp: statsam}). The remaining 97 clusters were not part of any {\sc XCS-Zoo} and so they have a heterogeneous selection. These 97 have limited use with regard to statistical studies based on X-ray selection, however they still have value for other purposes, e.g. the study of individual interesting clusters (such as those at high redshift or under going mergers) or the interpretation of optically selected cluster catalogs. We note that of these 97 clusters, 87 have $T_{\rm X}$ measurements in XCS-DR1  (whereas only 37 had a listed $T_{\rm X}$ value in BAX at the time of writing).


In Section~\ref{targetcheck} we described checks that were made to ensure that `target' clusters did not enter the XCS-DR1 sample; seven confirmed clusters were removed from XCS-DR1 as a result of these checks.

\subsection{The cluster catalogue}
\label{disc: cluster cat}

In Section~\ref{table columns}, we presented XCS-DR1 in the form of a table listing cluster attributes such as position, redshift and X-ray temperature. In an associated webpage (http://xcs-home.org/datareleases), we also provide similar information together with optical and X-ray (colour-composite and greyscale) images. In Section~\ref{best z} we described the hierarchy used to select the best redshift for a cluster, if more than one estimate was available for it. In Section~\ref{photoz error}, we provided an estimate of the CMR-redshift errors and catastrophic failure rates. We found the errors to be similar to those obtained by other authors using similar, single-colour, techniques. As demonstrated in S09, errors and failure rates at the measured values should not significantly impact our ability to derive meaningful cosmological or evolutionary parameters. 

We have found, using NED, that \nCCATe\, (approximately half) of the XCS-DR1 clusters have been matched with previously catalogued clusters. This large fraction is a reflection of the fact that much of our redshift follow-up to date (June 2011) has come from the SDSS, and there are by now several optically selected cluster catalogues covering the SDSS footprint, e.g. MaxBCG, \citet{Koester-2007b} and Cut-and-Enhance, \citet{2002AJ....123.1807G}. As our optical follow-up continues, the percentage of previously reported clusters will fall, as evidenced by the fact that less than 20 per cent  (\nCCATf\, of \nQCONTe) of the \textit{bronze} objects from {\sc XCS-Zoo} have matches to previously known clusters. Figure~8 of S09 shows that XCS samples the cluster population well beyond the redshifts ($z<0.6$) accessible to SDSS photometry, so the \textit{bronze} category from \textit{Zoo}$^{\rm DR7}$ likely contains many of the distant clusters detected by XCS. It is noteworthy that few of the previously catalogued clusters in XCS-DR1 have been confirmed as X-ray clusters before.  For example, we have compared XCS-DR1 to the BAX\footnote{http://bax.ast.obs-mip.fr/} database and found only \nCCATg\, matches. Of these, on 56 (roughly a third) have previously published temperatures, i.e. XCS-DR1 presents $T_{\rm X}$ values for  \nCCATh\, clusters for the first time. Indeed, there are only 394 clusters in BAX within the XCS-DR1 redshift ($\nABSd<z<\nABSe$) and temperature ($\nABSfi < T_{\rm X} < \nABSf$ keV) ranges, compared to the \nABSc\, values released herein.

\begin{figure}
\includegraphics[width=9.0cm]{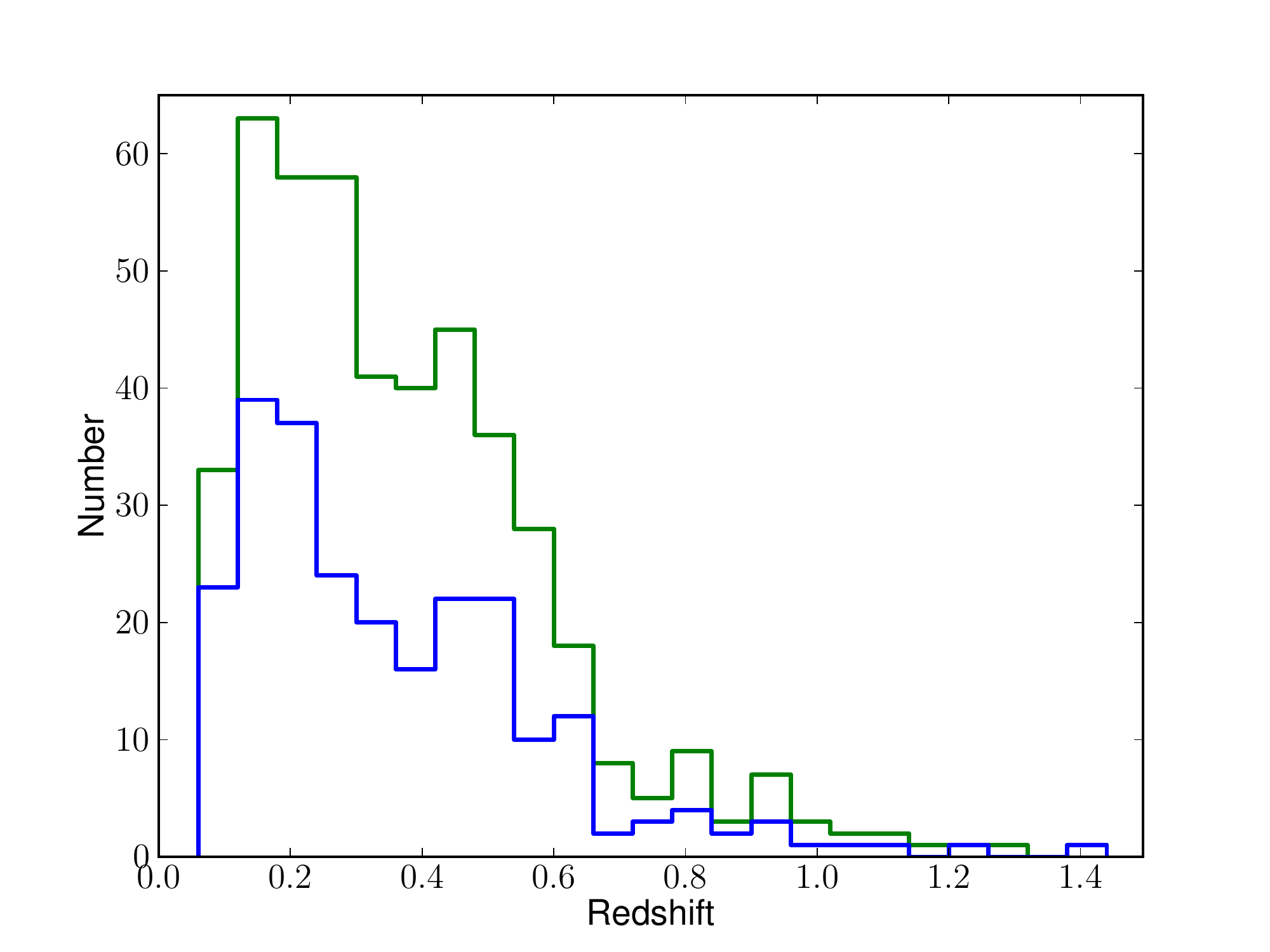}
\caption{The redshift distribution for the \nABSb\, clusters with measured redshifts in XCS-DR1. The green line represents the total sample, while the blue line represents clusters$^{300}$.}
\label{zfunc-new}
\end{figure}
\begin{figure}
\includegraphics[width=9.0cm]{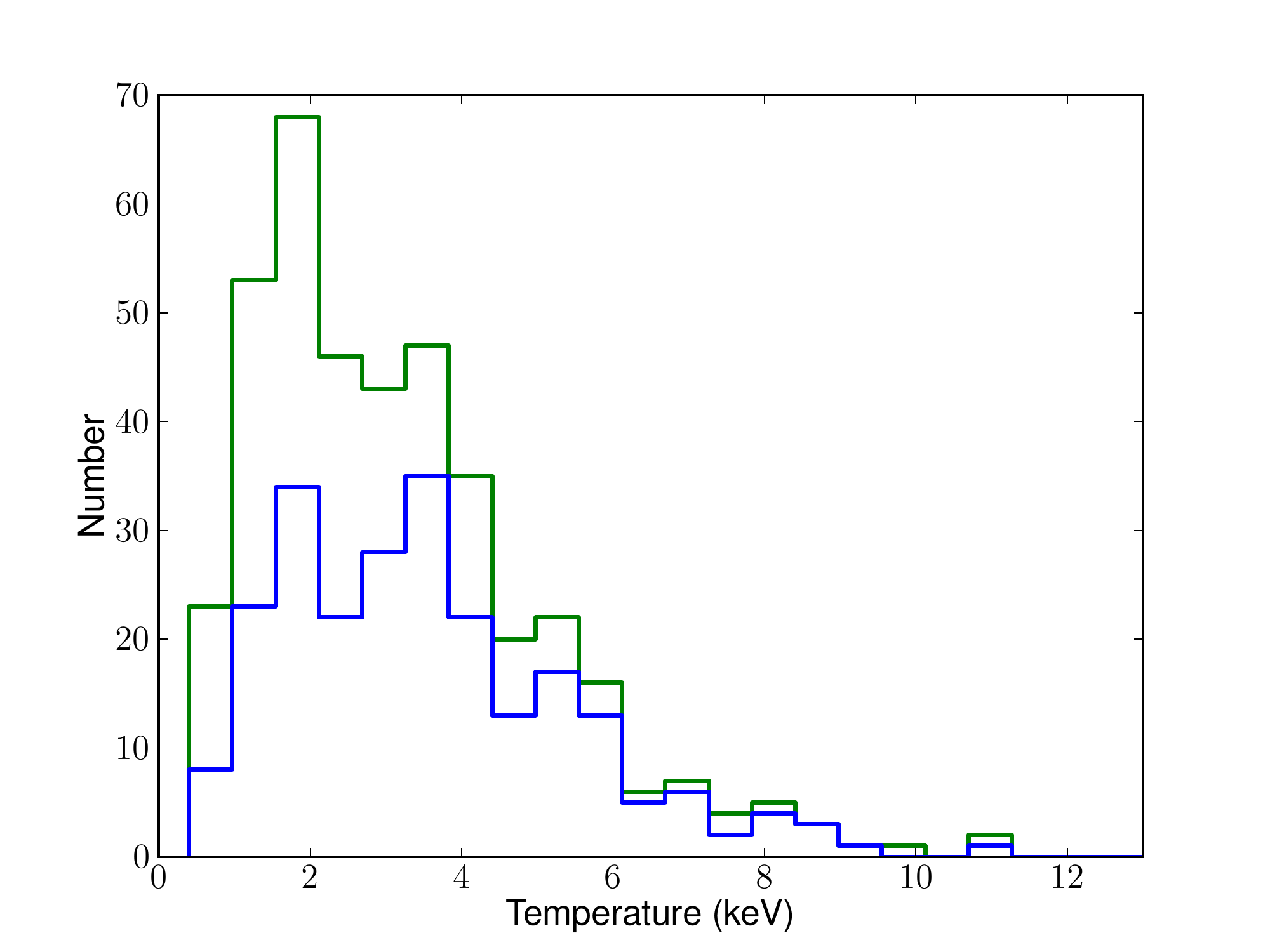}
\caption{The temperature distribution for the \nABSc\, clusters with measured X-ray temperatures in XCS-DR1. The green line represents the total sample, while the blue line represents clusters$^{300}$.}
\label{tfunc-new}
\end{figure}
\begin{figure*}
\includegraphics[scale=0.65]{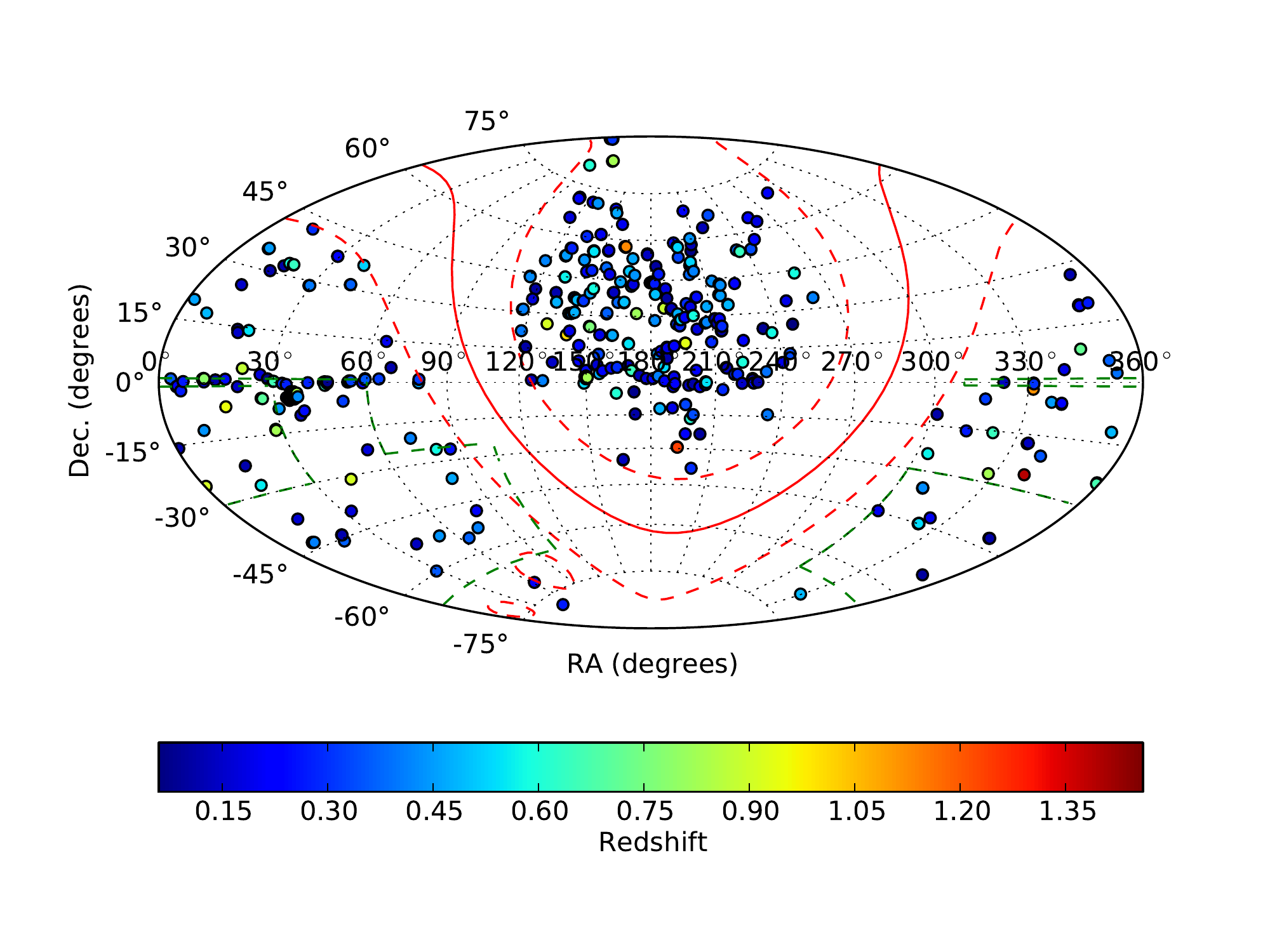}
\caption{The distribution on the sky of the \nABSb\, clusters with measured redshifts in XCS-DR1. The green hashed lines represent the footprint of the Dark Energy Survey. The colours of the dots represent the redshift of the cluster, as indicated.}
\label{zfunc-aitoff}
\end{figure*}
\begin{figure*}
\includegraphics[scale=0.65]{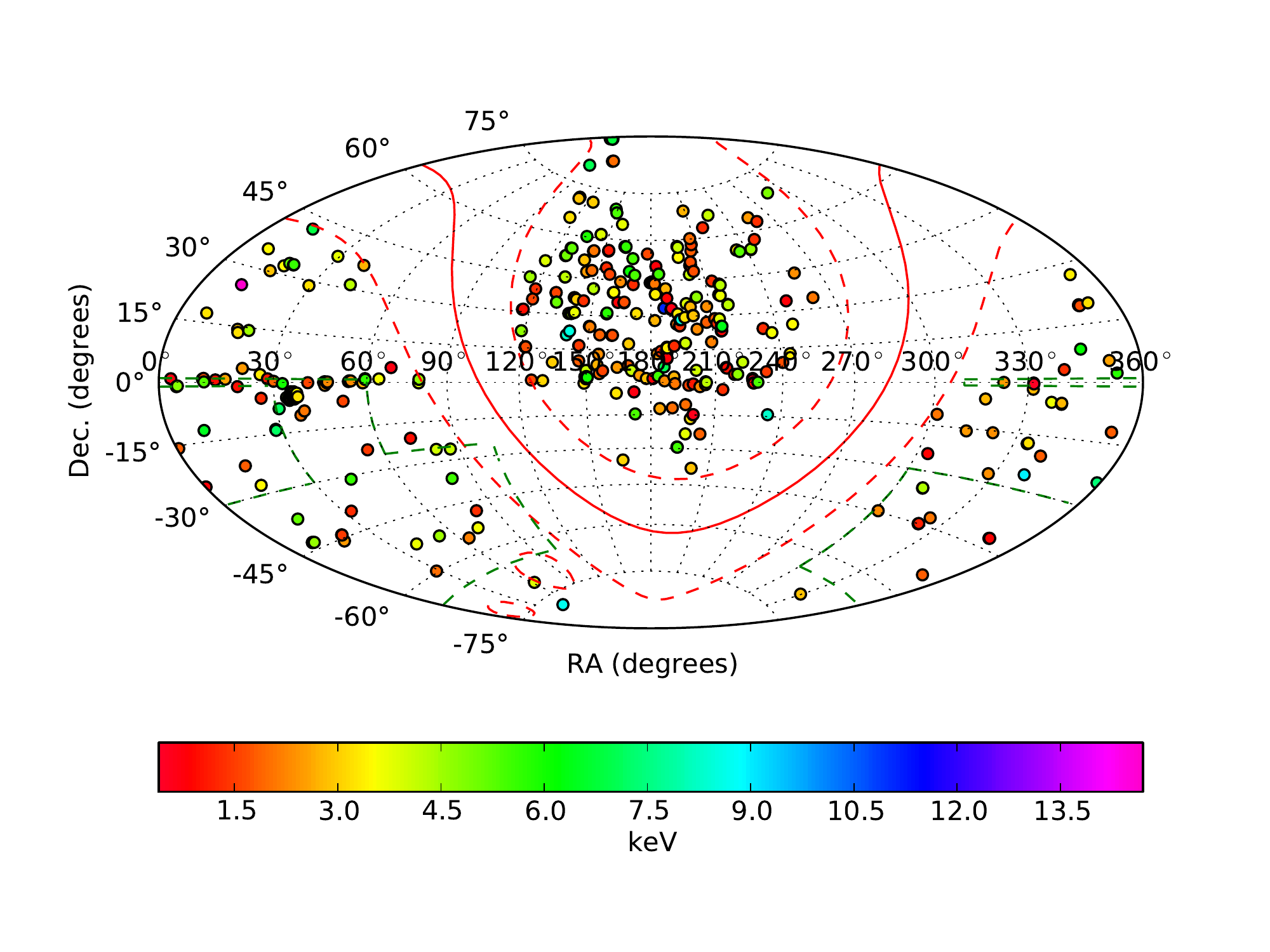}
\caption{The distribution on the sky of the \nABSc\, clusters with measured X-ray temperatures in XCS-DR1. The green hashed lines represent the footprint of the Dark Energy Survey. The colours of the dots represent the temperature of the cluster, as indicated.}
\label{tfunc-aitoff}
\end{figure*}
\begin{figure}. 
\includegraphics[width=9.0cm]{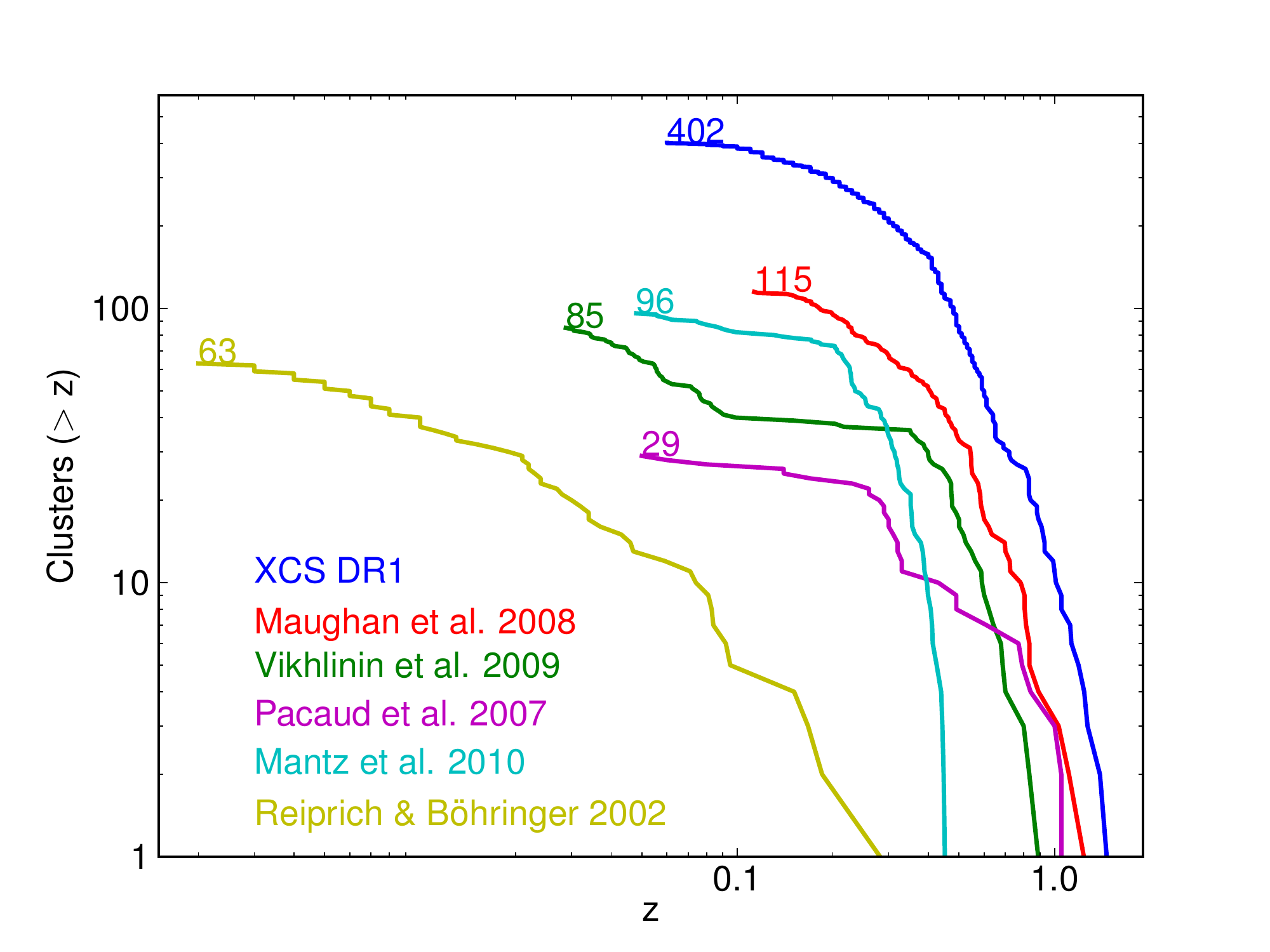}
\caption{A comparison of the redshift distributions of XCS-DR1 clusters with previous data releases of X-ray cluster temperatures. The coloured numbers indicate the total number of clusters in the respective sample.}
\label{z-ns}
\end{figure}
\begin{figure}
\includegraphics[width=9.0cm]{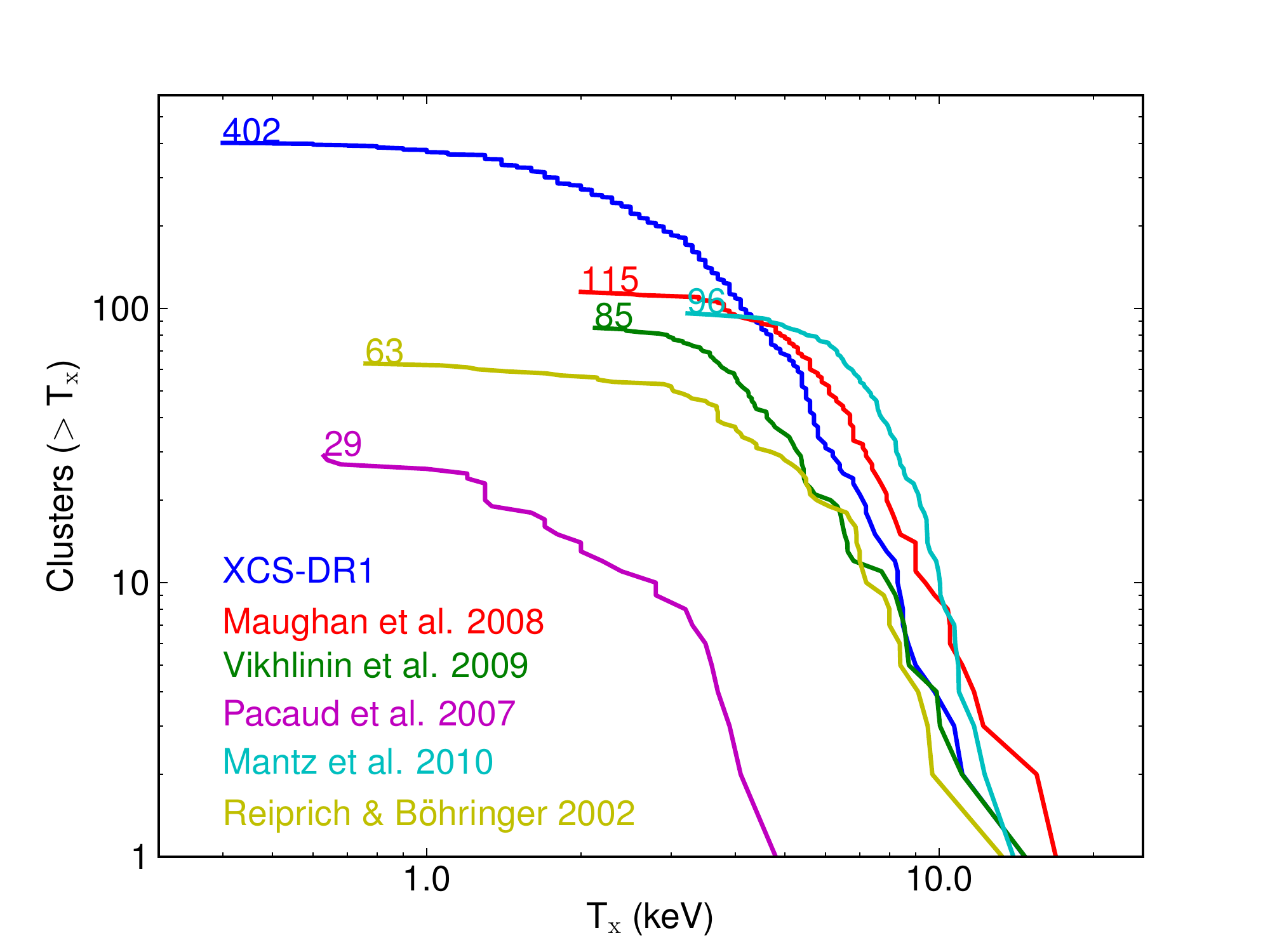}
\caption{As for Fig.~\ref{z-ns}, but for the temperature distributions.}
\label{t-ns}
\end{figure}
\begin{figure}
\includegraphics[width=9.0cm]{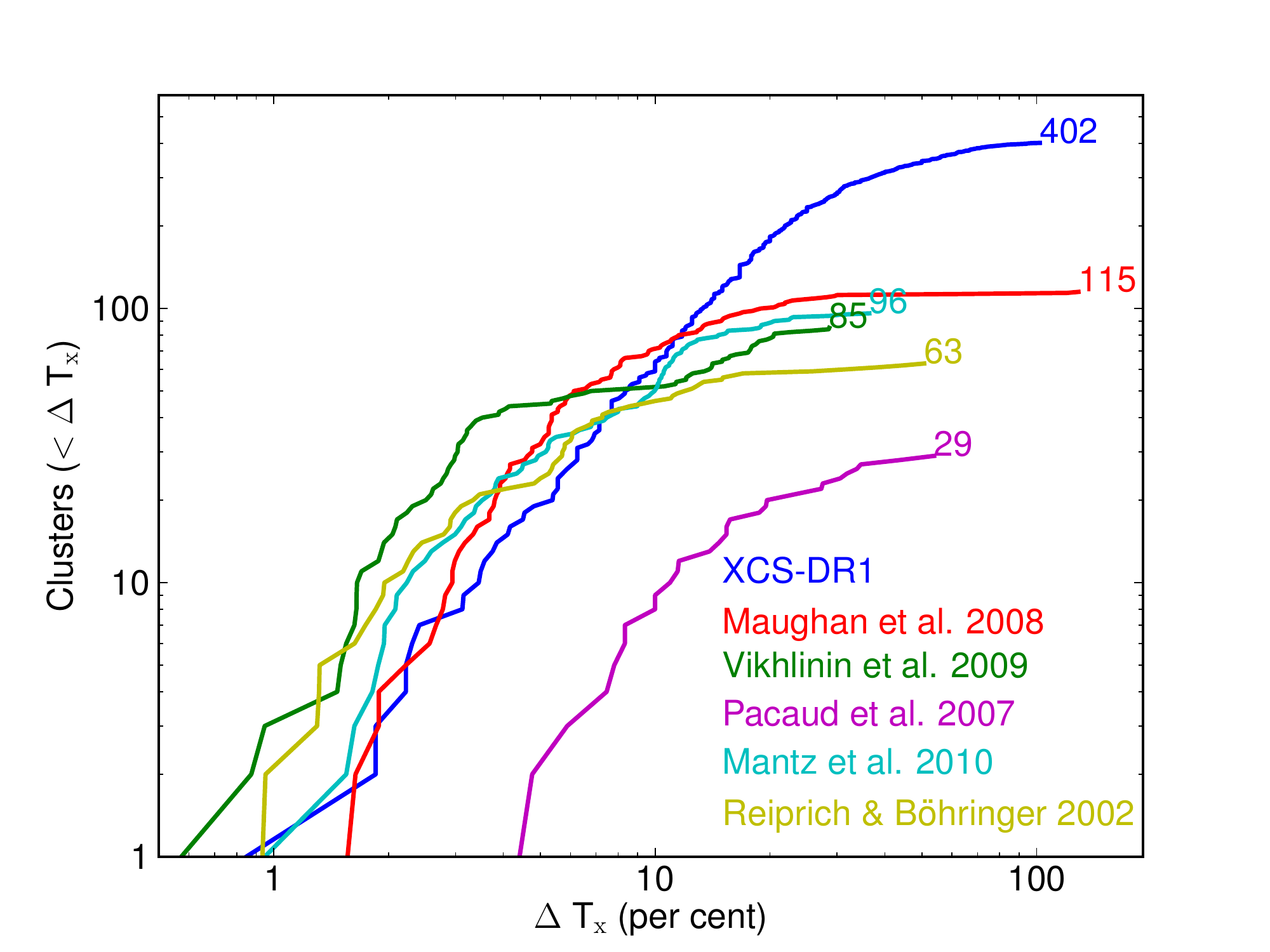}
\caption{As for Fig.~\ref{z-ns}, but for the distribution of 1$\sigma$ temperature errors.}
\label{dt-ns}
\end{figure}

The XCS-DR1 cluster sample is distributed across the entire extra-Galactic sky and spans a wide range of redshift (Fig.\ref{zfunc-new}, and \ref{tfunc-new}, ~\ref{zfunc-aitoff}, and \ref{tfunc-aitoff}). In Fig.~\ref{z-ns},~\ref{t-ns}, and \ref{dt-ns} we compare the XCS-DR1 redshifts and temperatures (values and errors) to previous data releases that have included X-ray cluster temperatures  (\citealt{Reiprich2002, Pacaud07, Maughan08, Vikhlinin09,  Mantz10}). These show that XCS-DR1 not only contains many more cluster temperatures than previous work, but that it also probes higher redshifts and has a greater fraction of lower temperature systems.

With the list of \nABSa\, confirmed clusters in place, we are continuing to improve the parameters we measure for each of them. For example, we are gathering additional spectroscopic data, and investigating the use of multi-colour CMR techniques, to improve the accuracy of the cluster redshifts. We also continue to monitor the literature to ensure we are using the best available published redshifts (rather than relying only on automated NED searches). Obtaining additional \textit{XMM} observations of both those clusters$^{300}$ with large (i.e. $>20$ per cent) $T_{\rm X}$  errors, and those 15 clusters$^{300}$ that failed the $T_{\rm X}$-pipeline entirely, will improve our X-ray temperature measurements. Follow-up of clusters that lie over CCD chip boundaries or on the edge of the field of view (see examples in Fig.~\ref{mt-montage}), would allow us to improve luminosity measurements. 


We acknowledge that some XCS-DR1 clusters will suffer from contamination from line-of-sight, or embedded, point sources (including AGN and cool cores).  In practice, contamination by high signal-to-noise point sources is not a significant problem for XCS-DR1, because such sources are distinguished by {\sc Xapa} from the cluster emission, e.g. in XMMXCS J011140.3$-$453908.0 and XMMXCS J022726.7$-$043209.1. Rather, it is contamination by low signal-to-noise point sources that concerns us; we have previously demonstrated (in \citealt{Hilton-2010}, using \textit{Chandra} follow-up) that the flux from XCS-DR1 cluster XMMXCS J221559.6$-$173816.2 ($z=1.46$) was contaminated at the 15 per cent level by two point sources that had not been detected by {\sc Xapa} (\textit{Chandra} is significantly more sensitive to point sources than \textit{XMM}, although the reverse is true for extended emission). Correcting for those point sources decreased the measured temperature by 2.4 keV (to $T_{\rm X}=4.1^{+0.6}_{-0.9}$ keV) and the luminosity by 33 per cent. To determine how common such low-level contamination might be (and the typical impact it has on derived parameters), we are undertaking an exercise that will make use of observations of XCS-DR1 clusters  (and  \textit{bronze} candidates$^{300}$) in the \textit{Chandra} Data Archive\footnote{cxc.harvard.edu/cda}. Using the \textit{Chandra} Simple Image Access service, we have determined that \nDISCf\, (\nDISCw) of the XCS-DR1 clusters (\textit{bronze} candidates$^{300}$) are covered by \textit{Chandra} observations (through to the end of 2009). The \nDISCf\, include \nDISCg\,  clusters$^{300}$.

\subsection{Selected subsamples of XCS-DR1 clusters}
\label{disc: subsamples}

Finally, we describe seven subsamples of the XCS-DR1 clusters that have particular scientific applications. In each description we include methods that are being, or could be, used to improve and/or extend the respective subsample.

\subsubsection{High-redshift XCS-DR1 Clusters}
\label{subsamp: high z}

There are \nABSj\, clusters in Table~\ref{BigTable} with $z>\nABSk$. Of these clusters, all have been spectroscopically confirmed and are accompanied by $T_{\rm X}$ measurements. By comparison, \nDISCh\, (\nDISCi\, with $T_{\rm X}$ measurements) $z>\nABSk$ clusters were registered on the BAX database at the time of writing (June 2011). 

Although the BAX database is not completely up to date, e.g. the $z=1.56$ XDCP cluster (XMMU J1007.4$+$1237, \citealt{Fassbender11}) was not included at the time of writing, this comparison still demonstrates that the XCS-DR1 $z>1.0$ cluster sample is the largest based on a single selection technique (the \nDISCh\, BAX clusters had been compiled from \nDISCj\, different publications, with $T_{\rm X}$ measurements from seven different publications). Most of the XCS-DR1 $z>\nABSk$ clusters were previously known X-ray clusters, e.g. XMMXCS J223520.4$-$255742.1 at $z=1.39$ \citep{Mullis-2005, Rosati-2009, Jee-2009}. However we highlight the two XCS discoveries:  XMMXCS J221559.6$-$173816.2 ($z=1.46$, previously published by us in \citealt{Stanford:2006}) and XMMXCS J091821.9$+$211446.0 ($z=1.01$, a spectroscopic redshift based on 16 galaxies, Fig.~\ref{J0918}). 

We plan to exploit the high-redshift XCS-DR1 clusters to extend our previous (\citealt{collins-2009-458,Stott:2010,Hilton-2010}) studies of galaxy, particularly BCG, evolution. Given the importance of high-$z$ clusters to evolution studies, we will request additional X-ray follow-up (\textit{XMM} and \textit{Chandra}) of some of these clusters, in order to improve the signal to noise and the spatial resolution. We are also working to extend the size of the XCS high-$z$ sample. We are doing this in a number of ways, including: continuing the spectroscopic follow-up, using the Keck and Gemini telescopes, of promising high-$z$ candidates that were highlighted during {\sc XCS-Zoo} (regardless of the number of detected counts); using UKIDSS-DXS and VISTA-Video surveys to select additional high-$z$ candidates; and exploiting redshifts measured by other teams, such as XDCP, for \textit{XMM} clusters as they enter the literature.

\subsubsection{High-temperature XCS-DR1 clusters}
\label{subsamp: high T}

There are \nABSl\, clusters in XCS-DR1 with $T_{\rm X}>\nABSm$\,keV. This is a much smaller number than available on BAX (\nDISCo\, at the time of writing, June 2011); however these clusters are still useful to the Sunyaev--Zeldovich community (SZ, \citealt{SZ-1972}) because typically it is only the $T_{\rm X}>\nABSm$\,keV clusters that can be detected via their SZ signal (using current instrumentation).

There are more $T_{\rm X}>\nABSm$ keV clusters in BAX than in XCS-DR1 at all redshifts, but we note that at $z>\nDISCk$, the numbers are comparable:  \nDISCl\, clusters from XCS-DR1, compared to \nDISCn\, in BAX. Of these, most (\nDISCm) were not previously catalogued by BAX. Moreover, the \nDISCn\, BAX clusters were drawn from a large number of different publications, whereas the XCS-DR1 sample is based on a single selection technique and a single $T_{\rm X}$ analysis method. 

The XCS collaboration does not have direct access to SZ experiments, but targets have been supplied to the APEX and AMI teams \citep{Schwan-2010,Zwart-2008}, and several XCS-DR1 clusters have already been studied by the AMI SZ experiment (AMI Consortium: Shimwell et al. 2011, in prep). In a separate publication (Viana et al. 2011, in prep.) we present a subsample of XCS-DR1 clusters (including some at $T_{\rm X}<\nABSm$ keV) that we predict will be detectable by the Planck SZ survey \citep{PlanckSZ}.

Given the importance of high temperature clusters to SZ studies, and the fact that the accuracy of  $T_{\rm X}$ measurements drops with increasing temperature (see Figure~16 in LD10), it would be worthwhile to obtain additional X-ray observations of this sub-sample (regardless of the {\sc Xapa} count value), in order to increase the $T_{\rm X}$ precision. Re-observing off-axis clusters at the \textit{XMM} aim point, and/or with \textit{Chandra}, will also increase the spatial resolution and allow us to correct for point source contamination (uncorrected AGN contamination will artificially raise the measured $T_{\rm X}$).

\subsubsection{Low-temperature XCS-DR1 clusters}
\label{subsamp: low T}

There are \nABSn\, clusters in XCS-DR1 with $\nABSfi<T_{\rm X}<\nABSo$ keV in the redshift range $\nABSd<z<\nABSe$. This is dramatically more than in BAX (which lists only \nDISCp\, such systems). We are already exploiting this unique data set to investigate AGN-ICM feedback mechanisms.

\subsubsection{High signal-to-noise XCS-DR1 clusters}

\label{subsamp: high SN}

\begin{figure*}
\includegraphics[scale=1.1]{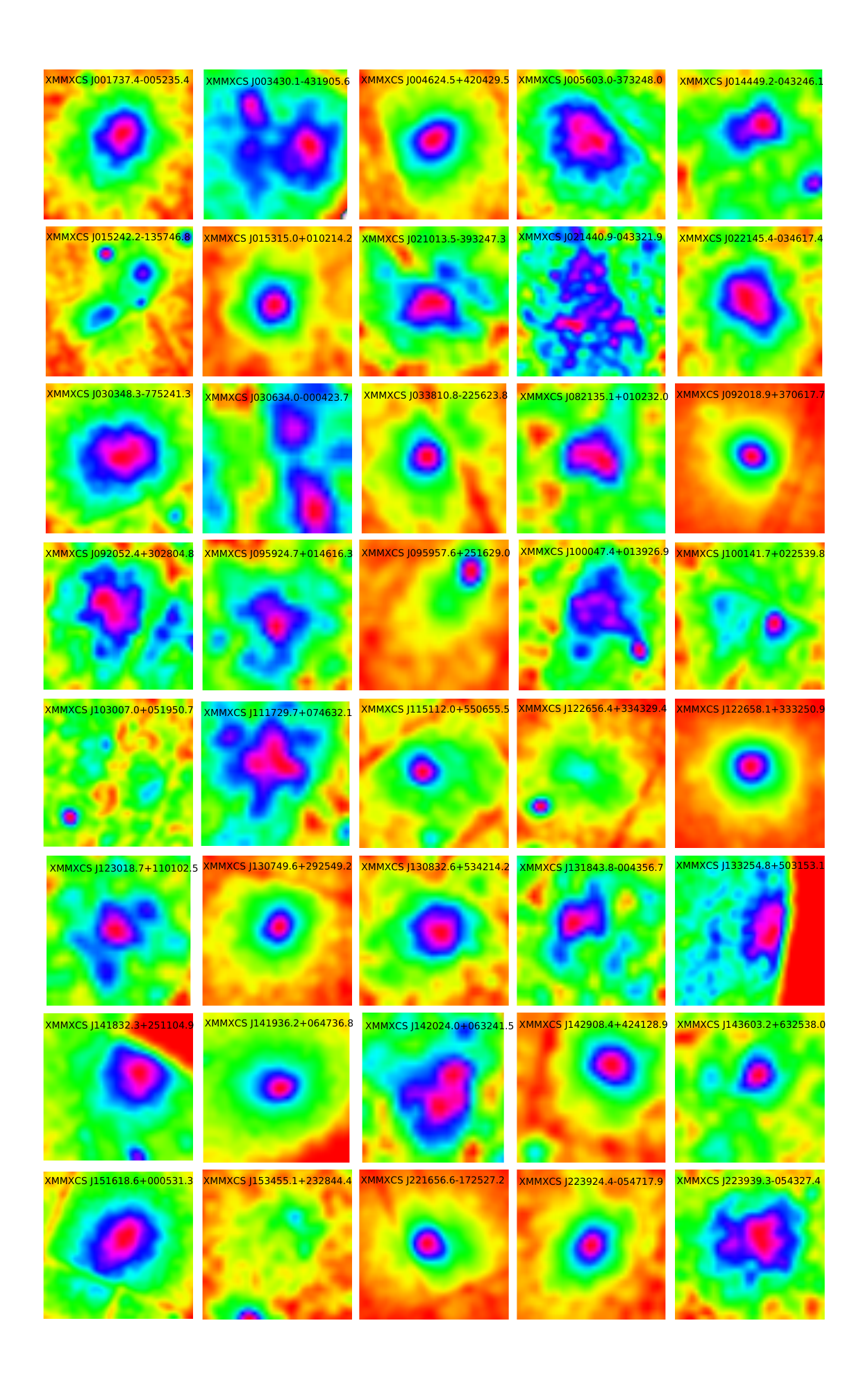}
\caption{Forty XCS-DR1 clusters that were detected with sufficient counts that it should be possible measure X-ray temperature profiles. The names of the clusters are as indicated within the sub-panels. Note that these are count, rather than count rate maps, i.e. they have not been corrected for variations in the exposure map. Several of the clusters lie over chip gaps, while others fall close to the edge of the field of view, hence some of the morphologies are artificially distorted. The ObsID images on the XCS-DR1 webpage provide a clearer impression of the exposure map variations.}
\label{mt-montage}
\end{figure*}

The XCS-DR1 $T_{\rm X}$ values come primarily from the discovery data (the exception being those clusters that are both \textit{XMM} targets and serendipitous detections, Section~\ref{targetcheck}) and so the $T_{\rm X}$ errors tend to be larger than the comparison samples, especially at the high-temperature/low-count end (Fig.~\ref{dt-ns}). However,  this is not always the case and, in particular, we note that there are \nABSr\, systems (Fig.~\ref{mt-montage}) from which it should be possible to measure $T_{\rm X}$, to better than 15 per cent accuracy, in three or more radial bins (where these expectations are based on the results in LD10 concerning  $T_{\rm X}$ accuracy as a function of number of counts and $T_{\rm X}$). The 40 clusters featured in Fig.~\ref{mt-montage} comprise of 23 clusters ranging in temperature from $2.6 <T_{\rm X}<11.1$\,keV, and 17 groups ranging from $1.0 <T_{\rm X}<1.8$\,keV.

The measurement of $T_{\rm X}$ profiles will permit the measurement of cluster masses, under the assumption of hydrostatic equilibrium. This is important since mass calibration will be required before XCS-DR1 clusters can be used to constrain cosmological parameters (S09). We hope to increase the number of XCS-DR1 clusters with $T_{\rm X}$ profiles by securing additional \textit{XMM} observations. In particular, we wish to re-observe some systems that are representative of the $L_{\rm X}-T_{\rm X}$ relation in the $1.8<T_{\rm X}<2.6$ keV range (i.e. where there is a gap in the current high signal-to-noise sample).

\subsubsection{XCS-DR1 clusters in the Stripe 82 footprint}
\label{subsamp: Stripe82}

There are \nDISCq\, clusters in XCS-DR1 that fall within the Stripe 82 (S82) co-add region of SDSS DR7. Of these, \nABSp\, have measured X-ray temperatures. The S82 region is of interest to many cluster scientists, primarily because it provides public high-quality imaging in several colours ($\sim$2 magnitudes deeper than regular SDSS) over a considerable area of sky. Optical cluster catalogues have been produced from S82, e.g. \citet{Geach-2011}, but prior to XCS-DR1, there were only been a handful of known X-ray clusters to use for S82 catalogue validation.

The Stripe 82 region will be re-observed by DES, to similar depths and in the same bands. Therefore, the \nABSp\, XCS-DR1 clusters with $T_{\rm X}$ measurements can be used immediately to investigate how DES-like cluster richnesses will correlate with the more reliable mass proxies.

\subsubsection{XCS-DR1 clusters in the DES footprint}
\label{subsamp: DES}

In total, there are \nDISCs\, XCS-DR1 clusters in the DES footprint (including the \nDISCq\, in S82). Of these, \nABSq\, have measured X-ray temperatures. All these clusters are worthy of further study (optical and X-ray) in order to support DES cluster science. The value of X-ray information about DES clusters has been demonstrated by \citet{Wu-2010}, who showed that with 200 follow-up observations, the dark energy figure of merit could be improved by 50 per cent. Although, with XCS-DR1, we have only been able to provide half that number, we note that there are several hundred more XCS clusters candidates in the DES footprint that have yet to be identified (as clusters or contaminants). Most of them have not been included in {\sc XCS-Zoo} before, but their identification should be straightforward once DES photometry (which reaches a depth comparable to S82) is available. Given that DES photometry will cover several bands, it should be possible to measure accurate CMR-redshifts for all DES-identified clusters. 

We cannot at this stage predict how many of the DES-identified clusters will yield $T_{\rm X}$ values, but we note that the number of unidentified candidates$^{300}$ in the DES footprint is \nDISCt. In addition, we are applying the XCS X-ray analysis pipelines to an additional $\simeq 40$ clusters with \textit{XMM} detections. These additional clusters are not in our candidate list, because they were the targets of their respective ObsID, however, they are still very useful to DES because they will help us reach the \citet{Wu-2010} target of 200. These 40 are likely to be particularly useful to DES since many will have been detected with sufficient signal-to-noise to yield mass estimates via the hydrostatic equilibrium method.

\subsubsection{XCS-DR1 clusters for statistical studies}
\label{subsamp: statsam}

\begin{figure*}
\begin{center}
{    
\includegraphics[scale=0.40]{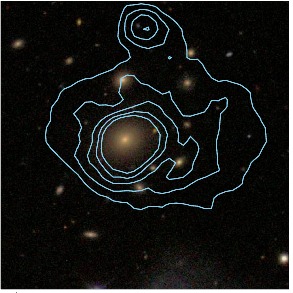}
}
\subfigure
{
\includegraphics[scale=0.40]{figures/XMMXCSJ010858_7+132557_7_3by3_sdss_contours.jpg}
}
\subfigure
{
\includegraphics[scale=0.40]{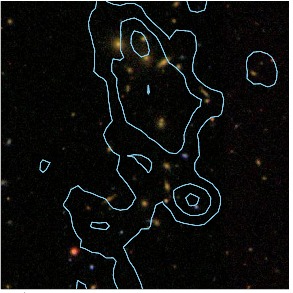}
}
\subfigure
{
\includegraphics[scale=0.40]{figures/XMMXCSJ001737_4-005235_4_3by3_sdss_contours2.jpg}
} 
\\
{
\includegraphics[scale=0.40]{figures/XMMXCSJ092018_9+370617_7_3by3_sdss_contours.jpg}
}
\subfigure
{
\includegraphics[scale=0.40]{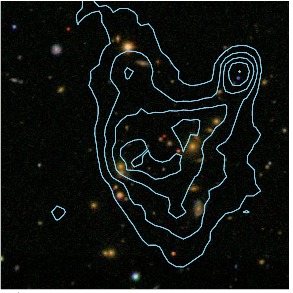}
}
\subfigure
{
\includegraphics[scale=0.40]{figures/XMMXCSJ083454_8+553420_9_3by3_sdss_contours.jpg}
}
\subfigure
{
\includegraphics[scale=0.40]{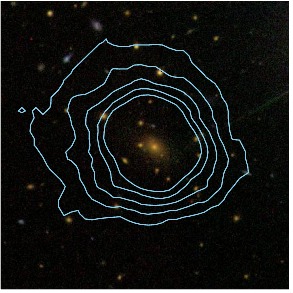}
} 
\\    
{
\includegraphics[scale=0.40]{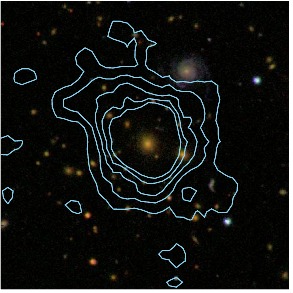}
}
\subfigure
{
\includegraphics[scale=0.40]{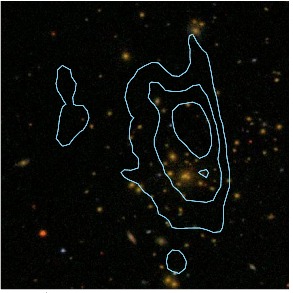}
}
\subfigure
{
\includegraphics[scale=0.40]{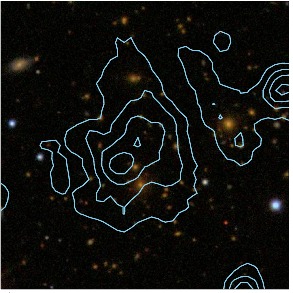}
}
\subfigure
{
\includegraphics[scale=0.40]{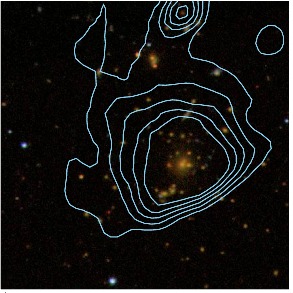} 
}
\\
{
\includegraphics[scale=0.40]{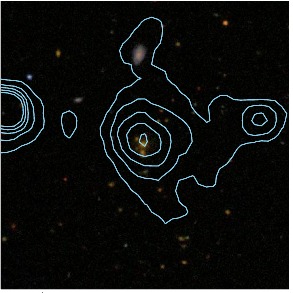}
}
\subfigure
{
\includegraphics[scale=0.40]{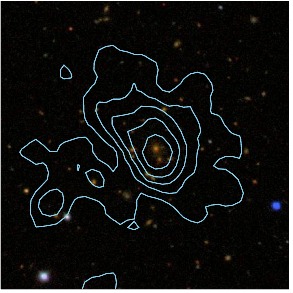}
}
\subfigure
{
\includegraphics[scale=0.40]{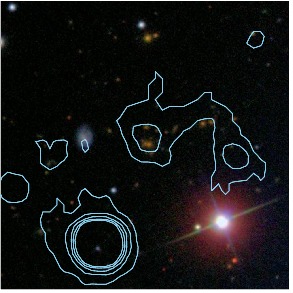}
}
\subfigure
{
\includegraphics[scale=0.40]{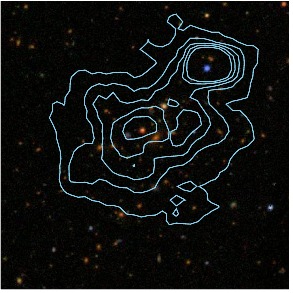} 
}
\caption{A selection of optically confirmed XCS clusters as imaged by SDSS DR7 and classified as \textit{gold} in \textit{Zoo}$^{\rm DR7}$ (Section~\ref{ClusterZoo}). False colour-composite images are $3'\times3'$ with X-ray contours overlaid in blue. From left to right and top to bottom, the compilation displays the clusters: XMMXCS J115112.0$+$550655.5 at $z=0.08$; XMMXCS J010858.7$+$132557.7 at $z=0.15$; XMMXCS J123019.6$+$161634.1 at $z=0.20$; XMMXCS J001737.4$-$005235.4 at $z=0.21$; XMMXCS J092018.9$+$370617.7 at $z=0.21$; XMMXCS J100047.4$+$013926.9 at $z=0.22$; XMMXCS J083454.8$+$553420.9 at $z=0.24$; XMMXCS J130749.6$+$292549.2 at $z=0.24$; XMMXCS J170041.9$+$641257.9 at $z=0.24$; XMMXCS J133254.8$+$503153.1 at $z=0.28$; XMMXCS J092052.4$+$302804.8 at  $z=0.29$; XMMXCS J141832.3$+$251104.9 at $z=0.29$; XMMXCS J105318.5$+$572043.7 at $z=0.34$; XMMXCS J115824.6$+$440533.9 at $z=0.41$; XMMXCS J153629.7$+$543920.8 at $z=0.41$; and XMMXCS J111515.6$+$531949.5 at $z=0.47$.} 
\label{sdssgoldcompilation}
\end{center}
\end{figure*}

Although the optical follow-up of XCS$^{300}$ is still ongoing, we have nevertheless been able to define a subsample of XCS-DR1 clusters that should be sufficiently complete to be suitable for preliminary statistical studies. For this we have only used clusters$^{300}$ that were classified by {\sc XCS-Zoo} and optically confirmed. We confined the sample to the redshift range that should yield CMR-redshifts from the respective imaging (i.e $z<0.3$ for \textit{Zoo}$^{\rm DR7}$ and $z<0.6$ for  \textit{Zoo}$^{\rm NXS}$ and \textit{Zoo}$^{\rm S82}$ --- refer to Fig.~\ref{fig:nxscompilation}, ~\ref{fig:SDSSBronze2Gold} and \ref{sdssgoldcompilation} to see how the galaxy density changes with cluster redshift and survey depth). We also impose a lower redshift cut of $z>0.1$, because this is the minimum allowed by the CMR algorithm.

Setting these limits, we have selected a total of \nABSs\, clusters$^{300}$, of which \nDISCv\, come from \textit{Zoo}$^{\rm DR7}$ and an additional \nDISCu\, come from \textit{Zoo}$^{\rm NXS}$ or \textit{Zoo}$^{\rm S82}$. To put this sample into context, we can compare it to the influential study by \citet{Vikhlinin09} that used 88 clusters (49 in the range $0.025<z<0.2$ and 39 in the range $0.35<z$) to derive constraints on dark energy parameters. 


We hope the sample will be widely used by the community, and have indicated the \nABSs\, members in Table~\ref{BigTable}. However, we stress that it is not applicable for all types of statistical studies, because it is not complete: some clusters in the selected redshift range will not have been `confirmed' yet and so do not appear in XCS-DR1. We do not know how many such clusters there are, but we estimate the number to be $\simeq 50$; there are 40 \textit{bronze} clusters$^{300}$ and 9 \textit{other} candidates$^{300}$ with CMR-redshifts in the respective ranges (where the \textit{other} candidates$^{300}$ are in the sub-categories that  `require additional follow-up', Section~\ref{Zoo-other}). Assuming all 49 are clusters, then the current sample of  \nABSs\, is only 68 per cent complete. Moreover, this sample cannot be used without reference to the XCS survey selection function (LD10) for science applications that require the volume density to be known. In future publications, we will apply this sample to a variety of investigations, e.g. the derivation of cosmological parameters and the measurement of cluster scaling relations. The scaling relations we plan to examine are those between: X-ray luminosity and $T_{\rm X}$;  optical richness and $T_{\rm X}$; and halo occupation number and $T_{\rm X}$.


We look forward to increasing the size, and completeness level, of future `statistical' sub-samples of XCS clusters using the Dark Energy Survey. The combination of DES and XCS will yield a homogeneously selected set of confirmed X-ray clusters that is at least twice the size of the current statistical sub-sample. It is important to note, in this context, that the fraction of \textit{bronze} and \textit{other} systems in the similarly (to DES) deep S82 region is significantly lower than that in the DR7 region. Extrapolating from the \textit{Zoo}$^{S82}$ results, we expect completeness levels of XCS clusters samples in the DES region to be at least 80 per cent. If the Pan-Starrs4\footnote{pan-starrs.ifa.hawaii.edu} project is successful, then even larger samples could be gathered because that optical survey will cover the whole sky north of $-30$ deg, to similar depths to DES.

\section{Conclusions}
\label{conclusions}

We have presented the first data release from the XMM Cluster Survey (XCS-DR1). This consists of \nABSa\, optically confirmed X-ray clusters serendipitously detected in \textit{XMM} archival imaging. Optical confirmation was established in one or more of three ways: through multi-object spectroscopy; by matching XCS candidates to clusters in the literature; or by visual inspection of optical CCD images via an exercise referred to as {\sc XCS-Zoo}. Redshifts for the clusters were derived from a variety of spectroscopic and photometric sources, namely public archives, the literature, and our own follow-up campaigns. X-ray temperatures and luminosities were measured for those clusters detected with sufficient signal-to-noise using an automated pipeline. We have established whether the clusters (and/or their X-ray temperature measurements) are new to the literature using comparisons with NED and BAX. Compared to previous data releases of cluster samples with $T_{\rm X}$ information, XCS-DR1 contains more clusters (especially at low temperature) and probes to higher redshifts. The XCS-DR1 catalogue, together with optical (colour-composite and greyscale) and X-ray imaging for each of the XCS-DR1 clusters, is publicly available from http://xcs-home.org/datareleases.

Some key statistics for the XCS-DR1 catalogue of \nABSa\, clusters are as follows: 

\begin{enumerate}

\item Redshifts: \nABSb\, clusters are accompanied with redshift information ($\nABSd<z<\nABSe$). Of these, \nABSi\, are spectroscopic, with most of the remainder coming from the photometric CMR-redshift technique (applied to single-colour optical imaging). Ten of the redshifts exceed $z=\nABSk$ (these include a new spectroscopically-confirmed cluster at $z=1.01$). The CMR-redshift accuracy (and catastrophic failure rates) were found to be $\sigma_z=0.08$ ($\simeq$5 per cent), $\sigma_z=0.03$ ($\simeq$3 per cent) and $\sigma_z=0.03$ ($\simeq$7 per cent) from NXS, SDSS DR7 and Stripe 82 data, respectively. 


\item Temperatures: \nABSc\, clusters are accompanied with X-ray temperature information ($\nABSfi$\,keV$ < T_{\rm X} < \nABSf$\, keV). Of these, \nABSl\, clusters have temperatures above $T_{\rm X}=\nABSm$\,keV (these systems will be particularly useful for SZ studies), and \nABSn\, clusters have temperatures below $T_{\rm X}=\nABSo$\,keV (these systems can be applied to studies of cluster physics and BCG evolution).  A small subset, of \nABSr\, clusters, were detected with sufficient signal-to-noise that mass measurements can be made using temperature profiles (these will be important to cosmology studies, as they will aid the mass calibration of XCS). All clusters presented with $T_{\rm X}$ values are also presented with $L_{\rm X}$ measurements.

\item New discoveries/measurements:  \nABSg\, clusters were not previously catalogued in the literature, and \nABSh\, of the X-ray temperature measurements were not previously catalogued in BAX.

\item Preliminary statistical subsample: Of the \nABSa\, XCS-DR1 clusters, \nABSs\, can be used, in conjunction with the XCS selection function, for statistical applications such as the derivation of cosmological parameters and the measurement of cluster scaling relations (including those between X-ray luminosity and $T_{\rm X}$, and optical richness and $T_{\rm X}$).

\end{enumerate}

\section*{Acknowledgments}

NM thanks Chris Wegg for for useful discussions on the CMR-redshift method. We are most grateful to the support staff who operate the following observatories and to the organisations that fund them: The ESA XMM-Newton mission. The National Optical Astronomy Observatory (NOAO) consists of Kitt Peak National Observatory near Tucson, Arizona, Cerro Tololo Inter-American Observatory near La Serena, Chile, and the NOAO Gemini Science Center. NOAO is operated by the Association of Universities for Research in Astronomy (AURA) under a cooperative agreement with the National Science Foundation. The Gemini  Observatory, which is operated by the Association of Universities for Research in Astronomy, Inc., under a cooperative agreement with the NSF on behalf of the Gemini partnership: the National Science Foundation (United States), the Science and Technology Facilities Council (United Kingdom), the National Research Council (Canada), CONICYT (Chile), the Australian Research Council (Australia), MinistŽrio da Cincia e Tecnologia (Brazil) and Ministerio de Ciencia, Tecnolog'a e Innovaci—n Productiva  (Argentina). The Keck Observatory. The analysis pipeline used to reduce the DEIMOS data was developed at UC Berkeley with support from NSF grant AST-0071048. The authors wish to recognize and acknowledge the very significant cultural role and reverence that the summit of Mauna Kea has always had within the indigenous Hawaiian community; we are fortunate to have the opportunity to conduct observations from this mountain. The W. M.  Keck Observatory is a scientific partnership between the University of California and the California Institute of Technology, made  possible by a generous gift of the W. M. Keck Foundation. The NTT ESO Telescope at the La Silla Observatory under programmes 077.A-0437, 078.A-0325, 080.A-0024, 081.A-843. The William Herschel Telescope which is  operated on the island of La Palma by the Isaac Newton Group in the Spanish Observatorio del Roque de los Muchachos of  the Instituto de Astrof\'{i}sica de Canarias. 

We also acknowledge the following public archives, surveys and analysis tools: The Sloan Digital Sky Survey (SDSS). Funding for the SDSS and SDSS-II has been provided by the Alfred P. Sloan Foundation, and a wide variety of the Participating Institutions (see www.sdss.org for details). The Digitized Sky Surveys (DSS) produced at the Space Telescope Science Institute under U.S. Government grant NAG W-2166. The NASA/IPAC Extragalactic Database (NED) which is operated by the Jet Propulsion Laboratory, California Institute of Technology, under contract with the National Aeronautics and Space Administration. The \textit{Chandra} Data Archive (cxc.harvard.edu/cda). The IRAF reduction pipeline distributed by the National Optical Astronomy Observatories, which are operated by the Association of Universities for Research in Astronomy, Inc., under cooperative agreement with the National Science Foundation. The X-Ray Clusters Database (BAX) which is operated by the Laboratoire d'Astrophysique de Tarbes-Toulouse (LATT), under contract with the Centre National d'Etudes Spatiales (CNES). Montage, funded by the National Aeronautics and Space Administration's Earth Science Technology Office, Computation Technologies Project, under Cooperative Agreement Number NCC5-626 between NASA and the California Institute of Technology. Montage is maintained by the NASA/IPAC Infrared Science Archive.

Financial support for this project includes: 
The Science and Technology Facilities Council (STFC) through grants ST/F002858/1 and/or ST/I000976/1 (for EL-D, AKR, NM, MHo, ARL and MS), ST/H002391/1 and PP/E001149/1 (for CAC and JPS), ST/G002592/1 (for STK) and through studentships (for NM, HCC), ST/H002774/1 and ST/I001204/1 (for EME)
The University of KwaZulu-Natal (for MHi). 
The Leverhulme Trust (for MHi).
The University of Sussex (MHo, HCC, PD). 
FP7-PEOPLE- 2007-4Ð3-IRG n 20218 (for BH). 
Funda\c{c}\~{a}o para a Ci\^{e}ncia e a Tecnologia through the project PTDC/CTE-AST/64711/2006 (for PTPV).
The South East Physics Network (for RCN, END, WAW). 
The Swedish Research Council (VR) through the Oskar Klein Centre for Cosmoparticle Physics (for MS).  
The RAS Hosie Bequest and the University of Edinburgh (for MD). 
The  U.S. Department of Energy, National Nuclear Security Administration by the University of California, Lawrence Livermore National Laboratory under contract No. W-7405-Eng-48 (for SAS).  
The Greek State Scholarship Foundation, trustee of the Nik. D. Chrysovergis legacy (for LC).
Parts of the manuscript were written during a visit by AKR to the Aspen Physics Center.  


\bibliography{bib}

\appendix

\bsp
\label{lastpage}

\end{document}